\documentclass[12pt]{article}
\usepackage{amsmath, amsthm, latexsym, amssymb, mathrsfs}
\usepackage{amsfonts}
\usepackage{changepage}
\usepackage{fancyhdr}
\usepackage[hidelinks]{hyperref}
\usepackage{pgfplots}
\usepgfplotslibrary{groupplots}
\usepackage{caption}
\usepackage{blkarray}
\usepackage{titlesec}
\usepackage{enumerate}
\usepackage[section]{placeins}  
\usepackage{float}

\newcommand{\numberthis}[1]{\addtocounter{equation}{1}\tag{\theequation}\label{#1}}

\titlespacing*{\section}{0pt}{-10pt}{-5pt}
\titlespacing*{\subsection}{0pt}{-10pt}{-5pt}

\usepackage[backend=bibtex,backref=true]{biblatex}  
\renewbibmacro{in:}{}
\defbibheading{bibliography}{%
  \clearpage
  \thispagestyle{empty}
\begin{center}  
  {\large\textbf{REFERENCES}}%
\end{center}
  \markboth{}{}
}

\newenvironment{proof*}{{\noindent {\it Proof.  \ignorespaces}}}{\hfill$\blacksquare$}

\newtheorem{theorem}{Theorem}

\newtheoremstyle{example}
{3pt}
{3pt}
{}
{}
{\bfseries}
{.}
{0.5em}
{}
\theoremstyle{example}
\newtheorem{example}{Example}
\newtheoremstyle{defn}
{3pt}
{3pt}
{}
{}
{\bfseries}
{.}
{.5em}
{}
\theoremstyle{defn}
\newtheorem{defn}[example]{Definition}

\numberwithin{equation}{section}
\numberwithin{theorem}{section}
\numberwithin{example}{section}

\newcommand{\email}[1]{\href{mailto:#1}{#1}}

\title{A Study on a New Method of Dynamic Aperture Enlargement\thanks{This material is based in part upon work supported by the U.S. Department of Energy, Office of Science, Office of High Energy Physics under Award Number DE-SC0011831.}}
\date{September 10, 2018}
\author{Herman D. Schaumburg\footnotemark[2], Bela Erdelyi\footnotemark[2] 
}

\footnotetext[2]{Department of Physics, Northern Illinois University, DeKalb, IL 60115, (\email{hschaumburg2@niu.edu}).}

\newcommand{\transpose}{^\mathrm{T}}
\newcommand{\red}[1]{ {\color{red} #1} }

\begin{document}

\maketitle

\thispagestyle{empty}

\begin{abstract}
This report summarizes progress made towards a new approach for enlarging the dynamic aperture of particle accelerators.  Unlike prior methods which attempted to move the location of select resonances outward in phase space, our approach aims to move all resonances concurrently.  These resonances are in one-to-one correspondence with fixed points of symplectic maps, which in turn are in a one-to-one correspondence with the critical points of their generating function.  Thus in this approach, the problem of enlarging dynamic aperture boils down to an approximation problem: given a generating function, approximate it by a function whose critical points are outside a specified elliptical region.

In attempting to solve the generating function approximation problem, we employed stable polynomials.  Many stable polynomials have a determinantal representation that indicates stability.  However, it is an open question as to whether such a determinantal representation can be found for a given stable polynomial.  In seeking to answer this, we made progress towards constructing a symmetric determinantal representations of multivariable polynomials with the smallest sized linear pencil.  This report also contains brief surveys of topics including Clifford numbers, stable polynomials, and symmetric determinantal representations of polynomials.  We also explored using Gr{\"o}bner bases to find where the gradient of a multivariable polynomial is zero. 

\end{abstract}

\clearpage

\tableofcontents

\clearpage

\section{Introduction}

The goal of this project was to implement a new approach for increasing the volume of the region of space where the orbital stability of particles is dynammically stable in particle accelerators.  This region is called the dynamic aperture (DA), and it plays a fundamental role in many practical applications of particle accelerators \cite{Berz:427002}.  Durring this project, new mathematical techniques were developed to make progress towards the solution of this problem. The mathematical tools employed for this project originate in a wide variety of fields ranging from symplectic geometry to algebra, and from numerical analysis to optimization.  While progress has been made towards the goal of improving the DA, the work to complete this new approach is ongoing.  

Earlier attempts to improve the DA mostly failed due to the seemingly insurmountable difficulties faced in the realistic, multi-dimensional, highly complicated phase spaces corresponding to the models of such systems: weakly nonlinear, but highly complex, Hamiltonian dynamical systems \cite{Channell:1991:SSM} \cite{PhysRevE.69.056501} \cite{Wan:2001nn} \cite{Wan:1998zz}. The methods followed some variant of the following idea: 
\begin{enumerate}[(i)]
\item model the physical systems of interest as Hamiltonian dynamical systems; 
\item find the location of resonances (i.e. particles with commensurable oscillation frequencies) in phase space from their numerical integration; 
\item use some numerical optimization method, during which some system parameters are fit to move the location of a select number of resonances outwards.
\end{enumerate}

However, symplectic maps contain redundant information about the system. This is clear in the sense that the system is completely determined by a scalar field (the Hamiltonian), while the solution (the symplectic map) is a vector field in phase space. Finding all resonances is a challenge in itself. Brute-force numerical optimization to move them had very limited success.  Typically, realistic objectives include moving some resonances at the expense of others. More precisely, the usual outcome of these attempts is to move some resonances outwards, while others inadvertently move inwards; it is like a puzzle, where some of the pieces never quite fit. It became clear that a more systematic approach is necessary that considers the resonance set as a whole, and a method that is able to move them in unison.  

We seek to revamp the main idea of DA enlargement with a new way of thinking about DA, namely reformulating it as a problem in symplectic geometry, and employing the concepts of stable polynomials and determinantal representation of arbitrary polynomials to enact the theory. The development of our revamp follows the steps: 
\begin{enumerate}[(1)]
\item Model particle accelerators mathematically as periodic Hamiltonian dynamical systems. 
\item Relate the size of the DA to the location of fixed points of the iterates of the time-1 maps of the flows of these Hamiltonian systems.

\item Devise methods that push outwards in space (away from the origin) as many fixed points as possible.  As alluded to, it is currently unknown how to do this systematically. The two crucial concepts that we employed are stable polynomials \cite{wagner_multivariate_2011} and determinantal representations of arbitrary polynomials \cite{quarez_symmetric_2012}. 
\end{enumerate}

Step (1) has no obstacles, it can be accomplished using standard methods \cite{Berz:427002}.  Step (2) is accomplished by reformulating the problem in the language of symplectic geometry \cite{Hofer:2011} \cite{McDuff:1995}.  Resonances are in one-to-one correspondence with fixed points of symplectic maps \cite{Erdelyi:2006:IJPAM} \cite{Erdelyi:2004:IJPAM}. The DA is usually correlated with the region of space were large-scale chaotic behavior is absent \cite{1979:Chirikov}. Chaotic motion is absent in regions free of resonances, hence also of fixed points. Furthermore, fixed points of symplectic maps are in one-to-one correspondence with critical points of their generating functions \cite{Erdelyi:2006:IJPAM} \cite{McDuff:1995}. This way we can simplify the more challenging problem of finding fixed points with the easier one of finding critical points. The remaining obstacle to accomplishing step (2) is to easily find the fixed points of iterates of symplectic maps without actually computing the iterates. In other words, given only the generating function of a symplectic map, find the generating function of its iterates. A partial result is known, for certain specific types of generating functions \cite{2012:Wolski}. However, there are infinitely many different types, and a general result is missing.

This study concerns accomplishing step (3).  Let’s assume that given a canonical, real analytic, multivariable, periodic Hamiltonian system, that closely models an actual physical system (a particle accelerator), we obtained the truncated Taylor expansion of its time-1 (one-turn) flow, which is a truncated symplectic map (a symplectic jet) \cite{Berz:427002}. 
We also computed its EXPO generating function, which is one kind of generating function among infinitely many different types \cite{Erdelyi:2004:IJPAM}. 
EXPO stands for Extended POincare. Although in principle any kind of generating function that exists for the symplectic map in question could be used to state the main problem, we found that the EXPO type makes the computational problem the simplest possible. Also, the EXPO generating function is a polynomial in this approximation. Well-developed computational methods and codes are readily available to enact this program \cite{Makino:2006:NIM}.
 The methods allow to make this approximation as good as needed, since the degree of the truncations is limited only by the amount of memory available. Therefore, we can take this state as the ``initial condition'' for the main problem statement:

\begin{adjustwidth}{0.5in}{0.5in}
  \textit{For a given (EXPO) generating function $g:\mathbb{R}^{2n} \rightarrow \mathbb{R}$, closely approximate $g$ by a polynomial $\hat{g}$ such that the gradient of $\hat{g}$ is nonzero within an elliptical region $E\subset \mathbb{R}^{2n}$, which is as large as possible.  
  }
\end{adjustwidth}

In our preliminary work, presented in Section \ref{sec:Prelim}, we gave a framework for solving the main problem.   This framework is constructing a polynomial $\hat{g}$ as a composition of a stable polynomial and a mapping from $E$ to $\mathbb{H}^{2n}$, where $\mathbb{H}^{2n}$ is the subset of $\mathbb{C}^{2n}$ in which all components of each element have positive imaginary part.    Our choice of mapping is given in Section \ref{sec:Map}.  The condition under which a stable polynomial and this mapping yield an approximation $\hat{g}$ with nonzero gradient is given in Theorem \ref{thm:nonzerograd}.  Sections \ref{sec:SymDet} through \ref{sec:OthDetRep} concern approximation of polynomials by stable polynomials.  Key to making this approximation are determinantal representations of polynomials.  Quarez gave a construction that gives a symmetric determinantal representation of a multivariable polynomial \cite{quarez_symmetric_2012}.  This representation has the smallest size known for the general case.  Under our approximation scheme, Quarez's construction does not give good results.  In attempting to find other symmetric determinantal representations that give better polynomial approximations, we show in several cases that the size in Quarrez's construction can be reduced.  The remaining open problem for approximating polynomials by stable polynomials is whether each stable polynomial has a determinantal representation that demonstrates the stability of the polynomial as in Theorem \ref{thm:PSDStable}, and whether such a determinantal representation can be constructed.  Thus the remaining obstacles to Step (3) are finding a stability demonstrating determinantal representation and possibly finding a mapping with more a lax condition than that of Theorem \ref{thm:nonzerograd}.

\section{Preliminary work}
\label{sec:Prelim}

\begin{figure}[ht]
\includegraphics[scale=1]{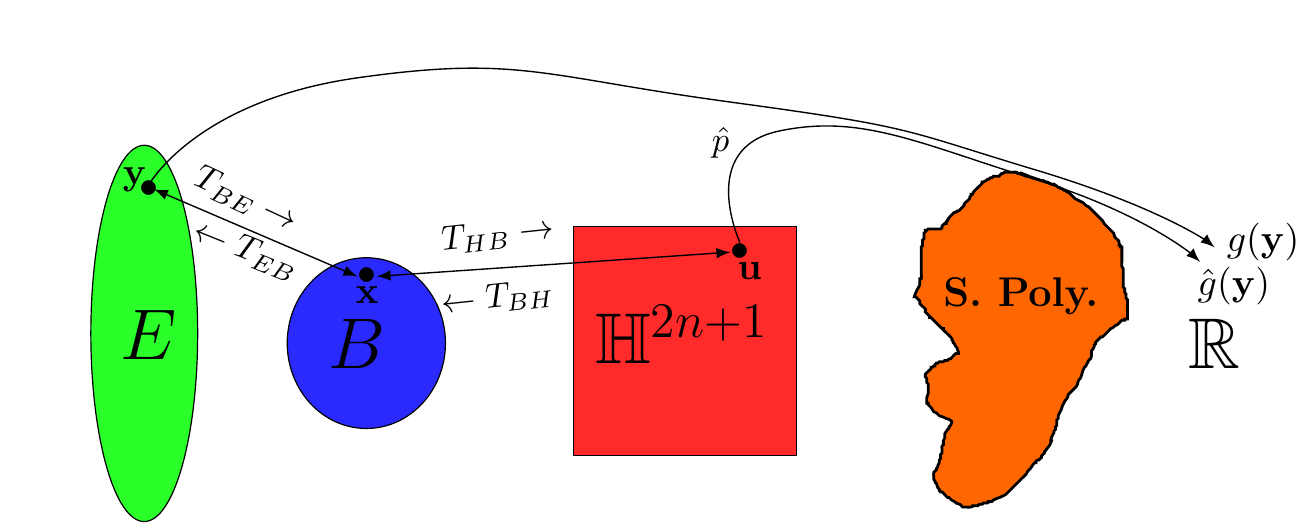}
\caption{Diagram of problem approach.}
\label{fig:ProblemApproach}
\end{figure}

The preliminary work established an approach to the main problem, which we follow throughout.  One key component of the approach is to find a mapping $T_{HB}$ as shown in Figure \ref{fig:ProblemApproach}.  A mapping was proposed in the preliminary work based on higher dimensional Mobi{\"u}s transformation, which was not suitable for this purpose.  However, we include this mapping, since the concept of Clifford numbers employed by the transformation may prove useful for later attempts.  

\subsection{Problem approach summary}

Given a polynomial $g: \mathbb{R}^{2n} \rightarrow \mathbb{R}$, our goal is an approximating polynomial $\hat{g}$ that has nonzero gradient on an elliptical domain $E\subseteq\mathbb{R}^{2n}$ and that is as close to $g$ as possible.  

The following approach is based upon the preliminary work.
\begin{enumerate}
\item Find invertible transform from $E$ to the unit ball, $T_{BE}$.  
\item Find inverse $T_{BE}^{-1}=T_{EB}$.  
\item Find invertible transform from $B$ to the ``half plane'' $\mathbb{H}^{2n}$, $T_{HB}$.  (See \ref{sec:Map} \nameref{sec:Map}.)
\item Find inverse $T_{HB}^{-1}=T_{BH}$.
\item Approximate $p=g \circ T_{EB} \circ T_{BH}$ by a stable polynomial $\hat{p}$.
\item Form approximating function $\hat{g}=\hat{p} \circ T_{HB} \circ T_{BE}$ with nonvanishing gradient on $E$.
\end{enumerate}

To approximate $g: E \rightarrow \mathbb{R}$, we first find $p: \mathbb{H}^{2n} \rightarrow \mathbb{R}$ defined by
\[p=g \circ T_{EB} \circ T_{BH}.\]

Ideally, polynomial $p$ is a stable polynomial, in which case $g$ has nonzero gradient inside $E$.  If $p$ is not stable, it is approximated by stable polynomial $\hat{p}$ which is as ``close'' to $p$ as possible.  
The main difficulties in following this plan are to find a mapping $T_{HB}$ and approximating polynomial $\hat{p}$.  The mapping $T_{BE}$ is more straightforward.

Transformations between $E$ and $B$ are discussed in Appendix \ref{app:Ellipsoid}.  A transformation from $B$ to $E$ could follow three steps: a dilation $(x/a,y/b/,z/c)$, a rotation $R_z(\gamma)R_y(\beta)R_z(\gamma)$, and possibly a translation.  Each of these may be inverted to give one mapping $T_{BE}$.

Because the mapping $T_{BE}$ should be more straightforward, throughout we consider the case when $E=B$.  When considering the $n=0$ case, Mobi{\"u}s transformations between $\mathbb{H}$ and $B$ were our first thought in our search for a mapping $T_{HB}$.   For higher dimensions, we considered Clifford numbers.  While we ultimately used a different mapping in Section \ref{sec:Map}, we include Clifford numbers here since there may be some way to use them to find an improved mapping $T_{HB}$.

\subsection{Higher Dimensional Mobi{\"u}s Transformations}

In the preliminary work, the idea of using a higher dimensional Mobi{\"u}s transformation for $T_{HB}$ was explored.  These transformations are given in terms of Clifford numbers.  The Mobi{\"u}s transformation considered does not map $B$ to $\mathbb{H}^{2n+1}$.  Most of what follows in this subsection is explained in more detail in \cite{Ahlfors:1985}.

\begin{defn}
The \emph{Clifford algebra} $\mathscr{C}_n$ is the associative algebra over the real numbers generated by $n-1$ elements $i_1,i_2,\dots,i_{n-1}$ subject to the relations
\[ i_ji_k=-i_ki_j \quad \text{and} \quad i_j^2=-1\]
with $j\neq k$.  The elements of $\mathscr{C}_n$ are \emph{Clifford numbers}.
\end{defn}

Each Clifford number $a$ has a unique representation:
\[a=\sum_{y \in \mathcal{Y}} c_y y,\]
where $\mathcal{Y}$ is the set of all products $i_{v_1}i_{v_2}\cdots i_{v_p}$ with $1\leq v_1<\cdots<v_p \leq n-1$, including the empty product $i_0$.  Note that $\mathcal{Y}$ has $2^{n-1}$ elements.  The Clifford algebra is a real vector space of dimension $2^{n-1}$.  Clifford numbers of the form
\[x=\sum_{j=0}^{n-1} x_j i_j\]
are called vectors.  The set of these vectors $V_n$ are a subspace of the Cliford algebra.  

There are three involutions, $'$, $^*$ , and \! $\bar{}$, \! defined on these vectors:
\begin{enumerate}
\item The involution $'$ replaces each $i_h$ appearing in $a$ with $-i_h$.
\item For $y=\prod_{j=1}^{n-1}i_{v_j}\in\mathcal{Y}$,  the product in reverse order is the involution $^*$.  That is, $y^*=\prod_{j=1}^{n-1}i_{v_{n-j}}$.  For $a\in\mathscr{C}_n$, $a^*=\sum_{y \in Y} c_m y^*$.
\item $\bar{a}=a^{{*}^{'}}$
\end{enumerate}
For a vector $x \in V^n$, $x^*=x$ and so $x'=\bar{x}$.  The usual euclidean two norm is 
\[x\bar{x}=\sum_{j=0}^{n-1} x_j^2 = ||x||_2^2.\]

Each vector $x \in V^n$ has multiplicative inverse $x^{-1}=||x||_2^{-2} \, \bar{x}$.  So, the set of all products of vectors forms a group called the \emph{Clifford Group}, denoted $\Gamma_n$.  

For $R > 0$ define the set $\Omega=\{\mathbf{z} \in \mathbb{C}^n \ | \ ||\mathbf{z}||_2\leq R\}$.  The mapping $T$ with domain $\Omega$ defined by 
\[ T(\mathbf{x})=(\mathbf{x}+Ri_{n-1})(\mathbf{x}-Ri_{n-1})^{-1}(-2Ri_{n-1})\]
maps $\Omega$ to the set $\overline{H}^n=\{\mathbf{x} \in V^n \ | \ x_{n-1} \geq 0\}$ instead of $\overline{\mathbb{H}}^{n-1}=\{\mathbf{x} \in V^n \ | \, x_{i-1} \geq 0, \ i=1,2,3,\dots,n\}$.

\section{Stable polynomials survey}

Denote the \emph{upper half plane} of the complex plane by $\mathbb{H}=\{ z \in \mathbb{C} : \mathrm{Im}(z)>0 \}$.  
{$\mathbb{H}^n$}{The upper half plane, $\{ z \in \mathbb{C} : \mathrm{Im}(z)>0 \}$}%
{$\mathbb{H}^n$}{The Cartesian product of $\mathbb{H}\times\mathbb{H}\times \cdots \times \mathbb{H}$}%

\begin{defn}
A polynomial $f \in \mathbb{C}[\mathbf{x}]$ and $f:\mathbb{C}^n \rightarrow \mathbb{C}$ is \emph{stable} if either $f$ is identically zero, or $f(\mathbf{z})\neq 0$ for $\mathbf{z}\in \mathbb{H}^n$.  A stable polynomial with real coefficients is called \emph{real stable}.  A subset of stable polynomials are real zero polynomials.  A polynomial $p \in \mathbb{R}[x_1,\dots, x_n]$ is a \emph{real zero polynomial} if for
each $x \in \mathbb{R}^n$ and $\mu \in \mathbb{C}$,
\[p(\mu x) = 0 \]
implies $\mu$ is real.
\end{defn}

Recall that a \emph{Hermitian matrix} is a matrix $A$ such that $A=A^*$, where $A^*$ is the conjugate transpose of $A$.  The following are properties of Hermitian matrices:
\begin{enumerate}
\item $z^*Az$ is real.
\item The eigen values are real.
\item The eigenvectors are orthogonal provided they correspond to different eigenvalues.
\end{enumerate}
Also, a Hermitian $m \times m$ matrix $A$ is called  \emph{positive semidefinite} when $x^*Ax \geq 0$ for all $x \neq 0$.  

\begin{defn}
An \emph{affine linear} pencil $\mathcal{A}$ is
\[\mathcal{A}=A_0+A_1x_1+A_2x_2+\dots+A_n x_n.\]
\label{defn:PencilRZPoly}
\end{defn}

\begin{theorem}
Let $A_0 , \dots , A_n$ be Hermitian $m \times m$ matrices. If $A_1 , \dots , A_n$ are positive semi definite (PSD), then
\begin{equation}
p( x ) = \det(A_0 + A_1 x_1 + \cdots + A_n x_n )
\label{eqn:PSDStable}
\end{equation}
is stable.
\label{thm:PSDStable}
\end{theorem}
The following theorem of Helton and Vinnikov gives that all degree $d$ real zero polynomials in two variables can be written as a determinant of a $d \times d$ matrix pencil in \eqref{eqn:PSDStable} \cite{helton_linear_2007}.
\begin{theorem}[Helton--Vinnikov]
A polynomial $p$ on $\mathbb{R}^2$ is a real zero polynomial of degree $d$ if and only if there exists $d \times d$ PSD matrices $A_0$, $A_1$, and $A_2$ such that
\[p(x_1,x_2)=\det(A_0+A_1 x_1 + A_2 x_2).\]
\label{thm:LaxConj}
\end{theorem}

Helton and Vinnikov conjectured that their result is true in the multivariable case if one allows for matrices of size larger than $d$, however, Br{\"a}nd{\'e}n later disproved this conjecture by finding the counter example in Theorem \ref{thm:BrandedCountEx} \cite{branden_obstructions_2011}.  Wagner and Wei proved that $h_{V_8}(x)$ is a stable polynomial \cite{2009:Wagner}.

\begin{theorem} Let $p(x) = h_{V_8}
(x_1 + 1, \dots , x_8 + 1)$. Then
\begin{itemize}
\item $p(x)$ is a real zero polynomial.
\item There is no positive integer $N$ such that $p(x)$ has a determinantal representation
\[p(x)^N = \det(I + x_1A_1 + \cdots + x_8A_8),\]
\end{itemize}
where matrices $A_1$, $A_2$, $\dots$, and $A_8$ are symmetric.
\label{thm:BrandedCountEx}
\end{theorem}
This means that when $A_0$ is required to be positive semidefinite, there are cases of stable polynomials that have no symmetric determinantal representation.  However, when the condition that $A_0$ is positive semidefinite is dropped, a symmetric determinantal representation exists for any polynomial\cite{2006:Helton}.
\begin{theorem}[Helton, McCullough and Vinnikov]
It is always possible to represent a polynomial as
\[\det\left(A_0+A_1x_1+A_2x_2+\dots+A_n x_n\right),\]
for some matrix size if we require that $A_0$ is symmetric rather than positive definite, where $A_1$, $A_2$, $\dots$, $A_n$ are symmetric.
\end{theorem}
In Section \ref{sec:SymDet}, we present Quarez's construction \cite{quarez_symmetric_2012} which gives a symmetric determinantal representation for a general multivariable polynomial with real coefficients.  An open question is whether all stable polynomials have a determinantal representation where matrices $A_1$, $A_2$, $\dots$, and $A_n$ are PSD.

Several properties of stable polynomials make them an ideal tool for solving the main problem.  Chief among these is that the partial derivatives of a stable polynomial are stable polynomials.  
\begin{theorem}  The following operations preserve stability of polynomials in $\mathbb{C}[\mathbf{x}]$ \cite{wagner_multivariate_2011}.
\begin{enumerate}[(i)]
\item Permutation: For any permutation of the coordinates of $\mathbf{x}$, $\sigma$, $p(\mathbf{x}) \mapsto p(\sigma(\mathbf{x}))$.
\item Scaling:  For and $c\in \mathbb{C}$ and $\mathbf{a} \in \mathbb{R}^n$ with $a_j>0$ For $j=1,2,\dots,n$, $p \mapsto c p(\mathbf{a} \cdot \mathbf{x})$.
\item Diagonalization: For $\{i,j\}\subseteq\{1,2,\dots,n\}$, $p \mapsto \left.p(\mathbf{x})\right|_{x_i=x_j}$.
\item Specialization:  For $a \in \overline{\mathbb{H}}$, $p \mapsto \left.p(\mathbf{x})\right|_{x_1=a}$.
\item Inversion:  If $\deg_1(p)=d$, $p \mapsto x_1^d p(-x_1^{-1},x_2,\dots,x_n)$.
\item Differentiation (or Contraction): $p \mapsto \partial_1 p(\mathbf{x})$.
\end{enumerate}
\label{thm:ListStab}
\end{theorem}

\section{\protect Mappings $T_{HB}$ and $T_{BH}$}
\label{sec:Map}

Our goal here is to supply a mapping $T_{HB}$ from $\mathbf{x} \in B^{2n}$ to $\mathbf{u} \in \mathbb{H}^{2n+1}$ to help in  approximating $g(\mathbf{x})$, where $g:\mathbb{R}^{2n}\rightarrow\mathbb{R}$.  This mapping satisfies the following properties:
\begin{description}
\item[M1] $T_{HB}$ is one-to-one.
\item[M2] $g\circ T_{BH}(\mathbf{u})$ is a polynomial in $\mathbf{u}$ where $\mathbf{u} \in T_{HB}(B^{2n})$.
\item[M3] The inverse mapping $T_{BH}$ maps the boundary of $B^{2n}$ to the boundary of $\mathbb{H}^{2n+1}$.
\end{description}
The boundary of $\mathbb{H}^{2n+1}$ is the set $\{\mathbf{u} : \text{Im}(u_j)=0 \ \exists \ 1\leq j\leq n\} = \partial\mathbb{H}^{2n+1}$.  

Define the map $T_{HB}:\overline{B^{2n}} \rightarrow \overline{\mathbb{H}^{2n+1}}$ as $T_{HB}(\mathbf{x})=\mathbf{u}$, where
{$T$}{Mapping from closure of unit ball to a subset of $\mathbb{H}^n$}%
\begin{equation}
u_j=x_j+i (1-||\mathbf{x}||^2) \text{ for }1\leq j \leq 2n.
\label{eqn:FirstMap1}
\end{equation}
and
\begin{equation}
\label{eqn:FirstMap2}
u_{2n+1}=i (1-||\mathbf{x}||^2).
\end{equation}
Note that this mapping satisfies {\bf M1} and {\bf M3}.  Property {\bf M3} may be unnecessary.  However, it may be that if $g$ has critical points in $B^{2n}$, $\hat{g}$ may be a better approximation for $g$ if $\hat{g}$ may have critical points on the boundary of $B^{2n}$.

The map $T_{HB}$ has an inverse $T_{BH}(\mathbf{u})=\mathbf{x}$.  The $j$th component of inverse of $T_{HB}(\mathbf{u})$ is
\[x_j=u_j-u_{2n+1}.\]
Note that $g(u_1-u_{2n+1},u_2-u_{2n+1},\dots,u_{2n}-u_{2n+1})$ is a polynomial in $\mathbf{u}$, so $T_{HB}$ satisfies {\bf M2}.

With a stable polynomial $\hat{p}$ which is as close to $g(u_1-u_{2n+1},u_2-u_{2n+1},\dots,u_{2n}-u_{2n+1})$ as possible, we can approximate $g(\mathbf{x})$ as
\begin{equation}
g(\mathbf{x})=g(u_1-u_{2n+1},u_2-u_{2n+1},\dots,u_{2n}-u_{2n+1})\approx \hat{p}(\mathbf{u})=:\hat{g}(\mathbf{u}).
\label{Eqn:ApproxStrat}
\end{equation}
The approximating function $\hat{g}$ has nonvanishing gradient on $B^{2n}$ when $g$ and $p$ satisfy the conditions of Theorem \ref{thm:nonzerograd} below.

\begin{theorem}  Let $\hat{p}$ be a stable polynomial.  Suppose that for each $\mathbf{v}\in T_{HB}(B^{2n})$ there is some $j$ such that 
\[ \left.\frac{\partial \hat{p}}{\partial u_j}\right|_{\mathbf{u}=\mathbf{v}} \neq  \left. 2i(u_j-u_{2n+1})\sum_{k=1}^{2n+1} \frac{\partial \hat{p}}{\partial u_k}\right|_{\mathbf{u}=\mathbf{v}}.\]
Then, the composition $\hat{g}=\hat{p} \, \circ \, T_{HB}$ has nonzero gradient on $B^{2n}$.
\label{thm:nonzerograd}
\end{theorem}
\begin{proof*}
We give a proof by contrapositive.  Suppose that $\nabla \hat{g}=0$ for some $\mathbf{y} \in B^{2n}$.  By the chain rule 
\begin{equation}
\frac{\partial \hat{g}}{\partial x_j}= \sum_{k=1}^{2n+1} \frac{\partial \hat{p}}{\partial u_k} \frac{\partial u_k}{\partial x_j}
\label{eqn:AppChainRule}
\end{equation}
for $1\leq j \leq 2n$.  By \eqref{eqn:FirstMap1} and \eqref{eqn:FirstMap2}, $\partial u_k / \partial x_j = -2i x_j$ for $k\neq j$ and $\partial u_{j} / \partial x_j=1-2ix_j$.  Substituting these partial derivatives into the right-hand side of \eqref{eqn:AppChainRule} and zero into the left-hand side of \eqref{eqn:AppChainRule} gives
\[0=\frac{\partial \hat{p}}{\partial u_{j}}-2ix_j\sum_{k=1}^{2n+1} \frac{\partial \hat{p}}{\partial u_k},\]
when $\mathbf{x}=\mathbf{y}$ for $1\leq j \leq 2n$.  Since $x_j=u_j-u_{2n+1}$,
\[\frac{\partial \hat{p}}{\partial u_{j}}=2i(u_j-u_{2n+1})\sum_{k=1}^{2n+1} \frac{\partial \hat{p}}{\partial u_k},\]
for all $j$ with $1\leq j \leq 2n$ when $\mathbf{u}=T_{HB}(\mathbf{y})=\mathbf{v}$ .  
\end{proof*}

\section{Symmetric determinantal representations via Quarez's formula}
\label{sec:SymDet}

This section provides the details on how to find a symmetric determinantal representation of 
\[
g(\mathbf{x})=g(u_1-u_n,u_2-u_n,\dots,u_{n-1}-u_n)
\]
from \eqref{Eqn:ApproxStrat}.  In Section \ref{sec:StabApproxQuarrez}, this symmetric determinantal representation will is used to find a stable polynomial $\hat{p}(\mathbf{u})$ to approximate $g(\mathbf{u})$.

\begin{defn}
A polynomial $p$ of degree $d$ in $n$ variables has a \emph{determinantal representation} if $p$ can be written in the form
\[p(\mathbf{x})=\det \left( A_0 + \sum_{i=1}^n A_i x_i \right),\]
where $A_0$, $A_1$, $\dots$, $A_n$ are $N\times N$ matrices.  When each matrix $A_i$ is symmetric, $p(\mathbf{x})$ is said to have a symmetric determinantal representation.
\end{defn}

Determinantal representations are determinants of affine linear pencils.  An $M \times N$ \emph{linear pencil} is an expression of the form
\[L_M(\mathbf{x})=M_1x_1+M_2x_2+\cdots+M_n x_n,\]
and an $N\times N$ \emph{affine linear pencil} is a linear pencil of the form
\[L_A(\mathbf{x})=A_0 + \sum_{i=1}^n A_i x_i.\]
An affine linear pencil is said to be \emph{symmetric} when each matrix $A_i$ is symmetric.  The determinantal representation for a polynomial $p$ constructed by Quarez is related to a linear description for $p$.

\begin{defn}
We say a polynomial $p(\mathbf{x})$ has a \emph{linear description} if there is a linear pencil $L_A$, a signature matrix $J$, a row matrix $L$, and a column matrix $C$ such that
\[p(\mathbf{x})=L(J-L_A(\mathbf{x}))^{-1}C.\]
A linear description is called \emph{unitary} if $J$ is the identity.  A linear description is called \emph{unipotent} when $I+L_A(\mathbf{x})$ is unipotent (i.e. there is $n$ such that $L_A(\mathbf{x})^n=0$).  Finally, a linear description of $p(\mathbf{x})$ is said to be $S$-symmetric if there is an invertible symmetric matrix $S$ and matrices $L$ and $C$ with $SC=L^\mathrm{T}$ such that $SL_A=L_A\transpose S$ and $SC=L\transpose$.
\end{defn}

A relationship between linear descriptions and symmetric is described the following theorems of Quarez \cite{quarez_symmetric_2012}.

\begin{theorem}
If a polynomial $p(\mathbf{x})$ has an $S$-symmetrizable linear description for a given invertible and symmetric matrix $S$, then it has a symmetric linear description.
\label{thm:S-sym}
\end{theorem}
In the following two theorems, $P(\mathbf{x})$ is the \emph{homogenization} of $-p(\mathbf{x})+1$.  If $p(\mathbf{x})$ is of degree $d$ in $n$ variables and $d'$ is the smallest odd integer with $d\leq d'$, the homogenization of $p(x)$ is the $n+1$ variable homogeneous polynomial $P(x)$ of degree $d'$ such that $P(\mathbf{x},1)=p(x)$.  

\begin{example}  For polynomial $p(\mathbf{x})=4x_1^2x_2^2+7x_1^3+5x_1x_2^2+3x_1+5$, the homogenization of $p(\mathbf{x})$ is
\[P(\mathbf{x})=-4x_1^2x_2^2x_3-7x_1^3x_3^2-5x_1x_2^2x_3^2-3x_1x_3^4-4x_3^5.\]
\end{example}
\begin{theorem}
Assume that the polynomial $P(\mathbf{x})$ admits a symmetric linear unipotent description
\[P(\mathbf{x})=C\transpose (J-L_A(\mathbf{x}))^{-1}C,\]
where $J$ is a signature matrix (diagonal matrix with diagonal elements $\pm 1$) and $A$ is symmetric.  Then,
\[1-P(\mathbf{x})=\det(J)\det\left(J-CC\transpose - L_A(\mathbf{x})\right)\]
\label{thm:QuarezDesc}
\end{theorem}
\begin{theorem}
Let $p(x)$ be a polynomial of degree $d$ in $n$ variables over $\mathbb{R}$ such that $p(0)\neq 0$.  Then there are a signature matrix $J\in \mathbb{R}^{N \times N}$ and a $N \times N$ symetric linear pencil $L_A(\mathbf{x})$ such that
\[p(x)=p(0)\det(J)\det(J-L_A(\mathbf{x})),\]
where $N=2 {n+\lfloor d/2 \rfloor \choose n}$.
\label{thm:QuarezMain}
\end{theorem}

The statement of Theorem \ref{thm:QuarezMain} seems to have an error.  For the general case it should state $N=2 {n+1+\lfloor d/2 \rfloor \choose n+1}$ variables to correct the mistake.  It is the homogenization of $p(x)$ in the paper that has $n+1$ variables.  In Example \ref{ex:DetRep} gives a symmetric determinantal representation for a degree three polynomial in two variables with
\[N=2 {n+1+\lfloor d/2 \rfloor \choose n+1}=2{2+1+1 \choose 2+1}=8.\]
The details of the proofs in Quarez's paper are used to complete a symmetric determinatal representation in Example \ref{ex:DetRep}.  This example relies upon Examples \ref{ex:LinDesc1} and \ref{ex:LinDesc2}.

\subsection{Linear description examples}

\begin{example}  
\label{ex:LinDesc1}
Here we find a linear description related to 
\[p(x_1,x_2)=x_1^3+2x_1^2x_2+3x_1^2+4x_1x_2^2+5x_1x_2+6x_1+7x_2^3+8x_2^2+9x_2+11\]
Another variable is introduced to make a polynomial $P$ with each term having degree equal to $\mathrm{deg}(p)$ with $P(x_1,x_2,1)=1-p(x_1,x_2)$.  The homogenization of $1-p(x_1,x_2)$ is
\begin{multline*}
P(x_1,x_2,x_3)=-x_1^3-2x_1^2x_2-3x_1^2x_3-4x_1x_2^2-5x_1x_2x_3\\
-6x_1x_3^2-7x_2^3-8x_2^2x_3-9x_2x_3^2-10x_3^3.\end{multline*}
In example 4.1 in Quarez's paper, $L_{A_1}$ and $L_{A_2}$ of are   
\[L_{A_1}=\left[ \begin{array}{c}
x_1\\
x_2\\
x_3 
\end{array}
\right] \qquad \text{and} \qquad L_{A_2}= \left[ \begin{array}{ccc}
x_1 & 0 & 0 \\
0 & x_1 & 0\\
0 & 0 & x_1\\
0 & x_2 & 0\\
0 & 0 & x_2\\
0 & 0 & x_3
\end{array} \right]\]

The formulas for $\alpha_{i,k}$ and $\beta_{i,k}$ are
\[\beta_{i,k}={{n-i+k-1}\choose{n-i}} \qquad \alpha_{i,k}={{n+k-2}\choose{n-1}}-\beta_{i,k}\]
Using these,
\begin{align*}
\beta_{1,3}&={{4}\choose{2}}=6 & \alpha_{1,3}&={{4}\choose{2}}-\beta_{1,3}=0\\
\beta_{2,3}&={{3}\choose{1}}=3 & \alpha_{1,3}&={{4}\choose{2}}-\beta_{2,3}=3\\
\beta_{3,3}&={{2}\choose{0}}=1 & \alpha_{3,3}&={{4}\choose{2}}-\beta_{3,3}=5
\end{align*}
So, 
\[L_{A_3}=
\begin{pmatrix}
x_1 & 0 & 0 & 0 & 0 & 0\\
0 & x_1 & 0 & 0 & 0 & 0\\
0 & 0 & x_1 & 0 & 0 & 0\\
0 & 0 & 0 & x_1 & 0 & 0\\
0 & 0 & 0 & 0 & x_1 & 0\\
0 & 0 & 0 & 0 & 0 & x_1\\
0 & 0 & 0 & x_2 & 0 & 0\\
0 & 0 & 0 & 0 & x_2 & 0\\
0 & 0 & 0 & 0 & 0 & x_2\\
0 & 0 & 0 & 0 & 0 & x_3\\
\end{pmatrix}.
\]

The linear pencil $L_A(x)$ is
\[\left[\begin{array}{*{6}c}
0 & 0 & 0 & 0 & 0& 0\\
L_{A_1} & 0 & 0 & 0 & 0 & 0\\
0 & L_{A_2} & 0 & 0 & 0 & 0 \\
0 & 0 & L_{A_3} & 0 & 0& 0\\
\end{array}
\right]
\]
or

\noindent\resizebox{0.98\textwidth}{!}{
$\displaystyle L_A(x)=
\left(\begin{array}{*{20}c}
0 & 0 & 0 & 0 & 0 & 0 & 0 & 0 & 0 & 0 & 0 & 0 & 0 & 0 & 0 & 0 & 0 & 0 & 0 & 0\\
x_1 & 0 & 0 & 0 & 0 & 0 & 0 & 0 & 0 & 0 & 0 & 0 & 0 & 0 & 0 & 0 & 0 & 0 & 0 & 0\\
x_2 & 0 & 0 & 0 & 0 & 0 & 0 & 0 & 0 & 0 & 0 & 0 & 0 & 0 & 0 & 0 & 0 & 0 & 0 & 0\\
x_3 & 0 & 0 & 0 & 0 & 0 & 0 & 0 & 0 & 0 & 0 & 0 & 0 & 0 & 0 & 0 & 0 & 0 & 0 & 0\\
0 & x_1 & 0 & 0 & 0 & 0 & 0 & 0 & 0 & 0 & 0 & 0 & 0 & 0 & 0 & 0 & 0 & 0 & 0 & 0\\
0 & 0 & x_1 & 0 & 0 & 0 & 0 & 0 & 0 & 0 & 0 & 0 & 0 & 0 & 0 & 0 & 0 & 0 & 0 & 0\\
0 & 0 & 0 & x_1 & 0 & 0 & 0 & 0 & 0 & 0 & 0 & 0 & 0 & 0 & 0 & 0 & 0 & 0 & 0 & 0\\
0 & 0 & {x_2} & 0 & 0 & 0 & 0 & 0 & 0 & 0 & 0 & 0 & 0 & 0 & 0 & 0 & 0 & 0 & 0 & 0\\
0 & 0 & 0 & {x_2} & 0 & 0 & 0 & 0 & 0 & 0 & 0 & 0 & 0 & 0 & 0 & 0 & 0 & 0 & 0 & 0\\
0 & 0 & 0 & {x_3} & 0 & 0 & 0 & 0 & 0 & 0 & 0 & 0 & 0 & 0 & 0 & 0 & 0 & 0 & 0 & 0\\
0 & 0 & 0 & 0 & {x_1} & 0 & 0 & 0 & 0 & 0 & 0 & 0 & 0 & 0 & 0 & 0 & 0 & 0 & 0 & 0\\
0 & 0 & 0 & 0 & 0 & {x_1} & 0 & 0 & 0 & 0 & 0 & 0 & 0 & 0 & 0 & 0 & 0 & 0 & 0 & 0\\
0 & 0 & 0 & 0 & 0 & 0 & {x_1} & 0 & 0 & 0 & 0 & 0 & 0 & 0 & 0 & 0 & 0 & 0 & 0 & 0\\
0 & 0 & 0 & 0 & 0 & 0 & 0 & {x_1} & 0 & 0 & 0 & 0 & 0 & 0 & 0 & 0 & 0 & 0 & 0 & 0\\
0 & 0 & 0 & 0 & 0 & 0 & 0 & 0 & {x_1} & 0 & 0 & 0 & 0 & 0 & 0 & 0 & 0 & 0 & 0 & 0\\
0 & 0 & 0 & 0 & 0 & 0 & 0 & 0 & 0 & {x_1} & 0 & 0 & 0 & 0 & 0 & 0 & 0 & 0 & 0 & 0\\
0 & 0 & 0 & 0 & 0 & 0 & 0 & {x_2} & 0 & 0 & 0 & 0 & 0 & 0 & 0 & 0 & 0 & 0 & 0 & 0\\
0 & 0 & 0 & 0 & 0 & 0 & 0 & 0 & {x_2} & 0 & 0 & 0 & 0 & 0 & 0 & 0 & 0 & 0 & 0 & 0\\
0 & 0 & 0 & 0 & 0 & 0 & 0 & 0 & 0 & {x_2} & 0 & 0 & 0 & 0 & 0 & 0 & 0 & 0 & 0 & 0\\
0 & 0 & 0 & 0 & 0 & 0 & 0 & 0 & 0 & {x_3} & 0 & 0 & 0 & 0 & 0 & 0 & 0 & 0 & 0 & 0\\
\end{array}\right).
$}

\noindent The matrix $(I-L_A(x))^{-1}$ has an interesting form:

\noindent\resizebox{0.98\textwidth}{!}{
$\displaystyle(I-L_A(x))^{-1}=\left(\begin{array}{*{20}c}
1 & 0 & 0 & 0 & 0 & 0 & 0 & 0 & 0 & 0 & 0 & 0 & 0 & 0 & 0 & 0 & 0 & 0 & 0 & 0\\
{x_1} & 1 & 0 & 0 & 0 & 0 & 0 & 0 & 0 & 0 & 0 & 0 & 0 & 0 & 0 & 0 & 0 & 0 & 0 & 0\\
{x_2} & 0 & 1 & 0 & 0 & 0 & 0 & 0 & 0 & 0 & 0 & 0 & 0 & 0 & 0 & 0 & 0 & 0 & 0 & 0\\
{x_3} & 0 & 0 & 1 & 0 & 0 & 0 & 0 & 0 & 0 & 0 & 0 & 0 & 0 & 0 & 0 & 0 & 0 & 0 & 0\\
{{x}_{1}^{2}} & {x_1} & 0 & 0 & 1 & 0 & 0 & 0 & 0 & 0 & 0 & 0 & 0 & 0 & 0 & 0 & 0 & 0 & 0 & 0\\
{x_1}\, {x_2} & 0 & {x_1} & 0 & 0 & 1 & 0 & 0 & 0 & 0 & 0 & 0 & 0 & 0 & 0 & 0 & 0 & 0 & 0 & 0\\
{x_1}\, {x_3} & 0 & 0 & {x_1} & 0 & 0 & 1 & 0 & 0 & 0 & 0 & 0 & 0 & 0 & 0 & 0 & 0 & 0 & 0 & 0\\
{{x}_{2}^{2}} & 0 & {x_2} & 0 & 0 & 0 & 0 & 1 & 0 & 0 & 0 & 0 & 0 & 0 & 0 & 0 & 0 & 0 & 0 & 0\\
{x_2}\, {x_3} & 0 & 0 & {x_2} & 0 & 0 & 0 & 0 & 1 & 0 & 0 & 0 & 0 & 0 & 0 & 0 & 0 & 0 & 0 & 0\\
{{x}_{3}^{2}} & 0 & 0 & {x_3} & 0 & 0 & 0 & 0 & 0 & 1 & 0 & 0 & 0 & 0 & 0 & 0 & 0 & 0 & 0 & 0\\
{{x}_{1}^{3}} & {{x}_{1}^{2}} & 0 & 0 & {x_1} & 0 & 0 & 0 & 0 & 0 & 1 & 0 & 0 & 0 & 0 & 0 & 0 & 0 & 0 & 0\\
{{x}_{1}^{2}}\, {x_2} & 0 & {{x}_{1}^{2}} & 0 & 0 & {x_1} & 0 & 0 & 0 & 0 & 0 & 1 & 0 & 0 & 0 & 0 & 0 & 0 & 0 & 0\\
{{x}_{1}^{2}}\, {x_3} & 0 & 0 & {{x}_{1}^{2}} & 0 & 0 & {x_1} & 0 & 0 & 0 & 0 & 0 & 1 & 0 & 0 & 0 & 0 & 0 & 0 & 0\\
{x_1}\, {{x}_{2}^{2}} & 0 & {x_1}\, {x_2} & 0 & 0 & 0 & 0 & {x_1} & 0 & 0 & 0 & 0 & 0 & 1 & 0 & 0 & 0 & 0 & 0 & 0\\
{x_1}\, {x_2}\, {x_3} & 0 & 0 & {x_1}\, {x_2} & 0 & 0 & 0 & 0 & {x_1} & 0 & 0 & 0 & 0 & 0 & 1 & 0 & 0 & 0 & 0 & 0\\
{x_1}\, {{x}_{3}^{2}} & 0 & 0 & {x_1}\, {x_3} & 0 & 0 & 0 & 0 & 0 & {x_1} & 0 & 0 & 0 & 0 & 0 & 1 & 0 & 0 & 0 & 0\\
{{x}_{2}^{3}} & 0 & {{x}_{2}^{2}} & 0 & 0 & 0 & 0 & {x_2} & 0 & 0 & 0 & 0 & 0 & 0 & 0 & 0 & 1 & 0 & 0 & 0\\
{{x}_{2}^{2}}\, {x_3} & 0 & 0 & {{x}_{2}^{2}} & 0 & 0 & 0 & 0 & {x_2} & 0 & 0 & 0 & 0 & 0 & 0 & 0 & 0 & 1 & 0 & 0\\
{x_2}\, {{x}_{3}^{2}} & 0 & 0 & {x_2}\, {x_3} & 0 & 0 & 0 & 0 & 0 & {x_2} & 0 & 0 & 0 & 0 & 0 & 0 & 0 & 0 & 1 & 0\\
{{x}_{3}^{3}} & 0 & 0 & {{x}_{3}^{2}} & 0 & 0 & 0 & 0 & 0 & {x_3} & 0 & 0 & 0 & 0 & 0 & 0 & 0 & 0 & 0 & 1\end{array}\right).$
}

\noindent Observe that it is the matrix
\[(I-L_A(x))^{-1}=
\begin{pmatrix}
1 & 0 & \hdots\\
L_{A_1} & I & \hdots\\
L_{A_2}L_{A_1} & L_{A_2} & I & \hdots\\
L_{A_3}L_{A_2}L_{A_1} & L_{A_3}L_{A_2} &  & I & \hdots \\
 & & L_{A_3} & & I & \hdots
\end{pmatrix}.
\]
Importantly, the bottom left corner is the product $L_{A_3}L_{A_2}L_{A_1}$ column vector whose components are a basis for homogeneous polynomials of degree three in thee variables.  This may be used to get a linear description of our example polynomial $P(x)$, namely
\[
[\mathbf{0}_{10} \ -1\ -2\ -3\ -4\ -5\ -6\ -7\ -8\ -9\ -10](I-L_{A})^{-1}[1 \ \mathbf{0}_{19}]^\mathrm{T}=P(x),\\
\]
where $\mathbf{0}_k$ is a $k$-dimensional zero row vector.
\end{example}

\begin{example}
\label{ex:LinDesc2}
Now we try for a symmetric unipotent linear description for $P$ in the last example.  This linear description can be used to find a determinantal representation for $1-P(\mathbf{x})$ as in Theorem \ref{thm:QuarezDesc}.  Since our polynomial is in three variables, we use greek letters for ordered triples.  For $\gamma=(\gamma_1,\gamma_2,\gamma_3)$, we define $b_\gamma$ as the coefficient of $x_1^{\gamma_1} x_2^{\gamma_2} x_3^{\gamma_3}$ in polynomial $P(x)$:

\begin{table}[ht]
\resizebox{\textwidth}{!}{
\begin{minipage}{1.3 \textwidth}
\noindent\begin{tabular}{c|*{10}{l}}
$b_\gamma$ & -1 & -2 & -3 & -4 & -5 & -6 & -7 & -8 & -9 & -10\\
$\gamma$ & (3,0,0) & (2,1,0) & (2,0,1) & (1,2,0) & (1,1,1) & (1,0,2) & (0,3,0) & (0,2,1) & (0,1,2) & (0,0,3)
\end{tabular}
\end{minipage}
}
\caption{Coefficients of $P$ and corresponding triples.}
\label{tab:PCoeff}
\end{table}
\noindent Also, define $x^\gamma=x_1^{\gamma_1} x_2^{\gamma_2} x_3^{\gamma_3}$.

For $3=2e+1$, $e=1$, we set 
$L_{B_i}=L_{A_i}$ for $i=1=e$, $L_{B_i}=L_{A_i}^\mathrm{T}$ for $i=3=e+2$ and $L_{B_2}$ yet to be determined.
\[L_{B_2}=(\phi_{\alpha,\beta})_{|\alpha|=|\beta|=e=1}.\]
with
\[\phi_{\alpha,\beta}=\sum_{i=1}^n \lambda_{\alpha,\beta}^{(i)}b_{\alpha+\beta+\delta^{(i)}}x_i,\]
where the $\lambda_{\alpha,\beta}^{(i)}$s are to be determined and $\delta_i$ is the tuple with 1 in the $i$th coordinate and other coordinates $0$.  The $3$-tuples with $|\alpha|=1$ (sum equal to one) are put into lexicographic order.  This ordering gives the indeces for $L_{B_2}$.
\begin{center}
\begin{tabular}{c|c}
Order & $3$-tuple $\alpha$\\
\hline
1 & (1,0,0)\\
2 & (0,1,0)\\
3 & (0,0,1)\\
\end{tabular}
\end{center}
We write the $\phi$s and $\lambda$s abusing notation a bit by indexing the subscript tuples by their order in the table above:
\begin{align*}
\phi_{1,1}&=\lambda_{1,1}^{(1)}b_{(3,0,0)}x_1+\lambda_{1,1}^{(2)}b_{(2,1,0)}x_2+\lambda_{1,1}^{(3)}b_{(2,0,1)}x_3 & \alpha+\beta&=(2,0,0)\\
\phi_{1,2}&=\lambda_{1,2}^{(1)}b_{(2,1,0)}x_1+\lambda_{1,2}^{(2)}b_{(1,2,0)}x_2+\lambda_{1,2}^{(3)}b_{(1,1,1)}x_3 & \alpha+\beta&=(1,1,0)\\
\phi_{1,3}&=\lambda_{1,3}^{(1)}b_{(2,0,1)}x_1+\lambda_{1,3}^{(2)}b_{(1,1,1)}x_2+\lambda_{1,3}^{(3)}b_{(1,0,2)}x_3 & \alpha+\beta&=(1,0,1)\\
\phi_{2,1}&=\lambda_{2,1}^{(1)}b_{(2,1,0)}x_1+\lambda_{2,1}^{(2)}b_{(1,2,0)}x_2+\lambda_{2,1}^{(3)}b_{(1,1,1)}x_3 & \alpha+\beta&=(1,1,0)\\
\phi_{2,2}&=\lambda_{2,2}^{(1)}b_{(1,2,0)}x_1+\lambda_{2,2}^{(2)}b_{(0,3,0)}x_2+\lambda_{2,2}^{(3)}b_{(0,2,1)}x_3 & \alpha+\beta&=(0,2,0)\\
\phi_{2,3}&=\lambda_{2,3}^{(1)}b_{(1,1,1)}x_1+\lambda_{2,3}^{(2)}b_{(0,2,1)}x_2+\lambda_{2,3}^{(3)}b_{(0,1,2)}x_3 & \alpha+\beta&=(0,1,1)\\
\phi_{3,1}&=\lambda_{3,1}^{(1)}b_{(2,0,1)}x_1+\lambda_{3,1}^{(2)}b_{(1,1,1)}x_2+\lambda_{3,1}^{(3)}b_{(1,0,2)}x_3 & \alpha+\beta&=(1,0,1)\\
\phi_{3,2}&=\lambda_{3,2}^{(1)}b_{(1,1,1)}x_1+\lambda_{3,2}^{(2)}b_{(0,2,1)}x_2+\lambda_{3,2}^{(3)}b_{(0,1,3)}x_3 & \alpha+\beta&=(0,1,1)\\
\phi_{3,3}&=\lambda_{3,3}^{(1)}b_{(1,0,2)}x_1+\lambda_{3,3}^{(2)}b_{(0,1,2)}x_2+\lambda_{3,3}^{(3)}b_{(0,0,3)}x_3 & \alpha+\beta&=(0,0,2)\\
\end{align*}
These must satisfy
\[
P(x)=\left( x_1 \ x_2 \ x_3 \right)\begin{pmatrix}
\phi_{1,1} & \phi_{1,2} & \phi_{1,3}\\
\phi_{2,1} & \phi_{2,2} & \phi_{2,3}\\
\phi_{3,1} & \phi_{3,2} & \phi_{3,3}\\
\end{pmatrix}\begin{pmatrix}
x_1\\
x_2\\
x_3
\end{pmatrix},\]
which is
\begin{multline*}
P(x)=x_1 \phi_{1,1}x_1 + x_1\phi_{1,2}x_2 + x_1\phi_{1,3}x_3+x_2\phi_{2,1}x_1 + x_2\phi_{2,2}x_2\\
 + x_2\phi_{2,3}x_3+x_3\phi_{3,1}x_1 + x_3\phi_{3,2}x_2 + x_3\phi_{3,3}x_3.
\end{multline*}
Noting $e=1$,
\begin{align}
P(x)&=\sum_{|\alpha|=e,|\beta|=e}x^\alpha \phi_{\alpha,\beta}x^\beta \nonumber\\
&=\sum_{\gamma=2e}x^\gamma \sum_{\alpha+\beta=\gamma} \phi_{\alpha,\beta} \nonumber\\
&=\sum_{\gamma=2e+1} \left( 
\sum_{i \in \mathrm{Supp}(\gamma)} \sum_{\alpha+\beta=\gamma-\delta^{(i)}} \lambda_{\alpha,\beta}^{(i)}\right)b_\gamma x^\gamma \label{eqn:QuarezWeigths},
\end{align}
where $\mathrm{Supp}(\gamma)$ are the indices such that $\gamma_i\neq0$.

Set
\[\Lambda_\gamma^{(i)}=\sum_{\alpha+\beta=\gamma} \lambda_{\alpha,\beta}^{(i)},\]
which we consider in the context of the inner sum appearing in \eqref{eqn:QuarezWeigths}.  For $|\gamma|=2e=2$, the relevant $\Lambda$s follow:
\begin{align*}
\Lambda_{(2,0,0)}^{(1)}&=\lambda_{1,1}^{(1)} & \Lambda_{(2,0,0)}^{(2)}&=\lambda_{1,1}^{(2)} & \Lambda_{(2,0,0)}^{(3)}&=\lambda_{1,1}^{(3)}\\
\Lambda_{(1,1,0)}^{(1)}&=\lambda_{1,2}^{(1)}+\lambda_{2,1}^{(1)} & \Lambda_{(1,1,0)}^{(2)}&=\lambda_{1,2}^{(2)}+\lambda_{2,1}^{(2)} & \Lambda_{(1,1,0)}^{(3)}&=\lambda_{1,2}^{(3)}+\lambda_{2,1}^{(3)}\\
\Lambda_{(1,0,1)}^{(1)}&=\lambda_{1,3}^{(1)}+\lambda_{3,1}^{(1)} & \Lambda_{(1,0,1)}^{(2)}&=\lambda_{1,3}^{(2)}+\lambda_{3,1}^{(2)} & \Lambda_{(1,0,1)}^{(3)}&=\lambda_{1,3}^{(3)}+\lambda_{3,1}^{(3)}\\
\Lambda_{(0,2,0)}^{(1)}&=\lambda_{2,2}^{(1)} & \Lambda_{(0,2,0)}^{(2)}&=\lambda_{2,2}^{(2)} & \Lambda_{(0,2,0)}^{(3)}&=\lambda_{2,2}^{(3)}\\
\Lambda_{(0,1,1)}^{(1)}&=\lambda_{2,3}^{(1)}+\lambda_{3,2}^{(1)} & \Lambda_{(0,1,1)}^{(2)}&=\lambda_{2,3}^{(2)}+\lambda_{3,2}^{(2)} & \Lambda_{(0,1,1)}^{(3)}&=\lambda_{2,3}^{(3)}+\lambda_{3,2}^{(3)}\\
\Lambda_{(0,0,2)}^{(1)}&=\lambda_{3,3}^{(1)} & \Lambda_{(0,0,2)}^{(2)}&=\lambda_{3,3}^{(2)} & \Lambda_{(0,0,2)}^{(3)}&=\lambda_{3,3}^{(3)}\\
\end{align*}

We want to choose $\lambda_{\alpha,\beta}^{(i)}$ so that 
\[\sum_{i \in \mathrm{Supp}(\gamma)} \Lambda_{\gamma-\delta^{(i)}}^{(i)}=1,\]
for each $\gamma$ with $|\gamma|=2e+1=3$.  Each case for this sum is in the following table.

\begin{center}
\begin{tabular}{c|c|c}
$\gamma$ & $\mathrm{Supp}(\gamma)$ & $\sum_{i \in \mathrm{Supp}(\gamma)} \Lambda_{\gamma-\delta^{(i)}}^{(i)}=1$\\
(3,0,0) & 1 & $\Lambda_{(2,0,0)}^{(1)}$\\
(2,1,0) & 1,2 & $\Lambda_{(1,1,0)}^{(1)}+\red{\Lambda_{(2,0,0)}^{(2)}}$\\
(2,0,1) & 1,3 & $\Lambda_{(1,0,1)}^{(1)}+\red{\Lambda_{(2,0,0)}^{(3)}}$\\
(1,2,0) & 1,2 & $\Lambda_{(0,2,0)}^{(1)}+\red{\Lambda_{(1,1,0)}^{(2)}}$\\
(1,1,1) & 1,2,3 & $\Lambda_{(0,1,1)}^{(1)}+\red{\Lambda_{(1,0,1)}^{(2)}}+\red{\Lambda_{(1,1,0)}^{(3)}}$\\
(1,0,2) & 1,3 & $\Lambda_{(0,0,2)}^{(1)}+\red{\Lambda_{(1,0,1)}^{(3)}}$\\
(0,3,0) & 2 & $\Lambda_{(0,2,0)}^{(2)}$\\
(0,2,1) & 2,3 & $\Lambda_{(0,1,1)}^{(2)}+\red{\Lambda_{(0,2,0)}^{(3)}}$\\
(0,1,2) & 2,3 & $\Lambda_{(0,0,2)}^{(2)}+\red{\Lambda_{(0,1,1)}^{(3)}}$\\
(0,0,3) & 3 & $\Lambda_{(0,0,2)}^{(3)}$\\
\end{tabular}
\end{center}
Two steps give the solution selected by Quarez. The first step is to set $\Lambda^{(i)}_\epsilon=0$ when $i>\min(\mathrm{Supp}(\epsilon))$ with $\lambda_{\alpha,\beta}^{(i)}=0$ for all $\alpha$ and $\beta$ such that $\alpha+\beta=\epsilon$.  These are in \red{red} above.  This means that each of the following are zero: $\lambda_{1,1}^{(2)},\lambda_{1,1}^{(3)},\lambda_{1,2}^{(2)},\lambda_{2,1}^{(2)},\lambda_{1,3}^{(2)},\lambda_{3,1}^{(2)},\lambda_{1,2}^{(3)},\lambda_{2,1}^{(3)},\lambda_{1,3}^{(3)},\lambda_{3,1}^{(3)},\lambda_{2,2}^{(3)}$, $\lambda_{2,3}^{(3)},\lambda_{3,2}^{(3)}$.  The second step is for the case when $i\leq \min(\mathrm{Supp}(\epsilon))$.  Let $\alpha_0$ be the highest in lexicographic ordering such that there is $\beta_0$ with $\alpha_0+\beta_0=\epsilon$.  If $\alpha_0=\beta_0$, $\lambda_{\alpha_0,\beta_0}^{(i)}=1$ and otherwise $\lambda_{\alpha_0,\beta_0}^{(i)}=\lambda_{\beta_0,\alpha_0}^{(i)}=1/2$.  The other $\lambda_{\alpha,\beta}^{(i)}$ are zero.  The weights $\lambda_{1,1}^{(1)}$, $\lambda_{2,2}^{(2)}$, and $\lambda_{3,3}^{(3)}$ are one, and $\lambda_{2,2}^{(1)},\lambda_{3,3}^{(1)},\lambda_{3,3}^{(2)}$ are zero.  

\renewcommand{\arraystretch}{1.5}
\noindent\begin{tabular}{llp{1.8in}}
$\epsilon$ & $\alpha_0$ and $\beta_0$ & $\lambda_{\alpha_0,\beta_0}^{(j)}$s\\
\hline
$\epsilon=(2,0,0)$ & $\alpha_0=\beta_0=(1,0,0)$ & $1=\lambda_{1,1}^{(1)}$\\
$\epsilon=(1,1,0)$ & $\alpha_0=(1,0,0)$ and $\beta_0=(0,1,0)$ & $1/2=\lambda_{1,2}^{(1)}=\lambda_{2,1}^{(1)}$\\
$\epsilon=(1,0,1)$ & $\alpha_0=(1,0,0)$ and $\beta_0=(0,0,1)$ & $1/2=\lambda_{1,3}^{(1)}=\lambda_{3,1}^{(1)}=1/2$\\
$\epsilon=(0,2,0)$ & $\alpha_0=\beta_0=(0,1,0)$ & $1=\lambda_{2,2}^{(1)}=\lambda_{2,2}^{(2)}$\\
$\epsilon=(0,1,1)$ & $\alpha_0=(0,1,0)$ and $\beta_0=(0,0,1)$ & $1/2=\lambda_{2,3}^{(1)}=\lambda_{3,2}^{(1)}$ and $1/2=\lambda_{2,3}^{(2)}=\lambda_{3,2}^{(2)}$\\
$\epsilon=(0,0,2)$ & $\alpha_0=\beta_0=(0,0,1)$ & $1=\lambda_{3,3}^{(1)}=\lambda_{3,3}^{(2)}=\lambda_{3,3}^{(3)}$
\end{tabular}
\renewcommand{\arraystretch}{1}

Thus the following are zero: $\lambda_{1,1}^{(2)}$, $\lambda_{1,1}^{(3)}$, $\lambda_{1,2}^{(2)}$, $\lambda_{2,1}^{(2)}$, $\lambda_{1,2}^{(3)}$, $\lambda_{2,1}^{(3)}$, $\lambda_{1,3}^{(2)}$, $\lambda_{3,1}^{(2)}$, $\lambda_{1,3}^{(3)}$, $\lambda_{3,1}^{(3)}$, $\lambda_{2,2}^{(3)}$, $\lambda_{2,3}^{(3)}$, and $\lambda_{3,2}^{(3)}$.
\begin{small}
\begin{align*}
\phi_{1,1}&=(1)b_{(3,0,0)}x_1+(0)b_{(2,1,0)}x_2+(0)b_{(2,0,1)}x_3=b_{(3,0,0)}x_1 \\
\phi_{1,2}&=(1/2)b_{(2,1,0)}x_1+(0)b_{(1,2,0)}x_2+(0)b_{(1,1,1)}x_3=b_{(2,1,0)}/2 \ x_1\\
\phi_{1,3}&=(1/2)b_{(2,0,1)}x_1+(0)b_{(1,1,1)}x_2+(0)b_{(1,0,2)}x_3=b_{(2,0,1)}/2 \ x_1\\
\phi_{2,1}&=(1/2)b_{(2,1,0)}x_1+(0)b_{(1,2,0)}x_2+(0)b_{(1,1,1)}x_3=b_{(2,1,0)}/2 \ x_1\\
\phi_{2,2}&=(1)b_{(1,2,0)}x_1+(1)b_{(0,3,0)}x_2+(0)b_{(0,2,1)}x_3=b_{(1,2,0)}x_1+b_{(0,3,0)}x_2\\
\phi_{2,3}&=(1/2)b_{(1,1,1)}x_1+(1/2)b_{(0,2,1)}x_2+(0)b_{(0,1,2)}x_3=b_{(1,1,1)}/2 \ x_1+b_{(0,2,1)}/2 \ x_2\\
\phi_{3,1}&=(1/2)b_{(2,0,1)}x_1+(0)b_{(1,1,1)}x_2+(0)b_{(1,0,2)}x_3=b_{(2,0,1)}/2 \ x_1\\
\phi_{3,2}&=(1/2)b_{(1,1,1)}x_1+(1/2)b_{(0,2,1)}x_2+(0)b_{(0,1,3)}x_3=b_{(1,1,1)}/2 \ x_1+b_{(0,2,1)}/2 \ x_2\\
\phi_{3,3}&=(1)b_{(1,0,2)}x_1+(1)b_{(0,1,2)}x_2+(1)b_{(0,0,3)}x_3=b_{(1,0,2)}x_1+b_{(0,1,2)}x_2+b_{(0,0,3)}x_3\\
\end{align*}\end{small}
Making the substitutions from Table \ref{tab:PCoeff},
\[L_2=L_{e+1}=\begin{bmatrix}
-x_1 & -x_1 & -3/2x_1\\
-x_1 & -4x_1-7x_2 & -5/2 x_1-4 x_2\\
-3/2 x_1 & -5/2 x_1-4 x_2 & -6x_1-9x_2-10x_3
\end{bmatrix}.
\]
For 
\[L_A=\begin{bmatrix}
0 & 0 & 0 & 0 & 0 & 0 & 0 & 0\\
x_1 & 0 & 0 & 0 & 0 & 0 & 0 & 0\\
x_2 & 0 & 0 & 0 & 0 & 0 & 0 & 0\\
x_3 & 0 & 0 & 0 & 0 & 0 & 0 & 0\\
0 & -x_1 & -x_1 & -3/2x_1 & 0 & 0 & 0 & 0\\
0 & -x_1 & -4x_1-7x_2 & -5/2 x_1-4 x_2 & 0 & 0 & 0 & 0\\
0 & -3/2 x_1 & -5/2 x_1-4 x_2 & -6x_1-9x_2-10x_3 & 0 & 0 & 0 & 0\\
0 & 0 & 0 & 0 & x_1 & x_2 & x_3 & 0\\
\end{bmatrix},
\]
\[P(x)=L_0(I-L_A(x))^{-1}C_0\]
is a unipotent linear description for $P(x)$ such that $L_0=[0,0,0,0,0,0,0,1]$ and $C_0=[1,0,0,0,0,0,0,0]\transpose$.  Indeed, $L_A^4=0$.
\label{ex:UnipLD}
\end{example}

\subsection{Symmetric determinantal representation}

We now aim to ``symmetrize'' the linear pencil from the previous example.

\begin{example} We take Example \ref{ex:UnipLD} as our starting point.  Before substitution the matrix

\noindent\resizebox{0.98\textwidth}{!}{$L_A=\begin{bmatrix}
0 & 0 & 0 & 0 & 0 & 0 & 0 & 0\\
x_1 & 0 & 0 & 0 & 0 & 0 & 0 & 0\\
x_2 & 0 & 0 & 0 & 0 & 0 & 0 & 0\\
x_3 & 0 & 0 & 0 & 0 & 0 & 0 & 0\\
0 & b_{(3,0,0)}x_1 & b_{(2,1,0)}/2 \ x_1 & b_{(2,0,1)}/2 \ x_1 & 0 & 0 & 0 & 0\\
0 & b_{(2,1,0)}/2 \ x_1 & b_{(1,2,0)}x_1+b_{(0,3,0)}x_2 & b_{(1,1,1)}/2 \ x_1+b_{(0,2,1)}/2 \ x_2 & 0 & 0 & 0 & 0\\
0 & b_{(2,0,1)}/2 \ x_1 & b_{(1,1,1)}/2 \ x_1+b_{(0,2,1)}/2 \ x_2 & b_{(1,0,2)}x_1+b_{(0,1,2)}x_2+b_{(0,0,3)}x_3 & 0 & 0 & 0 & 0\\
0 & 0 & 0 & 0 & x_1 & x_2 & x_3 & 0\\
\end{bmatrix}$}

The linear description given by the matrix abouf and $L_0$ and $C_0$ is symmetrizable using matrix
\[S=\begin{bmatrix}0 & 0 & 0 & 0 & 0 & 0 & 0 & 1\\
0 & 0 & 0 & 0 & 1 & 0 & 0 & 0\\
0 & 0 & 0 & 0 & 0 & 1 & 0 & 0\\
0 & 0 & 0 & 0 & 0 & 0 & 1 & 0\\
0 & 1 & 0 & 0 & 0 & 0 & 0 & 0\\
0 & 0 & 1 & 0 & 0 & 0 & 0 & 0\\
0 & 0 & 0 & 1 & 0 & 0 & 0 & 0\\
1 & 0 & 0 & 0 & 0 & 0 & 0 & 0\end{bmatrix}.\]
That is $SL_A=L_A\transpose S$ and $SC_0=L_0\transpose$.  

The following matrices are related to $S$:
\[P=\begin{bmatrix}1 & 0 & 0 & 0 & 0 & 0 & 0 & 0\\
0 & 0 & 0 & 1 & 0 & 0 & 0 & 0\\
0 & 0 & 1 & 0 & 0 & 0 & 0 & 0\\
0 & 1 & 0 & 0 & 0 & 0 & 0 & 0\\
0 & 0 & 0 & 0 & 1 & 0 & 0 & 0\\
0 & 0 & 0 & 0 & 0 & 1 & 0 & 0\\
0 & 0 & 0 & 0 & 0 & 0 & 1 & 0\\
0 & 0 & 0 & 0 & 0 & 0 & 0 & 1\end{bmatrix},\]
\[Y=\begin{bmatrix}1 & 0 & 0 & 0 & 0 & 0 & 0 & 1\\
0 & 1 & 0 & 0 & 0 & 0 & 1 & 0\\
0 & 0 & 1 & 0 & 0 & 1 & 0 & 0\\
0 & 0 & 0 & 1 & 1 & 0 & 0 & 0\\
0 & 0 & 0 & 1 & -1 & 0 & 0 & 0\\
0 & 0 & 1 & 0 & 0 & -1 & 0 & 0\\
0 & 1 & 0 & 0 & 0 & 0 & -1 & 0\\
1 & 0 & 0 & 0 & 0 & 0 & 0 & -1\end{bmatrix},\]
and $U=1/sqrt{2} \ PY$.  The relation is $S=UJU\transpose$.  The matrix $U$ comes up in the proof of Theorem \ref{thm:S-sym}.

We next find the matrix pencil $L_{\tilde{A}}=JU\transpose L_A U^{-\mathrm{T}}$.  It is

\renewcommand{\arraystretch}{3}
\noindent\resizebox{0.98\textwidth}{!}{
$\begin{bmatrix}
0 & \dfrac{{x_3}}{2} & \dfrac{{x_2}}{2} & \dfrac{{x_1}}{2} & -\dfrac{{x_1}}{2} & -\dfrac{{x_2}}{2} & -\dfrac{{x_3}}{2} & 0\\
\dfrac{{x_3}}{2} 
& \dfrac{{b_{(1,0,2)}, {x_1}+{b_{(0,1,2)}}\, {x_2}+{b_{(0,0,3)}}\, {x_3}}}{2} 
& \dfrac{ {b_{(1,1,1)}}\, {x_1}+{b_{(0,2,1)}}\, {x_2}}{4} 
& \dfrac{{b_{(2,0,1)\, {x_1}}}}{4} 
& \dfrac{{b_{(2,0,1)\, {x_1}}}}{4} 
& \dfrac{ {b_{(1,1,1)}}\, {x_1}+{b_{(0,2,1)}}\, {x_2}}{4} 
& \dfrac{{b_{(1,0,2)}, {x_1}+{b_{(0,1,2)}}\, {x_2}+{b_{(0,0,3)}}\, {x_3}}}{2} 
& \dfrac{{x_3}}{2}\\
\dfrac{{x_2}}{2} 
& \dfrac{{b_{(1,1,1)}}\, {x_1}+{b_{(0,2,1)}}\, {x_2}}{4} 
& \dfrac{{b_{(1,2,0)}\, {x_1}}+{b_{(0,3,0)}}\, {x_2}}{2} 
& \dfrac{{b_{(2,1,0)}}\, {x_1}}{4} 
& \dfrac{{b_{(2,1,0)}}\, {x_1}}{4} 
& \dfrac{{b_{(1,2,0)}\, {x_1}}+{b_{(0,3,0)}}\, {x_2}}{2}
& \dfrac{{b_{(1,1,1)}}\, {x_1}+{b_{(0,2,1)}}\, {x_2}}{4} 
& \dfrac{{x_2}}{2}\\
\dfrac{{x_1}}{2} 
& \dfrac{{b_{(2,0,1)}\, {x_1}}}{4} 
& \dfrac{{b_{(2,1,0)}\, {x_1}}}{4} 
& \dfrac{{b_{(3,0,0)}}\, {x_1}}{2} 
& \dfrac{{b_{(3,0,0)}}\, {x_1}}{2} 
& \dfrac{{b_{(2,1,0)}\, {x_1}}}{4} 
& \dfrac{{b_{(2,0,1)}\, {x_1}}}{4}
& \dfrac{{x_1}}{2}\\
-\dfrac{{x_1}}{2} 
& \dfrac{{b_{(2,0,1)}\, {x_1}}}{4} 
& \dfrac{{b_{(2,1,0)}\, {x_1}}}{4} 
& \dfrac{{b_{(3,0,0)}}\, {x_1}}{2} 
& \dfrac{{b_{(3,0,0)}}\, {x_1}}{2} 
& \dfrac{{b_{(2,1,0)}\, {x_1}}}{4} 
& \dfrac{{b_{(2,0,1)}\, {x_1}}}{4}
& -\dfrac{{x_1}}{2}\\
-\dfrac{{x_2}}{2} 
& \dfrac{{b_{(1,1,1)}}\, {x_1}+{b_{(0,2,1)}}\, {x_2}}{4} 
& \dfrac{{b_{(1,2,0)}\, {x_1}}+{b_{(0,3,0)}}\, {x_2}}{2} 
& \dfrac{{b_{(2,1,0)}}\, {x_1}}{4} 
& \dfrac{{b_{(2,1,0)}}\, {x_1}}{4} 
& \dfrac{{b_{(1,2,0)}\, {x_1}}+{b_{(0,3,0)}}\, {x_2}}{2}
& \dfrac{{b_{(1,1,1)}}\, {x_1}+{b_{(0,2,1)}}\, {x_2}}{4} 
& -\dfrac{{x_2}}{2}\\
-\dfrac{{x_3}}{2} 
& \dfrac{{b_{(1,0,2)}, {x_1}+{b_{(0,1,2)}}\, {x_2}+{b_{(0,0,3)}}\, {x_3}}}{2} 
& \dfrac{ {b_{(1,1,1)}}\, {x_1}+{b_{(0,2,1)}}\, {x_2}}{4} 
& \dfrac{{b_{(2,0,1)\, {x_1}}}}{4} 
& \dfrac{{b_{(2,0,1)\, {x_1}}}}{4} 
& \dfrac{ {b_{(1,1,1)}}\, {x_1}+{b_{(0,2,1)}}\, {x_2}}{4} 
& \dfrac{{b_{(1,0,2)}, {x_1}+{b_{(0,1,2)}}\, {x_2}+{b_{(0,0,3)}}\, {x_3}}}{2} 
& -\dfrac{{x_3}}{2}\\
0 & \dfrac{{x_3}}{2} & \dfrac{{x_2}}{2} & \dfrac{{x_1}}{2} & -\dfrac{{x_1}}{2} & -\dfrac{{x_2}}{2} & -\dfrac{{x_3}}{2} & 0\end{bmatrix}$}
\renewcommand{\arraystretch}{1}

\noindent Now, define $\tilde{L}=L_0U^{-\mathrm{T}}$, where $L_0=[0 , 0 , 0 , 0 , 0 , 0 , 0 , 1]$.
This is 
$\tilde{L}=[1/{\sqrt{2}} , 0 , 0 , 0 , 0 , 0 , 0 , -1/\sqrt{2}]$.  We set $\tilde{C}=\tilde{L}\transpose$.  Now, 
\[\tilde{C}\tilde{C}\transpose=\begin{bmatrix}\frac{1}{2} & 0 & 0 & 0 & 0 & 0 & 0 & -\frac{1}{2}\\
0 & 0 & 0 & 0 & 0 & 0 & 0 & 0\\
0 & 0 & 0 & 0 & 0 & 0 & 0 & 0\\
0 & 0 & 0 & 0 & 0 & 0 & 0 & 0\\
0 & 0 & 0 & 0 & 0 & 0 & 0 & 0\\
0 & 0 & 0 & 0 & 0 & 0 & 0 & 0\\
0 & 0 & 0 & 0 & 0 & 0 & 0 & 0\\
-\frac{1}{2} & 0 & 0 & 0 & 0 & 0 & 0 & \frac{1}{2}\end{bmatrix}.\]

Note that $P(x)=\tilde{L}(J-L_{\tilde{A}})^{-1}\tilde{L}\transpose$.  The matrix pencil $J-\tilde{C}\tilde{C}\transpose-L_{\tilde{A}}$ equals 

\renewcommand{\arraystretch}{3}
\noindent\resizebox{0.98\textwidth}{!}{$\displaystyle
\begin{bmatrix}
\dfrac{1}{2} & -\dfrac{{x_3}}{2} & -\dfrac{{x_2}}{2} & -\dfrac{{x_1}}{2} & \dfrac{{x_1}}{2} & \dfrac{{x_2}}{2} & \dfrac{{x_3}}{2} & \dfrac{1}{2}\\
-\dfrac{{x_3}}{2} 
& 1-\dfrac{{b_{(1,0,2)}, {x_1}+{b_{(0,1,2)}}\, {x_2}+{b_{(0,0,3)}}\, {x_3}}}{2} 
& -\dfrac{ {b_{(1,1,1)}}\, {x_1}+{b_{(0,2,1)}}\, {x_2}}{4} 
& -\dfrac{{b_{(2,0,1)\, {x_1}}}}{4} 
& -\dfrac{{b_{(2,0,1)\, {x_1}}}}{4} 
& -\dfrac{ {b_{(1,1,1)}}\, {x_1}+{b_{(0,2,1)}}\, {x_2}}{4} 
& -\dfrac{{b_{(1,0,2)}, {x_1}+{b_{(0,1,2)}}\, {x_2}+{b_{(0,0,3)}}\, {x_3}}}{2} 
& -\dfrac{{x_3}}{2}\\
-\dfrac{{x_2}}{2} 
& -\dfrac{{b_{(1,1,1)}}\, {x_1}+{b_{(0,2,1)}}\, {x_2}}{4} 
& 1-\dfrac{{b_{(1,2,0)}\, {x_1}}+{b_{(0,3,0)}}\, {x_2}}{2} 
& -\dfrac{{b_{(2,1,0)}}\, {x_1}}{4} 
& -\dfrac{{b_{(2,1,0)}}\, {x_1}}{4} 
& -\dfrac{{b_{(1,2,0)}\, {x_1}}+{b_{(0,3,0)}}\, {x_2}}{2}
& -\dfrac{{b_{(1,1,1)}}\, {x_1}+{b_{(0,2,1)}}\, {x_2}}{4} 
& -\dfrac{{x_2}}{2}\\
-\dfrac{{x_1}}{2} 
& -\dfrac{{b_{(2,0,1)}\, {x_1}}}{4} 
& -\dfrac{{b_{(2,1,0)}\, {x_1}}}{4} 
& 1-\dfrac{{b_{(3,0,0)}}\, {x_1}}{2} 
& -\dfrac{{b_{(3,0,0)}}\, {x_1}}{2} 
& -\dfrac{{b_{(2,1,0)}\, {x_1}}}{4} 
& -\dfrac{{b_{(2,0,1)}\, {x_1}}}{4}
& -\dfrac{{x_1}}{2}\\
\dfrac{{x_1}}{2} 
& -\dfrac{{b_{(2,0,1)}\, {x_1}}}{4} 
& -\dfrac{{b_{(2,1,0)}\, {x_1}}}{4} 
& -\dfrac{{b_{(3,0,0)}}\, {x_1}}{2} 
& -1-\dfrac{{b_{(3,0,0)}}\, {x_1}}{2} 
& -\dfrac{{b_{(2,1,0)}\, {x_1}}}{4} 
& -\dfrac{{b_{(2,0,1)}\, {x_1}}}{4}
& \dfrac{{x_1}}{2}\\
\dfrac{{x_2}}{2} 
& -\dfrac{{b_{(1,1,1)}}\, {x_1}+{b_{(0,2,1)}}\, {x_2}}{4} 
& -\dfrac{{b_{(1,2,0)}\, {x_1}}+{b_{(0,3,0)}}\, {x_2}}{2} 
& -\dfrac{{b_{(2,1,0)}}\, {x_1}}{4} 
& -\dfrac{{b_{(2,1,0)}}\, {x_1}}{4} 
& -1-\dfrac{{b_{(1,2,0)}\, {x_1}}+{b_{(0,3,0)}}\, {x_2}}{2}
& -\dfrac{{b_{(1,1,1)}}\, {x_1}+{b_{(0,2,1)}}\, {x_2}}{4} 
& \dfrac{{x_2}}{2}\\
\dfrac{{x_3}}{2} 
& -\dfrac{{b_{(1,0,2)}, {x_1}+{b_{(0,1,2)}}\, {x_2}+{b_{(0,0,3)}}\, {x_3}}}{2} 
& -\dfrac{ {b_{(1,1,1)}}\, {x_1}+{b_{(0,2,1)}}\, {x_2}}{4} 
& -\dfrac{{b_{(2,0,1)\, {x_1}}}}{4} 
& -\dfrac{{b_{(2,0,1)\, {x_1}}}}{4} 
& -\dfrac{ {b_{(1,1,1)}}\, {x_1}+{b_{(0,2,1)}}\, {x_2}}{4} 
& -1-\dfrac{{b_{(1,0,2)}, {x_1}+{b_{(0,1,2)}}\, {x_2}+{b_{(0,0,3)}}\, {x_3}}}{2} 
& \dfrac{{x_3}}{2}\\
\dfrac{1}{2} & -\dfrac{{x_3}}{2} & -\dfrac{{x_2}}{2} & -\dfrac{{x_1}}{2} & \dfrac{{x_1}}{2} & \dfrac{{x_2}}{2} & \dfrac{{x_3}}{2} & -\dfrac{3}{2}\end{bmatrix}.$}
\renewcommand{\arraystretch}{1}

The determinant $\det(J)\det(J-\tilde{C}\tilde{C}\transpose-L_{\tilde{A}})$ equals $1-P(\mathbf{x})$.  When $x_3=1$, $p(x)=1-P(\mathbf{x})$.  In this case $\det(J)=1$.  In general, $\det(J)$ can be predetermined according to the parity of $N/2$.  For the case where $N/2$ is odd, we would set $P$ to be the homogenization of $-p(\mathbf{x})$.  After substituting $x_{3}=1$ in the matrix pencil $J-\tilde{C}\tilde{C}\transpose-L_{\tilde{A}}$,
\[p(x)=\left.\det(J-\tilde{C}\tilde{C}\transpose-L_{\tilde{A}})\right|_{ x_{3}=1}.\]
Thus $p(x)$ is the determinant of

\renewcommand{\arraystretch}{3}
\noindent\resizebox{0.98\textwidth}{!}{$\displaystyle
\begin{bmatrix}\dfrac{1}{2} & -\dfrac{{x_3}}{2} & -\dfrac{{x_2}}{2} & -\dfrac{{x_1}}{2} & \dfrac{{x_1}}{2} & \dfrac{{x_2}}{2} & \dfrac{{x_3}}{2} & \dfrac{1}{2}\\
-\dfrac{{x_3}}{2} & \dfrac{6 {x_1}+9 {x_2}+10 {x_3}+2}{2} & \dfrac{5 {x_1}+8 {x_2}}{4} & \dfrac{3 {x_1}}{4} & \dfrac{3 {x_1}}{4} & \dfrac{5 {x_1}+8 {x_2}}{4} & \dfrac{6 {x_1}+9 {x_2}+10 {x_3}}{2} & -\dfrac{{x_3}}{2}\\
-\dfrac{{x_2}}{2} & \dfrac{5 {x_1}+8 {x_2}}{4} & \dfrac{4 {x_1}+7 {x_2}+2}{2} & \dfrac{{x_1}}{2} & \dfrac{{x_1}}{2} & \dfrac{4 {x_1}+7 {x_2}}{2} & \dfrac{5 {x_1}+8 {x_2}}{4} & -\dfrac{{x_2}}{2}\\
-\dfrac{{x_1}}{2} & \dfrac{3 {x_1}}{4} & \dfrac{{x_1}}{2} & \dfrac{{x_1}+2}{2} & \dfrac{{x_1}}{2} & \dfrac{{x_1}}{2} & \dfrac{3 {x_1}}{4} & -\dfrac{{x_1}}{2}\\
\dfrac{{x_1}}{2} & \dfrac{3 {x_1}}{4} & \dfrac{{x_1}}{2} & \dfrac{{x_1}}{2} & \dfrac{{x_1}-2}{2} & \dfrac{{x_1}}{2} & \dfrac{3 {x_1}}{4} & \dfrac{{x_1}}{2}\\
\dfrac{{x_2}}{2} & \dfrac{5 {x_1}+8 {x_2}}{4} & \dfrac{4 {x_1}+7 {x_2}}{2} & \dfrac{{x_1}}{2} & \dfrac{{x_1}}{2} & \dfrac{4 {x_1}+7 {x_2}-2}{2} & \dfrac{5 {x_1}+8 {x_2}}{4} & \dfrac{{x_2}}{2}\\
\dfrac{{x_3}}{2} & \dfrac{6 {x_1}+9 {x_2}+10 {x_3}}{2} & \dfrac{5 {x_1}+8 {x_2}}{4} & \dfrac{3 {x_1}}{4} & \dfrac{3 {x_1}}{4} & \dfrac{5 {x_1}+8 {x_2}}{4} & \dfrac{6 {x_1}+9 {x_2}+10 {x_3}-2}{2} & \dfrac{{x_3}}{2}\\
\dfrac{1}{2} & -\dfrac{{x_3}}{2} & -\dfrac{{x_2}}{2} & -\dfrac{{x_1}}{2} & \dfrac{{x_1}}{2} & \dfrac{{x_2}}{2} & \dfrac{{x_3}}{2} & -\dfrac{3}{2}\end{bmatrix}.
$}
\renewcommand{\arraystretch}{1}

\label{ex:DetRep}
\end{example}

\section{Stable approximation based on Quarrez's construction}
\label{sec:StabApproxQuarrez}

Given $g(\mathbf{u})$, we computed an approximating stable polynomial $p(\mathbf{u})$ in four steps.  First, a symmetric determinantal representation is computed using Quarez's construction detailed in Section \ref{sec:SymDet}.  Second, the eigendecomposition of each matrix $A_1,A_2,\dots,A_n$ is computed using the QR algorithm with Wilkinson shift \cite{Golub:1996}.  Each symmetric matrix $A_j$ is factored into eigendecomposition $Q_j\Lambda_jQ_j^\mathrm{T}$, where $\Lambda$ is diagonal and $Q_j$ is unitary.  Third, each negative eigenvalue appearing on the diagonal of $\Lambda_j$ in the eigendecomposition of $A_j$ is replaced by zero.  The result of this substitution are the positive semidefinite matrices $B_j=Q_j\Lambda^{\geq 0}_jQ_j^\mathrm{T}$ for $1 \leq j \leq n$.  The matrix $B_j$ is the closest positive definite matrix to $A_j$ in the Froebinius norm\cite{higham_computing_1988}.  Last, the computed determinant of the linear pencil $\det(A_0+B_1x_1+B_2x_2+\cdots+B_nx_n)$ is the approximating stable polynomial $p(\mathbf{u})$.  This determinant is computed by Gaussian elimination.  

Overall, this method of approximation did not give good results.  When the input polynomial is stable, the approximating polynomial and stable input polynomial should ideally be identical.  However, the matrices in the construction of Quarez's symmetric matrix pencil are not PSD for any input polynomial.  Indeed, matrices in Quarez's symmetric linear pencil have zero diagonal entries in rows and columns that have nonzero entries, and such matrices are not PSD.  

Because of this obstacle, three kinds of tests were tried to improve the absolute error of the approximation.  The meaning of the different tests follows:
\begin{enumerate}[(I)]
\item  Result of replacing symmetric matrices with nearest PSD.  Method I approximates polynomial
\[\det\left(A_0+A_1x_1\right).\]
\item  Result of dividing the polynomial by its constant, then finding nearest PSD.  Method II approximates polynomial
\[p(0)\det\left(\tilde{A}_0+\tilde{A_1}x_1\right)\]
\item  Result of dividing the polynomial by its constant then converting pencil to the form below.  Method III approximates
\[p(0)\det(J)\det \left(J + V^{-1}\tilde{A}_1V^{-\mathrm{T}} x_1 \right),\]
where
\[V^{-1}=|D_0|^{-1/2}Q_0\transpose\]
with $\det V=\det V^{-1}=1$.
\end{enumerate}

\subsection{Single variable code testing}

\noindent{\bf First Test:} Input unstable polynomial $2(x^2+1)$.  Outputs are
\begin{enumerate}[(I)]
\item $-x+2$.
\item $0.25x^2-1.5x+2$.
\item $0.17586 {{x}^{2}}-1.23586 x+2$.
\end{enumerate}

\begin{figure}[H]
\hspace*{\fill}
\includegraphics[scale=1]{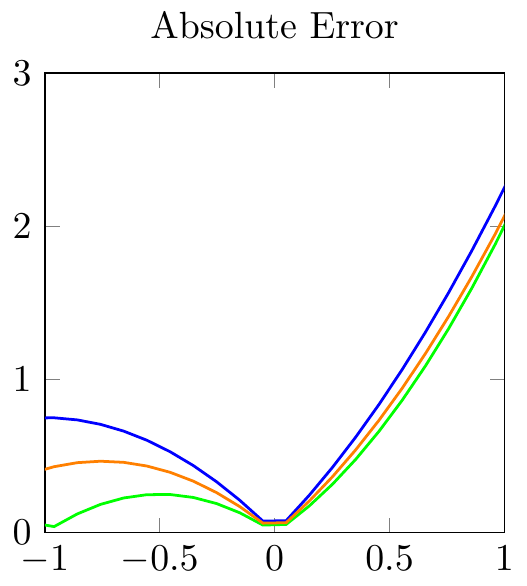}
\includegraphics[scale=1]{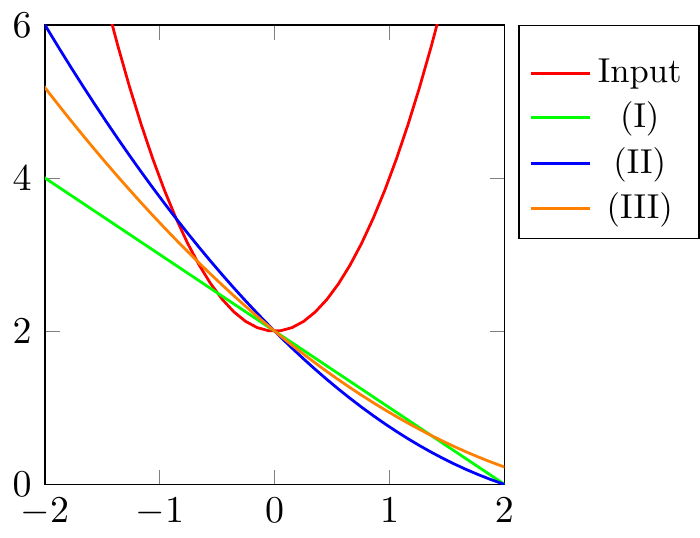}
\hspace*{\fill}
\caption{First test (unstable polynomial)}
\label{fig:1stTest}
\end{figure}

\noindent{\bf Second Test:} Input stable $(x-3)(x+5)=x^2+2x-15$.  Outputs are
\begin{enumerate}[(I)]
\item $0.8649186937918076x^2-0.5062305898757611x-15$.
\item $0.5295084x^2-9.50623058x-15$.
\item $2.84460 {{x}^{2}}-28.12852 x-15$.
\end{enumerate}

\begin{figure}[H]
\hspace*{\fill}
\includegraphics[scale=1]{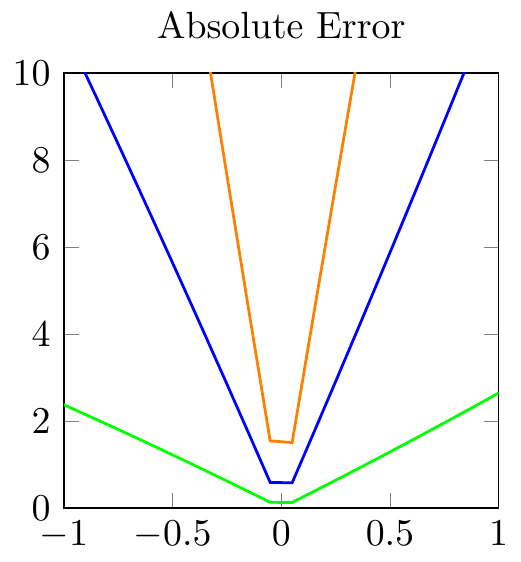}
\includegraphics[scale=1]{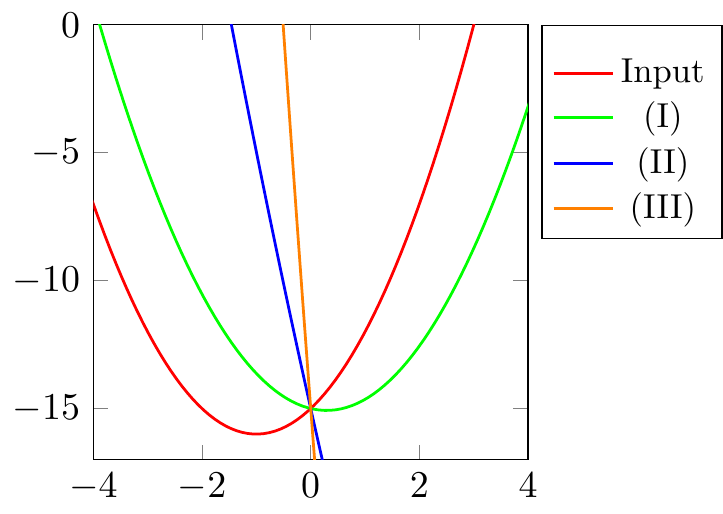}
\hspace*{\fill}
\caption{Second test (stable polynomial)}
\label{fig:2ndTest}
\end{figure}

\noindent{\bf Third Test:} Input stable $(3x+4)(2x-8)=6x^2-16x-32$.  Outputs are
\begin{enumerate}[(I)]
\item $11.33145636483897x^2-16.51731858291348x-32$.
\item $0.35884x^2+16.017318x-32$.
\item $0.00843999{{x}^{2}}+6.85904 x-32$.
\end{enumerate}
\begin{figure}[H]
\hspace*{\fill}
\includegraphics[scale=1]{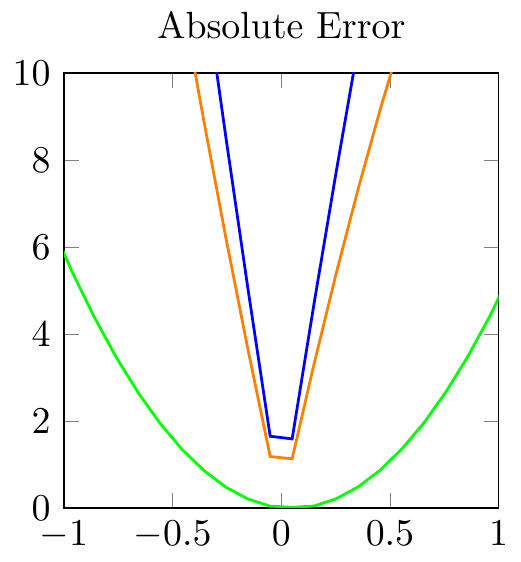}
\includegraphics[scale=1]{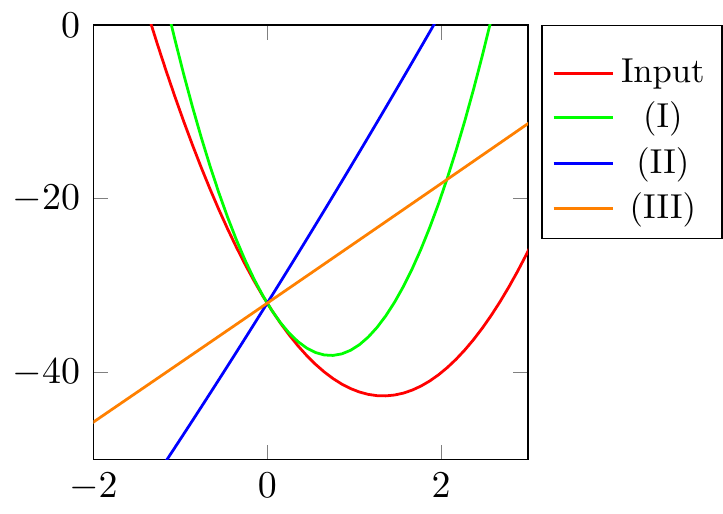}
\hspace*{\fill}
\caption{Third test (stable polynomial)}
\label{fig:3rdTest}
\end{figure}

\noindent{\bf Forth Test:} Input unstable polynomial $2(x^3+1)$.  Outputs are
\begin{enumerate}[(I)]
\item $-0.5x+2$.
\item $-x+2$.
\item $-2.0686569566293x+2$.
\end{enumerate}
\begin{figure}[H]
\hspace*{\fill}
\includegraphics[scale=1]{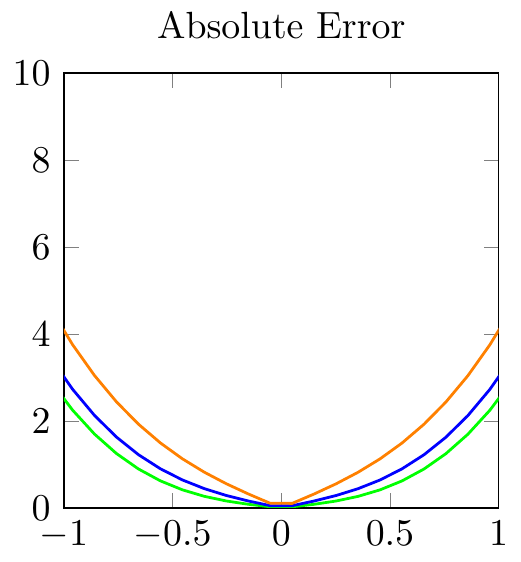}
\includegraphics[scale=1]{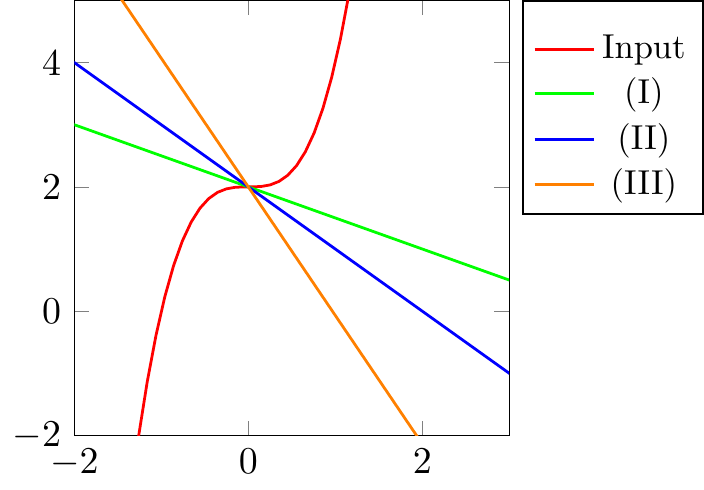}
\hspace*{\fill}
\caption{Fourth test (unstable polynomial)}
\label{fig:4thTest}
\end{figure}

\noindent{\bf Fifth Test:} Input stable polynomial $(x-2)^2(x+3)=x^3-x^2-8x+12$.  Outputs are
\begin{enumerate}[(I)]
\item $-.590553x^2 -8.503142x +12$.
\item $-.46780x^2-11.003142x+12$.
\item $-2.31755x^2-2.22928x+12$.
\end{enumerate}
\begin{figure}[H]
\hspace*{\fill}
\includegraphics[scale=1]{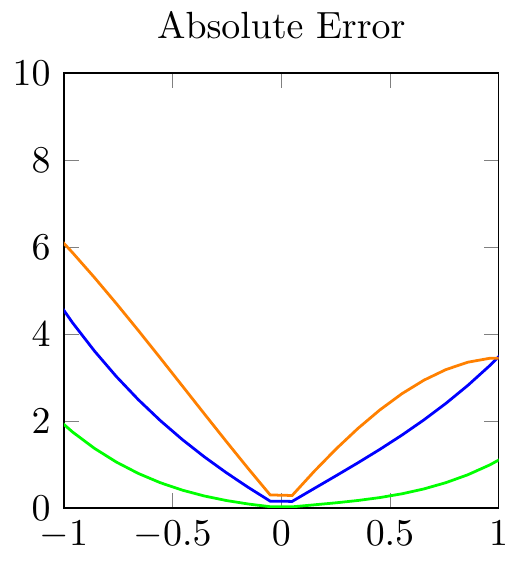}
\includegraphics[scale=1]{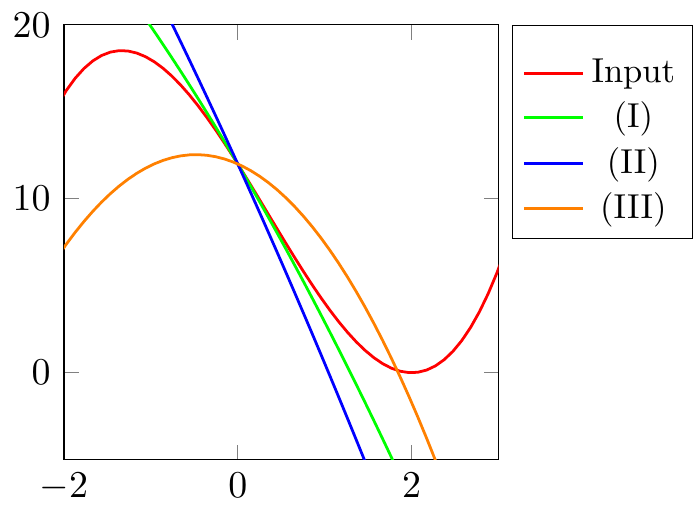}
\hspace*{\fill}
\caption{Fifth test (stable polynomial)}
\label{fig:5thTest}
\end{figure}

\noindent{\bf Sixth Test:} Input stable polynomial $(x-1)(x+2)(x+4)=x^3+5x^2+2x-8$.  Outputs are

\begin{enumerate}[(I)]
\item $3.025235x^2 -.9218419x -8$
\item $2.799269x^2+6.421841x-8$
\item $4.407706 x^2-8.2145986 x-8$
\end{enumerate}
\begin{figure}[H]
\hspace*{\fill}
\includegraphics[scale=1]{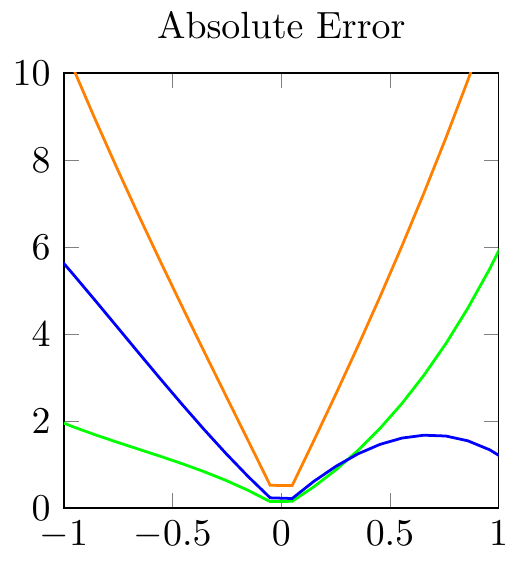}
\includegraphics[scale=1]{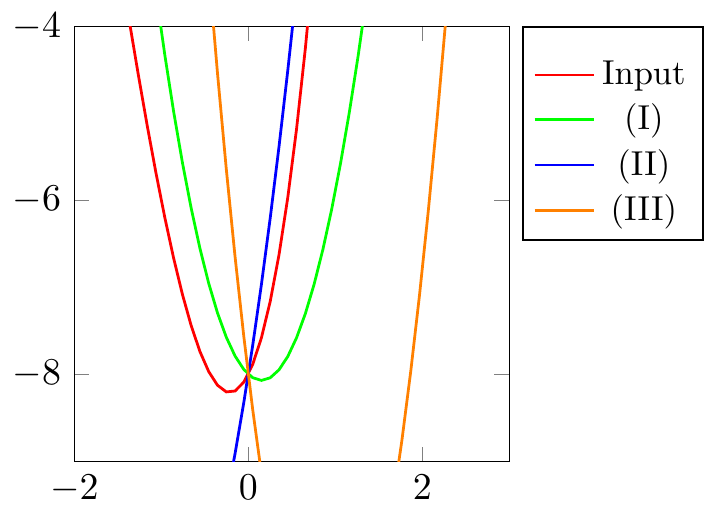}
\hspace*{\fill}
\caption{Sixth test (stable polynomial)}
\label{fig:6thTest}
\end{figure}

\noindent{\bf Seventh Test:} Input unstable polynomial $2(x^4+1)=2x^4+2$.  Outputs are
\begin{enumerate}[(I)]
\item $0.25x^4-0.5x^3+1.5x^2+x+2$
\item $0.1875x^4-.625x^3-x^2+2x+2$
\item $0.8005825 {{x}^{3}}-0.8005825 {{x}^{2}}-2x+2$
\end{enumerate}
\begin{figure}[H]
\hspace*{\fill}
\includegraphics[scale=1]{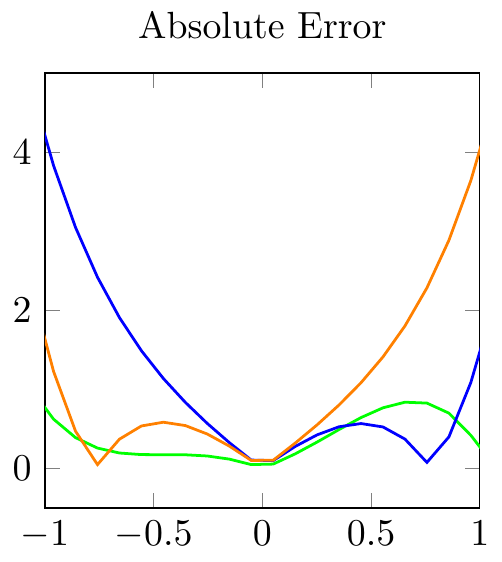}
\includegraphics[scale=1]{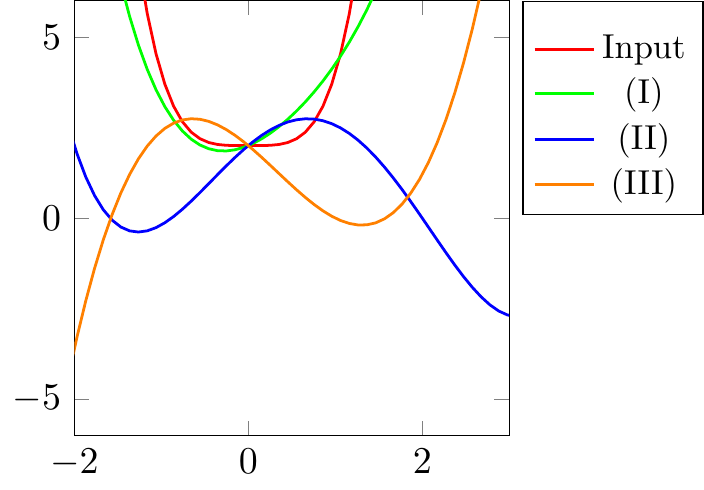}
\hspace*{\fill}
\caption{Seventh test (unstable polynomial)}
\label{fig:7thTest}
\end{figure}

\noindent{\bf Eighth Test:} Input stable polynomial $(x-2)^2(x+3)^2=x^4+2x^3-11x^2-12x+36$.  Outputs are
\begin{enumerate}[(I)]
\item $0.0159055x^5+9.981988x^4-0.7308409x^3-44.810846x^2+1.27689523x+36$
\item $0.00044182x^5+0.438715x^4-5.094134x^3-10.692309x^2+36.276895x+36$
\item $0.00016231x^5+0.0615861x^4-18.59065 x^3-23.85552 x^2+30.68338 x+36$
\end{enumerate}
\begin{figure}[ht]
\hspace*{\fill}
\includegraphics[scale=1]{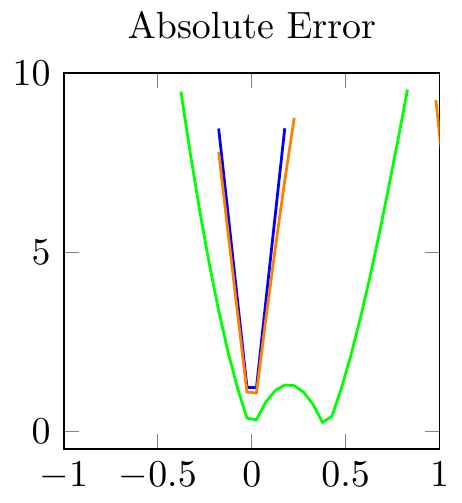}
\includegraphics[scale=1]{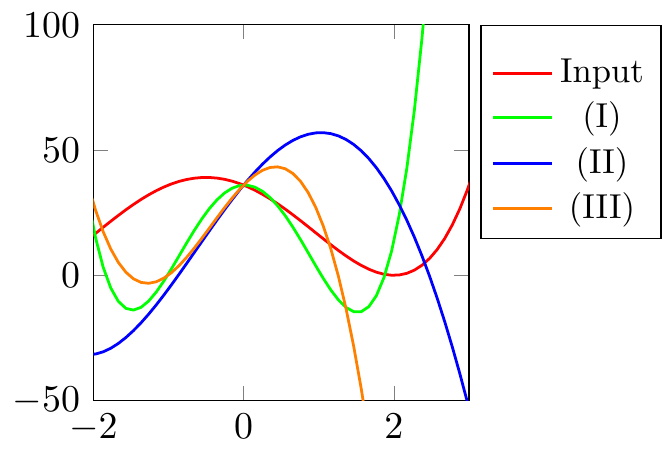}
\hspace*{\fill}
\caption{Eighth test (stable polynomial)}
\label{fig:8thTest}
\end{figure}

\noindent{\bf Ninth Test:} Input stable polynomial $(3x+4)(2x-3)(4x+5)(x-7)=24x^4-142x^3-235x^2+311x+420$.  Outputs are
\begin{enumerate}[(I)]
\item $45366.84581x^{12}-2885.25820x^{11}+111.31554x^{10}-4.24957x^9\newline+0.176432x^8 -0.005266458x^7
+0.0002677091x^6-.000005539157x^5\newline+1833.79782x^4-127.116239x^3-10232.78663x^2+317.570619x+420$
\item $35.716124x^4-179.4912389x^3-233.116026x^2+736.570619x+420$
\item $-144.33366x^4-466.80303x^3-216.00197x^2+520.24111 x+420$
\end{enumerate}
\begin{figure}[ht]
\hspace*{\fill}
\includegraphics[scale=1]{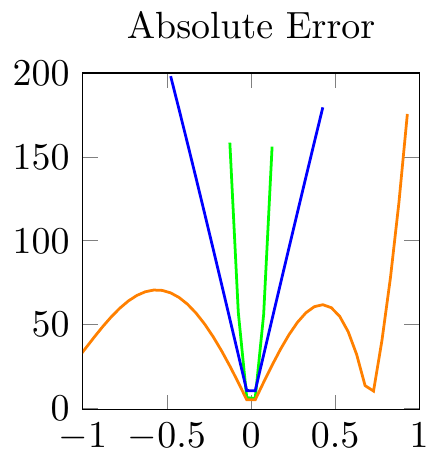}
\includegraphics[scale=1]{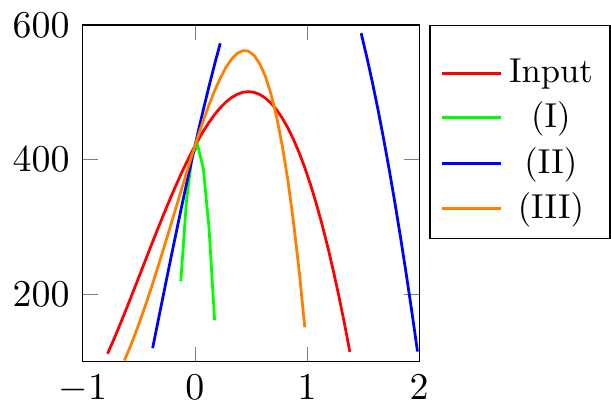}
\hspace*{\fill}
\caption{Ninth test (stable polynomial)}
\label{fig:9thTest}
\end{figure}

\subsection{Two variable code testing}

The same three kinds of tests were performed on bivariable polynomials.  The meaning of the different tests follows:

\begin{enumerate}[(I)]
\item  Result of replacing symmetric matrices with nearest PSD.  Method I approximates polynomial
\[\det\left(A_0+A_1x+A_2y\right)\]
\item  Result of dividing the polynomial by its constant, then finding the nearest PSD.  Method II approximates the polynomial
\[p(0)\det\left(\tilde{A}_0+\tilde{A_1}x+\tilde{A_2} y\right)\]
\item  Result of dividing the polynomial by its constant then converting the pencil to the form below.  Method III approximates
\[p(0)\det \left(J + V^{-1}\tilde{A_1}V^{-\mathrm{T}}x+ V^{-1}\tilde{A_2}V^{-\mathrm{T}}y \right),\]
where
\[V^{-1}=|D_0|^{-1/2}Q_0\transpose\]
with $\det V=\det V^{-1}=1$.
\end{enumerate}

\noindent{\bf First two variable test:} Unstable polynomial $p(x,y)=5(xy+1)$.  $(i,i)$ is a root in $\mathbb{H}^2$.  Outputs are

\begin{enumerate}[(I)]
\item $-0.3125x^2y-1.875xy+0.5y+1.75x+5$

\item $-0.3125x^2y+0.625xy+2.5y+3.75x+5$

\item $0.864922x^2y-2.178960xy-1.803055y-1.860956x^2+0.398135x+5$
\end{enumerate}
\hspace*{\fill}\includegraphics[scale=1]{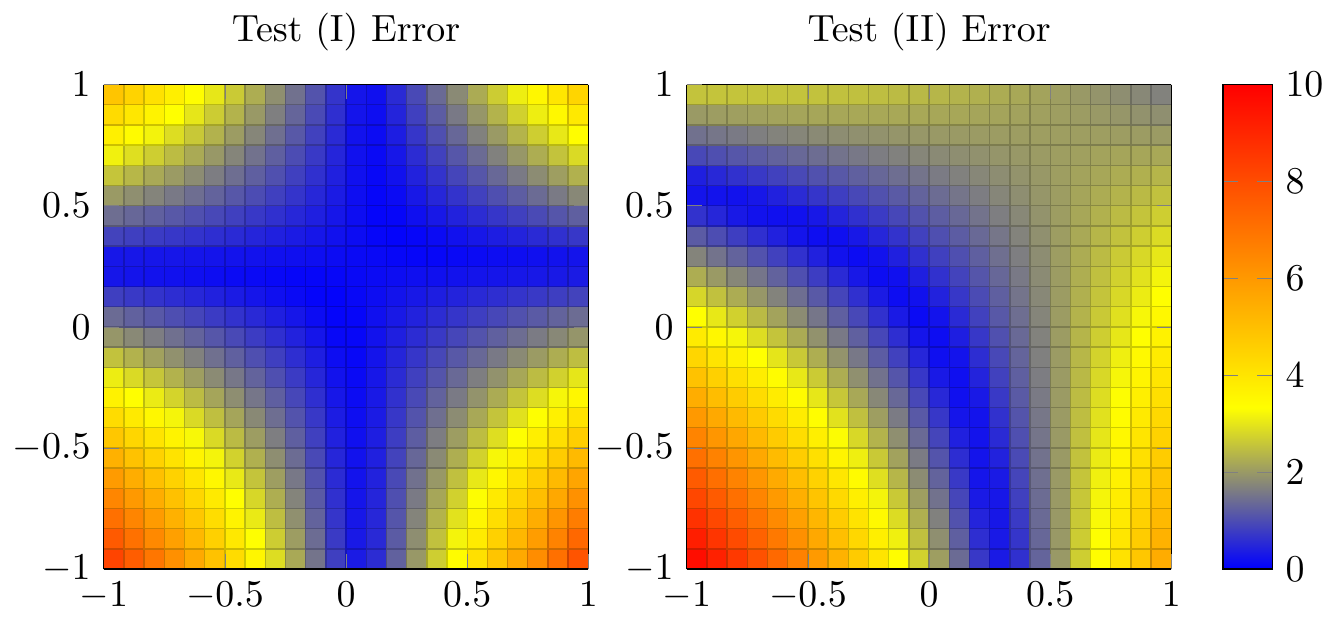}\hspace*{\fill}

\begin{figure}[H]

\hspace*{\fill}\includegraphics[scale=1]{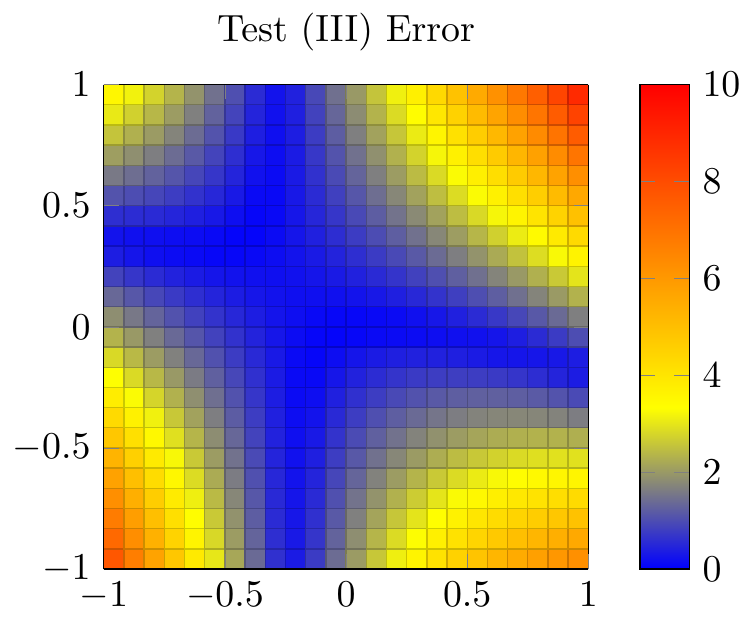} \hspace*{5.8cm}\hspace*{\fill}
\caption{First two variable test.}
\label{fig:1st2VTest}
\end{figure}

\noindent{\bf Second two variable test:} Unstable polynomial $p(x,y)=7(x^2-y^2-1)$.  One solution set contained in $\mathbb{H}^2$ is $(i\sqrt{a^2-1},ai)$ for $a>1$.  Output is 

\begin{enumerate}[(I)]

\item $-5.35937 x^2 y^2 -1.96875 xy^2 -1.96875x^2y+4.375y^2+7.875x^2+2.25y+2.25x-7$
\end{enumerate}

\begin{figure}[ht]
\hspace*{\fill}\includegraphics[scale=1]{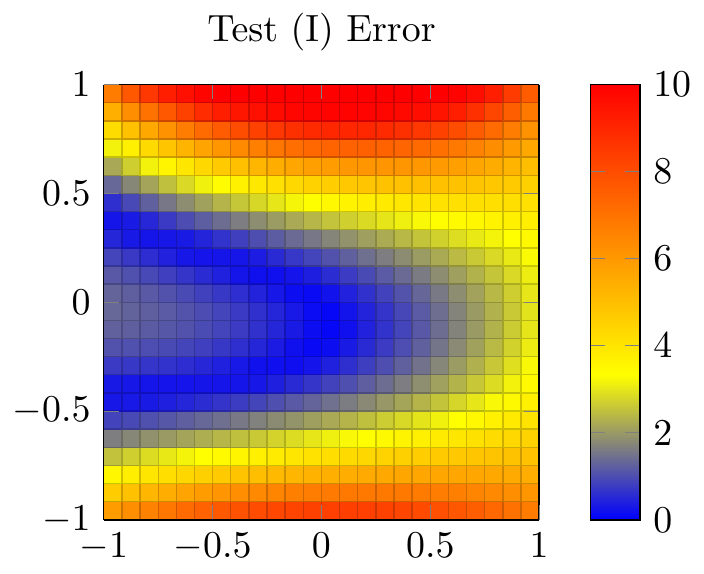}\hspace*{\fill}
\caption{Second two variable test.}
\label{fig:2nd2VTest}
\end{figure}

\section{Reducing the size of Quarrez's linear pencil}

Since the tests in Section \ref{sec:StabApproxQuarrez} did not provide satisfactory results for the case where a polynomial is stable, we continued our search for an approximation scheme in which the approximating polynomial equals the input polynomial when the input polynomial is stable.  A place to start was to reduce the size of the linear pencils in Quarrez's construction using elementary row operations.  The hope being that the negative eigenvalues inherent in Quarez's construction might be removed.  These eigenvalues of matrices appearing in the pencil occur due to zeros on the main diagonal occurring in rows which have nonzero entries.  Since these rows do not depend on the coefficients of the polynomial, they may be unnecessary.  An added bonus of reducing the size is to make the approximation method less expensive computationally.  

In this section, an example polynomial is studied for which a reduction in the size of the linear pencil in Quarrez's construction is possible.  For this example, the size of Quarrez's matrix pencil can be shrank by four in the single variable degree two case.  Examples of larger linear pencils being reduced are provided in Appendix \ref{app:LargeRed}.  These examples suggest that the size $N$ can be reduced from $2 {n+1+\lfloor d/2 \rfloor \choose n+1}$ to at most $2 {n+\lfloor d/2 \rfloor \choose n}$.  This is a reduction of $2 {n+\lfloor d/2 \rfloor \choose n+1}$.  Further evidence to support the plausibility of this conjectured size reduction is that in Quarez's construction there are $2 {n+\lfloor d/2 \rfloor \choose n+1}$ rows of the symmetric linear pencil that are independent of the coefficients of the polynomial.  

The first example we give here is reducing the size of the symmetric linear pencil given in Quarez's constuction for the stable polynomial
\[p(x)=4 {{x}^{2}}+9 x+1\]
with real roots $(-9\pm \sqrt{65})/8$.  The polynomial has a $2 \times 2$ symmetric determinantal representation,
\begin{equation} p(x)=\det\left(
\begin{bmatrix}5 & 11\\
11 & 25\end{bmatrix}x+
\begin{bmatrix}
1 & 3\\
3 & 10\end{bmatrix}
\right).
\label{OrgPen}
\end{equation}
The above representation demonstrates that $p$ is a stable polynomial---the matrix $\begin{bmatrix}5 & 11\\
11 & 25\end{bmatrix}$ is PSD.  Neither reducing the symmetric linear pencil from Quarez's construction based on the original choice of weights nor based on the diagonal choice of weights gave a determinantal representation that demostrates that $p$ is a stable polynomial.  We explored relationships between these two linear pencils and also attempted to relate them with the linear pencil in \eqref{OrgPen}.

\subsection{Original weights}
From Quarez's original method of choosing weights this polynomial can be represented as follows:

\noindent\resizebox{0.98\textwidth}{!}{
$\displaystyle p(x)=\det \left(
\begin{bmatrix}0 & 0 & -0.5 & 0.5 & 0 & 0\\
0 & -4.5 & -1 & -1 & -4.5 & 0\\
-0.5 & -1 & 0 & 0 & -1 & -0.5\\
0.5 & -1 & 0 & 0 & -1 & 0.5\\
0 & -4.5 & -1 & -1 & -4.5 & 0\\
0 & 0 & -0.5 & 0.5 & 0 & 0\end{bmatrix}x
+
\begin{bmatrix}0.5 & -0.5 & 0 & 0 & 0.5 & 0.5\\
-0.5 & 0 & 0 & 0 & -1 & -0.5\\
0 & 0 & 1 & 0 & 0 & 0\\
0 & 0 & 0 & -1 & 0 & 0\\
0.5 & -1 & 0 & 0 & -2 & 0.5\\
0.5 & -0.5 & 0 & 0 & 0.5 & -1.5\end{bmatrix}
\right)
$}

\noindent Denote the pencil $Ax+B$ by $P$.

\[P=\begin{bmatrix}0.5 & -0.5 & -0.5 x & 0.5 x & 0.5 & 0.5\\
-0.5 & -4.5 x & -x & -x & -4.5 x-1 & -0.5\\
-0.5 x & -x & 1 & 0 & -x & -0.5 x\\
0.5 x & -x & 0 & -1 & -x & 0.5 x\\
0.5 & -4.5 x-1 & -x & -x & -4.5 x-2 & 0.5\\
0.5 & -0.5 & -0.5 x & 0.5 x & 0.5 & -1.5\end{bmatrix}
\]

A series of seven row operations $M_7M_6M_5\cdots M_1$ yields
\[ \mathcal{M}= M_7M_6M_5\cdots M_1 = \begin{bmatrix}0.5 & 0.625 & 0.0 & 0.0 & -0.625 & 0.0\\
1.0 & 1.0 & 0.0 & 0.0 & 0.0 & 0.0\\
0.0 & -0.5 x & 0.75 & -0.25 & 0.5 x & 0.0\\
0.0 & x & -0.5 & 1.5 & -x & 0.0\\
1.0 & -0.75 & 0.0 & 0.0 & 0.75 & 0.0\\
-1.0 & 0.0 & 0.0 & 0.0 & 0.0 & 1.0\end{bmatrix}.\]
The row operations applied to $P$ are 
\[\mathcal{M} P \mathcal{M}^\mathrm{T} =
\begin{bmatrix}-0.5 & 0.0 & 0.0 & 0.0 & 0.0 & 0.0\\
0.0 & -4.5 x-0.5 & -1.0 x & 0.0 & 0.0 & 0.0\\
0.0 & -1.0 x & 0.5 & 0.0 & 0.0 & 0.0\\
0.0 & 0.0 & 0.0 & -2.0 & 0.0 & 0.0\\
0.0 & 0.0 & 0.0 & 0.0 & 2.0 & 0.0\\
0.0 & 0.0 & 0.0 & 0.0 & 0.0 & -2.0\end{bmatrix}.\]
By construction each row operations has $\det(M_i)=1$.  So,
\[ \det(P) = \det( \mathcal{M} P \mathcal{M}^\mathrm{T} )= (-2)^2(-0.5)(2) \det \left(
\begin{bmatrix}-4.5 x-0.5 & -x\\
-x & 0.5\end{bmatrix}
\right).\]
We may write
\begin{equation} p(x)=\det(P)=-4\det \left( 
\begin{bmatrix}-4.5 & -1\\
-1 & 0\end{bmatrix}x+\begin{bmatrix}-0.5 & 0\\
0 & 0.5\end{bmatrix}
\right).
\label{Pen1}
\end{equation}
The matrix 
\[\begin{bmatrix}-4.5 & -1\\
-1 & 0\end{bmatrix}\]
has eigenvalues $-4.71221445044902$, and $0.21221445044902$.  Unfortunately, the matrix pencil \eqref{Pen1} does not demonstrate that $p(x)$ is stable.  For that we need both to be positive ($-p$ is stable) or both negative ($p$ is stable).  We can rewrite \eqref{Pen1} as
\[p(x)=\det(P)=-\det\left( \begin{bmatrix}
2 & 0\\
0 & 2
\end{bmatrix} \right) \det \left( 
\begin{bmatrix}-4.5 & -1\\
-1 & 0\end{bmatrix}x+\begin{bmatrix}-0.5 & 0\\
0 & 0.5\end{bmatrix}
\right).
\]
or
\begin{equation}p(x)=\det(P)=-\det \left( 
\begin{bmatrix}-9 & -2\\
-2 & 0\end{bmatrix}x+\begin{bmatrix}-1 & 0\\
0 & 1\end{bmatrix}
\right).
\label{Pen1F}
\end{equation}

\subsection{Diagonal weights}
Quarez also provides a choice of weights that has a diagonal upper left inner {\color{blue} block} is

\noindent\resizebox{0.98\textwidth}{!}{
$\displaystyle p(x)= \det \left(
\begin{bmatrix}0 & 0 & -0.5 & 0.5 & 0 & 0\\
0 & {\color{blue}-4.5} & {\color{blue}0} & 0 & -4.5 & 0\\
-0.5 & {\color{blue}0} & {\color{blue}0} & 0 & 0 & -0.5\\
0.5 & 0 & 0 & 0 & 0 & 0.5\\
0 & -4.5 & 0 & 0 & -4.5 & 0\\
0 & 0 & -0.5 & 0.5 & 0 & 0\end{bmatrix}
x+
\begin{bmatrix}0.5 & -0.5 & 0 & 0 & 0.5 & 0.5\\
-0.5 & 0 & 0 & 0 & -1 & -0.5\\
0 & 0 & -1 & -2 & 0 & 0\\
0 & 0 & -2 & -3 & 0 & 0\\
0.5 & -1 & 0 & 0 & -2 & 0.5\\
0.5 & -0.5 & 0 & 0 & 0.5 & -1.5\end{bmatrix}
\right).
$}

Again, we use capitol $P$ to denote the pencil
\[P=\begin{bmatrix}0.5 & -0.5 & -0.5 {x} & 0.5 {x} & 0.5 & 0.5\\
-0.5 & -4.5x & 0 & 0 & -4.5x-1 & -0.5\\
-0.5 {x} & 0 & -1 & -2 & 0 & -0.5 {x}\\
0.5 {x} & 0 & -2 & -3 & 0 & 0.5 {x}\\
0.5 & -4.5x-1 & 0 & 0 & -4.5x-2 & 0.5\\
0.5 & -0.5 & -0.5 {x} & 0.5 {x} & 0.5 & -1.5\end{bmatrix}.\]

Ten row operations $M_{10}M_9M_8\cdots M_1$ yielded
\[\mathcal{M}=M_{10}M_9M_8\cdots M_1=
\begin{bmatrix}1 & 0.0 & 0 & 0 & 0.0 & -1\\
2 & 1.0 & 0 & 0 & -1.0 & 0\\
0 & 0.0 & 1 & 1 & 0.0 & 0\\
-1/9 & -0.5 x-1/9 & 5/8 & -3/8 & 0.5 x & 0\\
1 & 1.0 & 0 & 0 & 0.0 & 0\\
1 & 0.0 & 0 & 0 & 0.0 & 0\end{bmatrix}.
\]
Applied to $P$ this gives
\[\mathcal{M}P\mathcal{M}^\mathrm{T}
\begin{bmatrix}-2 & 0 & 0 & 0 & 0 & 0\\
0 & -2 & 0 & 0 & 0 & 0\\
0 & 0 & -8 & 0 & 0 & 0\\
0 & 0 & 0 & x/18+77/648 & 1/18 & 0\\
0 & 0 & 0 & 1/18 & -9/2x-1/2 & 0\\
0 & 0 & 0 & 0 & 0 & 1/2\end{bmatrix}.
\]
Since $\det(P) = \det( \mathcal{M} P \mathcal{M}^\mathrm{T} )$,
\[ \det(P) = (-8)(-2)^2(1/2) \det \left(
\begin{bmatrix}x/18+77/648 & 1/18\\
1/18 & -9x/2-1/2\end{bmatrix}
\right).\]
Thus,
\begin{equation} p(x)=\det(P)=-16\det \left( 
\begin{bmatrix}1/18 & 0\\
0 & -9/2\end{bmatrix}x+\begin{bmatrix}77/648 & 1/18\\
1/18 & -1/2\end{bmatrix}
\right).
\label{Pen2}
\end{equation}
Again, the matrix coefficient of $x$ in the above has mixed signed eigenvalues and the pencil appearing in \eqref{Pen2} does not indicate that  $p(x)$ is stable.  We can rewrite \eqref{Pen2} as
\[p(x)=\det(P)=-\det \left(\begin{bmatrix}
4 & 0\\
0 & 4
\end{bmatrix} \right)\det \left( 
\begin{bmatrix}1/18 & 0\\
0 & -9/2\end{bmatrix}x+\begin{bmatrix}77/648 & 1/18\\
1/18 & -1/2\end{bmatrix}
\right)\]
or
\begin{equation}p(x)=\det(P)=-\det \left( 
\begin{bmatrix}2/9 & 0\\
0 & -18\end{bmatrix}x+\begin{bmatrix}77/162 & 2/9\\
2/9 & -2\end{bmatrix}
\right)
\label{Pen2F}
\end{equation}

\subsection{Relationships between pencils}

The pencils from \eqref{Pen1F} and \eqref{Pen2F},
\begin{equation}\begin{bmatrix}-9 & -2\\
-2 & 0\end{bmatrix}x+\begin{bmatrix}-1 & 0\\
0 & 1\end{bmatrix}
\label{Pen1Mod}
\end{equation}
and
\[\begin{bmatrix}2/9 & 0\\
0 & -18\end{bmatrix}x+\begin{bmatrix}77/162 & 2/9\\
2/9 & -2\end{bmatrix}\]
 are related.  For 
\[\mathcal{M}=\begin{bmatrix}\sqrt{65}/\sqrt{2} & 73 \sqrt{2}/(9 \ {{65}^{3/2}})\\
-9 \sqrt{65}/\sqrt{2} & \sqrt{2}/\sqrt{65}-73 \sqrt{2}/{65}^{3/2}\end{bmatrix},\]

\[\mathcal{M} \begin{bmatrix}2 x/9+77/162 & -2 x-73/18\\
-2 x-73/18 & 65/2\end{bmatrix} \mathcal{M}^\mathrm{T} =\begin{bmatrix}-9 x-1 & -2 x\\
-2 x & 1\end{bmatrix}.
\]

It was difficult to relate back to the original pencil.  The original pencil after raising its dimension is
\[\begin{bmatrix}5 & 11 & 0\\
11 & 25 & 0\\
0 & 0 & 0
\end{bmatrix}x+
\begin{bmatrix}
1 & 3 & 0\\
3 & 10 & 0\\
0 & 0 & 1
\end{bmatrix}=\begin{bmatrix}5 x+1 & 11 x+3 & 0\\
11 x+3 & 25 x+10 & 0\\
0 & 0 & 1\end{bmatrix}.\]
Consider the pencil from \eqref{Pen1F}.  Raising its dimension and bringing in the negative sign gives
\[P=\begin{bmatrix}-1 & 0 & 0\\
0 & -9 x-1 & -2 x\\
0 & -2 x & 1\end{bmatrix}.\]
Initially, column and row operations described by the matrix
\[M_1=\begin{bmatrix}0 & 1 & 0\\
0 & 0 & 1\\
1 & 0 & 0\end{bmatrix}\]
and its transpose move the $-1$ to the lower right: 
\[M_1 P M_1^\mathrm{T}=\begin{bmatrix}-9 x-1 & -2 x & 0\\
-2 x & 1 & 0\\
0 & 0 & -1\end{bmatrix}.\]
To change the sign of this $-1$, the symmetry of the pencil will spoiled.  Later the symmetry can be restored.

The matrix
\[M_2=\begin{bmatrix}0 & -1 & 0\\
-1 & 0 & 0\\
0 & 0 & -1\end{bmatrix}\]
has determinant $1$.

\[M_2M_1 P M_1^\mathrm{T}=\begin{bmatrix}2 x & 9 x+1 & 0\\
-1 & 2 x & 0\\
0 & 0 & 1\end{bmatrix}\]
is no longer symmetric.  After a few choices, right multiplication by
\[M_3=\begin{bmatrix}17 & -4 & 0\\
-4 & 1 & 0\\
0 & 0 & 1\end{bmatrix}\]
with determinant $1$ yields a symmetric matrix
\[M_3M_2M_1 P M_1^\mathrm{T}
\begin{bmatrix}-8 x-1 & -34 x-4 & 0\\
-34 x-4 & -145 x-17 & 0\\
0 & 0 & 1\end{bmatrix}.\]
Row and column operations can be done to get
\[M_4=\begin{bmatrix}13/10 & -3/25 & 0\\
-5/2 & 1 & 0\\
0 & 0 & 1\end{bmatrix},\]
such that
\[M_4M_3M_2M_1 P M_1^\mathrm{T} M_4^\mathrm{T}=
\begin{bmatrix}-5 x-1717/2500 & -11 x-111/100 & 0\\
-11 x-111/100 & -25 x-13/4 & 0\\
0 & 0 & 1\end{bmatrix}\]
Right multiplication by
\[M_5=\begin{bmatrix}-1 & 0 & 0\\
0 & -1 & 0\\
0 & 0 & 1\end{bmatrix},\]
yields a matrix that shows the polynomial is stable:
\[M_5M_4M_3M_2M_1 P M_1^\mathrm{T} M_4^\mathrm{T}=
\begin{bmatrix}5 x+1717/2500 & 11 x+111/100 & 0\\
11 x+111/100 & 25 x+13/4 & 0\\
0 & 0 & 1\end{bmatrix}\]
Rewritten this is
\begin{equation}
\begin{bmatrix}5 & 11 & 0\\
11 & 25 & 0\\
0 & 0 & 0\end{bmatrix}x+\begin{bmatrix}1717/2500 & 111/100 & 0\\
111/100 & 13/4 & 0\\
0 & 0 & 1\end{bmatrix},
\label{RelPen}
\end{equation}
which has the same determinant as the original pencil and has the same ``coefficient'' matrix but different ``constant'' matrix.  That is,

\noindent\resizebox{0.98\textwidth}{!}{
$\displaystyle 
\det \left(\begin{bmatrix}5 & 11 & 0\\
11 & 25 & 0\\
0 & 0 & 0
\end{bmatrix}x+
\begin{bmatrix}
1 & 3 & 0\\
3 & 10 & 0\\
0 & 0 & 1
\end{bmatrix}\right)=
\det \left(
\begin{bmatrix}5 & 11 & 0\\
11 & 25 & 0\\
0 & 0 & 0\end{bmatrix}x+\begin{bmatrix}1717/2500 & 111/100 & 0\\
111/100 & 13/4 & 0\\
0 & 0 & 1\end{bmatrix}
\right).
$}

\section{Other determinantal representations}
\label{sec:OthDetRep}

The goal of our search for other determinantal representations was to find a determinantal representation that would determine whether a polynomial is stable.   A place to start is to replace the zeros that appear on the diagonals in some of the matrices in Quarez's construction.

\subsection{First attempt at a different representation}

Set

\[A_1=\begin{bmatrix}\frac{1}{2} & 0 & -\frac{1}{2} & \frac{1}{2} & 0 & \frac{1}{2}\\
0 & a & c & c & a & 0\\
-\frac{1}{2} & c & b & -b & c & -\frac{1}{2}\\
\frac{1}{2} & c & -b & b & c & \frac{1}{2}\\
0 & a & c & c & a & 0\\
\frac{1}{2} & 0 & -\frac{1}{2} & \frac{1}{2} & 0 & \frac{1}{2}\end{bmatrix}\]
and

\[A_2=\begin{bmatrix}\frac{1}{2} & -\frac{1}{2} & 0 & 0 & \frac{1}{2} & \frac{1}{2}\\
-\frac{1}{2} & f+1 & 0 & 0 & f & -\frac{1}{2}\\
0 & 0 & 1 & 0 & 0 & 0\\
0 & 0 & 0 & -1 & 0 & 0\\
\frac{1}{2} & f & 0 & 0 & f-1 & \frac{1}{2}\\
\frac{1}{2} & -\frac{1}{2} & 0 & 0 & \frac{1}{2} & -1.5\end{bmatrix}\]

Then

\[
\det(A_1x+A_2)=8 b\, {{c}^{2}}\, {{x}^{3}}-4 c\, {{x}^{2}}+\left( -2 a-1\right)  x-2 f-1
\]	

Making substitutions $f=-1$, $a=-5$, $c=-1$, and $b=0$, we obtain
\[
\det(A_1x+A_2)=4x^2+9x+1.
\]
The matrix 
\[A_1=\begin{bmatrix}\frac{1}{2} & 0 & -\frac{1}{2} & \frac{1}{2} & 0 & \frac{1}{2}\\
0 & -5 & -1 & -1 & -5 & 0\\
-\frac{1}{2} & -1 & 0 & 0 & -1 & -\frac{1}{2}\\
\frac{1}{2} & -1 & 0 & 0 & -1 & \frac{1}{2}\\
0 & -5 & -1 & -1 & -5 & 0\\
\frac{1}{2} & 0 & -\frac{1}{2} & \frac{1}{2} & 0 & \frac{1}{2}\end{bmatrix}\]
is not positive semidefinite.

\subsection{Second attempt at a different representation}

Set

\[A_1=\begin{bmatrix}\frac{1}{2} & 0 & -\frac{1}{2} & \frac{1}{2} & 0 & \frac{1}{2}\\
0 & a & c & c & a & 0\\
-\frac{1}{2} & c & b & b & c & -\frac{1}{2}\\
\frac{1}{2} & c & b & b & c & \frac{1}{2}\\
0 & a & c & c & a & 0\\
\frac{1}{2} & 0 & -\frac{1}{2} & \frac{1}{2} & 0 & \frac{1}{2}\end{bmatrix}\]
and

\[A_2=\begin{bmatrix}\frac{1}{2} & -\frac{1}{2} & 0 & 0 & \frac{1}{2} & \frac{1}{2}\\
-\frac{1}{2} & f+1 & 0 & 0 & f & -\frac{1}{2}\\
0 & 0 & 1 & 0 & 0 & 0\\
0 & 0 & 0 & -1 & 0 & 0\\
\frac{1}{2} & f & 0 & 0 & f-1 & \frac{1}{2}\\
\frac{1}{2} & -\frac{1}{2} & 0 & 0 & \frac{1}{2} & -1.5\end{bmatrix}\]

Then

\[
\det(A_1x+A_2)=2 b\, {{x}^{3}}-4 c\, {{x}^{2}}+\left( -2 a-1\right)  x-2 f-1
\]	

Making substitutions $f=-1$, $a=-5$, $c=-1$, and $b=0$, we obtain
\[
\det(A_1x+A_2)=4x^2+9x+1.
\]
The matrix 
\[A_1=\begin{bmatrix}\frac{1}{2} & 0 & -\frac{1}{2} & \frac{1}{2} & 0 & \frac{1}{2}\\
0 & -5 & -1 & -1 & -5 & 0\\
-\frac{1}{2} & -1 & 0 & 0 & -1 & -\frac{1}{2}\\
\frac{1}{2} & -1 & 0 & 0 & -1 & \frac{1}{2}\\
0 & -5 & -1 & -1 & -5 & 0\\
\frac{1}{2} & 0 & -\frac{1}{2} & \frac{1}{2} & 0 & \frac{1}{2}\end{bmatrix}\]
is not positive semidefinite.  

Lets try to use $(x+1)(4x^2+9x+1)=4 {{x}^{3}}+13 {{x}^{2}}+10 x+1$ instead.  Making substitutions $f=-1$, $a=-5.5$, $c=-1$, and $b=-2$, gives
\[
\det(A_1x+A_2)=4 {{x}^{3}}+13 {{x}^{2}}+10 x+1.
\]
The matrix 
\[A_1=\begin{bmatrix}\frac{1}{2} & 0 & -\frac{1}{2} & \frac{1}{2} & 0 & \frac{1}{2}\\
0 & -5.5 & -\frac{13}{4} & -\frac{13}{4} & -5.5 & 0\\
-\frac{1}{2} & -\frac{13}{4} & -2 & -2 & -\frac{13}{4} & -\frac{1}{2}\\
\frac{1}{2} & -\frac{13}{4} & -2 & -2 & -\frac{13}{4} & \frac{1}{2}\\
0 & -5.5 & -\frac{13}{4} & -\frac{13}{4} & -5.5 & 0\\
\frac{1}{2} & 0 & -\frac{1}{2} & \frac{1}{2} & 0 & \frac{1}{2}\end{bmatrix}\]
is not positive semidefinite.  

\subsection{Third attempt}

Set
\[A_1=\begin{bmatrix}\frac{1}{2} & 0 & \frac{1}{2} & -\frac{1}{2} & 0 & \frac{1}{2}\\
0 & -\frac{a}{2}-\frac{1}{2} & \frac{c}{4} & \frac{c}{4} & -\frac{a}{2}-\frac{1}{2} & 0\\
\frac{1}{2} & \frac{c}{4} & -\frac{b}{2} & -\frac{b}{2} & \frac{c}{4} & \frac{1}{2}\\
-\frac{1}{2} & \frac{c}{4} & -\frac{b}{2} & -\frac{b}{2} & \frac{c}{4} & -\frac{1}{2}\\
0 & -\frac{a}{2}-\frac{1}{2} & \frac{c}{4} & \frac{c}{4} & -\frac{a}{2}-\frac{1}{2} & 0\\
\frac{1}{2} & 0 & \frac{1}{2} & -\frac{1}{2} & 0 & \frac{1}{2}\end{bmatrix}\]
and
\[A_2=\begin{bmatrix}\frac{1}{2} & -\frac{1}{2} & 0 & 0 & \frac{1}{2} & \frac{1}{2}\\
-\frac{1}{2} & f+1 & 0 & 0 & f & -\frac{1}{2}\\
0 & 0 & 1 & 0 & 0 & 0\\
0 & 0 & 0 & -1 & 0 & 0\\
\frac{1}{2} & f & 0 & 0 & f-1 & \frac{1}{2}\\
\frac{1}{2} & -\frac{1}{2} & 0 & 0 & \frac{1}{2} & -1.5\end{bmatrix}\]
Then,
\[
\det(A_1x+A_2)=b\, {{x}^{3}}+c\, {{x}^{2}}+a x-2 f-1
\]	

This time we will try to use this on stable polynomial
\[ \det (I_3x+I_3)={{x}^{3}}+3 {{x}^{2}}+3 x+1.\]
Substitutions are $b=1$, $c=a=3$, and $f=-1$.  In this case,
\[A_1=\begin{bmatrix}\frac{1}{2} & 0 & \frac{1}{2} & -\frac{1}{2} & 0 & \frac{1}{2}\\
0 & -2 & \frac{3}{4} & \frac{3}{4} & -2 & 0\\
\frac{1}{2} & \frac{3}{4} & -\frac{1}{2} & -\frac{1}{2} & \frac{3}{4} & \frac{1}{2}\\
-\frac{1}{2} & \frac{3}{4} & -\frac{1}{2} & -\frac{1}{2} & \frac{3}{4} & -\frac{1}{2}\\
0 & -2 & \frac{3}{4} & \frac{3}{4} & -2 & 0\\
\frac{1}{2} & 0 & \frac{1}{2} & -\frac{1}{2} & 0 & \frac{1}{2}\end{bmatrix}\]
is not positive semidefinite.

\subsection{Attempt using a size reduced generic linear pencil}

Here we start with matrices
\[A=\begin{bmatrix}0 & 0 & -0.5 & 0.5 & 0 & 0\\
0 & a & e & e & a & 0\\
-0.5 & e & b & b & e & -0.5\\
0.5 & e & b & b & e & 0.5\\
0 & a & e & e & a & 0\\
0 & 0 & -0.5 & 0.5 & 0 & 0\end{bmatrix}\]
and
\[B=\begin{bmatrix}-0.5 & -0.5 & 0 & 0 & 0.5 & 0.5\\
-0.5 & c & f & f & c & -0.5\\
0 & f & d & d & f & 0\\
0 & f & d & d & f & 0\\
0.5 & c & f & f & c & 0.5\\
0.5 & -0.5 & 0 & 0 & 0.5 & -0.5\end{bmatrix}.\]
The polynomial we associate with these is
\[p(x)=\det(J+Ax+B)=-2 b\, {{x}^{3}}+\left( -4 e-2 d\right) \, {{x}^{2}}+\left( -4 f-2 a\right)  x-2 c-1.\]
The pencil $P$ is
\[P=\begin{bmatrix}0.5 & -0.5 & -0.5 x & 0.5 x & 0.5 & 0.5\\
-0.5 & a x+c+1 & e x+f & e x+f & a x+c & -0.5\\
-0.5 x & e x+f & b x+d+1 & b x+d & e x+f & -0.5 x\\
0.5 x & e x+f & b x+d & b x+d-1 & e x+f & 0.5 x\\
0.5 & a x+c & e x+f & e x+f & a x+c-1 & 0.5\\
0.5 & -0.5 & -0.5 x & 0.5 x & 0.5 & -1.5\end{bmatrix}.\]
A series of row operations (each having determinant $1$) produces

\noindent\resizebox{0.98\textwidth}{!}{
$\displaystyle \mathcal{M}=\begin{bmatrix}1.0 & 0.0 & 0.0 & 0.0 & 0.0 & -1.0\\
1.0 & 1.25 & 0.0 & 0.0 & -1.25 & 0.0\\
-1.0 & 0.5 a x-0.5 x-1.0 & 0.5-0.5 a & 0.5 a-0.5 & 0.5 x-0.5 a x & 0.0\\
0.0 & -1.0 e x & 1.0 e+0.5 & 0.5-1.0 e & 1.0 e x & 0.0\\
1.0 & -0.5 a x-0.5 x+1.0 & 0.5 a+0.5 & -0.5 a-0.5 & 0.5 a x+0.5 x & 0.0\\
-0.5 & 0.375 & 0.0 & 0.0 & -0.375 & 0.0\end{bmatrix}$.}

\noindent For which,

\noindent\resizebox{0.98\textwidth}{!}{
$\displaystyle
\mathcal{M}P\mathcal{M}^\mathrm{T}=\begin{bmatrix}-2.0 & 0.0 & 0.0 & 0.0 & 0.0 & 0.0\\
0.0 & -2.0 & 0.0 & 0.0 & 0.0 & 0.0\\
0.0 & 0.0 & 1.0 x+1.0 c+0.5 & -1.0 f-0.5 a+0.5 & -1.0 c-0.5 & 0.0\\
0.0 & 0.0 & -1.0 f-0.5 a+0.5 & 1.0 b x+2.0 e+1.0 d & 1.0 f+0.5 a+0.5 & 0.0\\
0.0 & 0.0 & -1.0 c-0.5 & 1.0 f+0.5 a+0.5 & -1.0 x+1.0 c+0.5 & 0.0\\
0.0 & 0.0 & 0.0 & 0.0 & 0.0 & 0.5\end{bmatrix}
$.}

\noindent So,

\noindent\resizebox{0.98\textwidth}{!}{
$\displaystyle p(x)=(-2)^2(0.5)\det\left(\begin{bmatrix}
1.0 x+1.0 c+0.5 & -1.0 f-0.5 a+0.5 & -1.0 c-0.5\\
-1.0 f-0.5 a+0.5 & 1.0 b x+2.0 e+1.0 d & 1.0 f+0.5 a+0.5\\
-1.0 c-0.5 & 1.0 f+0.5 a+0.5 & -1.0 x+1.0 c+0.5\\
\end{bmatrix}\right)$,}

\noindent or

\noindent\resizebox{0.98\textwidth}{!}{
$\displaystyle p(x)=(-2)^2(0.5)\det\left(\begin{bmatrix}1.0  & 0.0 & 0.0\\
0.0  & 1.0b  & 0.0\\
0.0 & 0.0 & -1.0\end{bmatrix}x+
\begin{bmatrix}
1.0 c+0.5 & -1.0 f-0.5 a+0.5 & -1.0 c-0.5\\
-1.0 f-0.5 a+0.5 & 2.0 e+1.0 d & 1.0 f+0.5 a+0.5\\
-1.0 c-0.5 & 1.0 f+0.5 a+0.5 & 1.0 c+0.5\\
\end{bmatrix}\right)$.}

\noindent The first matrix appearing above has eigenvalues $\pm 1$ and $b$.  So it is not possible to determine that $p$ is stable.  

\subsection{Another kind of linear pencil}
\label{sec:AnoKin}
Since all three methods tried gave bad results, we considered a determinantal representation that arose from applying our method of approximating polynomials by stable polynomials from the previous section.  We then ask is there a transformation from a determinantal representation that demonstrates the stability of a polynomial to Quarez's construction.  This attempt was not successful.

The result of trying stable approximation based using Quarez's construction on the polynomial $(x+1)^3$ yields the stable approximating polynomial $p_{\ref{sec:AnoKin}}=1/8(3 {{x}^{3}}-8 {{x}^{2}}-28 x-8)$ with determinantal representation
\begin{equation}
\resizebox{0.90\textwidth}{!}{
\begin{minipage}{1.35\textwidth}$\displaystyle \det(A_1x+A_0)=
\det \left(\begin{bmatrix}0.25 x+0.5 & -0.5 & -0.25 x & 0.25 x & 0.5 & 0.25 x+0.5\\
-0.5 & 1.5 x+1.0 & 0.75 x & 0.75 x & 1.5 x & -0.5\\
-0.25 x & 0.75 x & 0.75 x+1.0 & 0.25 x & 0.75 x & -0.25 x\\
0.25 x & 0.75 x & 0.25 x & 0.75 x-1.0 & 0.75 x & 0.25 x\\
0.5 & 1.5 x & 0.75 x & 0.75 x & 1.5 x-1.0 & 0.5\\
0.25 x+0.5 & -0.5 & -0.25 x & 0.25 x & 0.5 & 0.25 x-1.5\end{bmatrix}
\right).$\end{minipage}
}
\label{eqn:StabPen}
\end{equation}
In this case, the matrix $A_1$ is PSD and the determinatal representation \eqref{eqn:StabPen} demonstrates that $p_{\ref{sec:AnoKin}}$ is stable.  Via Quarez's constuction, the polynomial $p_{\ref{sec:AnoKin}}$ has determinantal representation 
\begin{equation}
\resizebox{0.90\textwidth}{!}{
\begin{minipage}{1.35\textwidth}$\displaystyle \det(B_1x+B_0)=
\det\left(\begin{bmatrix}0.5 & -0.5 & -0.5 x & 0.5 x & 0.5 & 0.5\\
-0.5 & 1.75 x+1.0 & 0.25 x & 0.25 x & 1.75 x & -0.5\\
-0.5 x & 0.25 x & 1.0-0.1875 x & -0.1875 x & 0.25 x & -0.5 x\\
0.5 x & 0.25 x & -0.1875 x & -0.1875 x-1.0 & 0.25 x & 0.5 x\\
0.5 & 1.75 x & 0.25 x & 0.25 x & 1.75 x-1.0 & 0.5\\
0.5 & -0.5 & -0.5 x & 0.5 x & 0.5 & -1.5\end{bmatrix}\right)$.\end{minipage}
}
\label{eqn:QuarPen}
\end{equation}
A series of row operations and column operations on \eqref{eqn:StabPen} yields matrix
\[M=\begin{bmatrix}1 & 0.125 x & 0 & 0 & -0.125 x & 0\\
1 & 0.25 x+1 & 0 & 0 & -0.25 x & 0\\
0 & -0.25 x & 1 & 0 & 0.25 x & 0\\
0 & 0.25 x & 0 & 1 & -0.25 x & 0\\
-2 & -0.25 x-1 & 0 & 0 & 0.25 x+1 & 0\\
-1 & 0 & 0 & 0 & 0 & 1\end{bmatrix}\]
with $\det(M)=1$ and
\begin{equation}
\resizebox{0.90\textwidth}{!}{\begin{minipage}{1.05\textwidth}
$\displaystyle M(A_1x+A_0)M^{\mathrm{T}}=\begin{bmatrix}0.5 & 0.0 & 0.0 & 0.0 & 0.0 & 0.0\\
0.0 & 1.75 x+0.5 & 0.5 x & 1.0 x & 0.0 & 0.0\\
0.0 & 0.5 x & 0.75 x+1.0 & 0.25 x & 0.0 & 0.0\\
0.0 & 1.0 x & 0.25 x & 0.75 x-1.0 & 0.0 & 0.0\\
0.0 & 0.0 & 0.0 & 0.0 & -2.0 & 0.0\\
0.0 & 0.0 & 0.0 & 0.0 & 0.0 & -2.0
\end{bmatrix}$.
\end{minipage}}
\label{eqn:StabPenRed}
\end{equation}
Similarly, a series of row operations on \eqref{eqn:QuarPen} yields matrix
\[
N=
\begin{bmatrix}1 & 0 & 0 & 0 & 0 & 0\\
1 & 1 & 0 & 0 & 0 & 0\\
0 & -0.5 x & 1 & 0 & 0.5 x & 0\\
0 & 0.5 x & 0 & 1 & -0.5 x & 0\\
-2 & -1 & 0 & 0 & 1 & 0\\
-1 & 0 & 0 & 0 & 0 & 1\end{bmatrix}
\]
with $\det(N)=1$ and
\begin{equation}
\noindent\resizebox{0.90\textwidth}{!}{
$\displaystyle 
N(B_1x+B_0)N^{\mathrm{T}}=\begin{bmatrix}0.5 & 0.0 & 0.0 & 0.0 & 0.0 & 0.0\\
0.0 & 1.75 x+0.5 & -0.25 x & 0.75 x & 0.0 & 0.0\\
0.0 & -0.25 x & 1.0-0.1875 x & -0.1875 x & 0.0 & 0.0\\
0.0 & 0.75 x & -0.1875 x & -0.1875 x-1.0 & 0.0 & 0.0\\
0.0 & 0.0 & 0.0 & 0.0 & -2.0 & 0.0\\
0.0 & 0.0 & 0.0 & 0.0 & 0.0 & -2.0\end{bmatrix}.$}
\label{eqn:QuarPenRed}
\end{equation}
We were not successful in transforming the matrix pencil in \eqref{eqn:QuarPenRed} to \eqref{eqn:StabPenRed}.  An open question is whether there is a general purpose way to transform a sized reduced matrix pencil from Quarez's consturction such as \eqref{eqn:QuarPenRed} to one that demonstrates stability like \eqref{eqn:StabPenRed}.

\subsection{\protect $LDL^\mathrm{T}$ factorization of pencils}

Real symmetric matrices admit an $LDL^\mathrm{T}$ factorization where $L$ is unit lower triangular (It can be found using elementary row operations).  Here, we investigate the $LDL^\mathrm{T}$ factorization of matrix pencils.  Matrix $D$ will contain a factorization of $p(x)=\det(D)$.  Some entries of $D$ may be rational functions.  The matrix $D$ in the example $LDL^\mathrm{T}$ factorizations is not helpful for determining the stability of the pencil.

For the first pencil with extra dimension,
\[P_1=\begin{bmatrix}5 x+1 & 11 x+3 & 0\\
11 x+3 & 25 x+10 & 0\\
0 & 0 & 1\end{bmatrix}\]
multiplication by 
\[M_1=\begin{bmatrix}1 & 0 & 0\\
\dfrac{-11 x-3}{5 x+1} & 1 & 0\\
0 & 0 & 1\end{bmatrix}\]
yields an upper triangular matrix
\[M_1P_1=\begin{bmatrix}5 x+1 & 11 x+3 & 0\\
0 & \dfrac{-\left( 11 x+3\right)^2 }{5 x+1}+25 x+10 & 0\\
0 & 0 & 1\end{bmatrix}.\]
Right mutliplication by the transpose of $M_1$ gives a diagonal matrix

\noindent\resizebox{0.98\textwidth}{!}{
$\displaystyle 
D_1=M_1P_1M_1^\mathrm{T}=\begin{bmatrix}5 x+1 & 0 & 0\\
0 & \dfrac{-\left( 11 x+3\right)^2 }{5 x+1}+25 x+10 & 0\\
0 & 0 & 1\end{bmatrix}
=\begin{bmatrix}5 x+1 & 0 & 0\\
0 & \dfrac{p(x)}{5 x+1} & 0\\
0 & 0 & 1\end{bmatrix}.
$}

\noindent Setting 
\[L_1=M_1^{-1}=\begin{bmatrix}1 & 0 & 0\\
-\dfrac{-11 x-3}{5 x+1} & 1 & 0\\
0 & 0 & 1\end{bmatrix},\]
(Note difference between $M_1$ and $M_1^{-1}$ is change of sign for one entry) $P=L_1DL_1^\mathrm{T}$.  The same process of Gaussian elimination for the other two pencils 
\[P_2=\begin{bmatrix}-1 & 0 & 0\\
0 & -9 x-1 & -2 x\\
0 & -2 x & 1\end{bmatrix}\]
and
\[P_3=\begin{bmatrix}-1 & 0 & 0\\
0 & 2 x/9+77/162 & 2/9\\
0 & 2/9 & -18 x-2\end{bmatrix}
\]
have factorizations

\noindent\resizebox{0.98\textwidth}{!}{
\begin{minipage}{1.1 \textwidth}
\begin{multline*}P_2=\begin{bmatrix}1 & 0 & 0\\
0 & 1 & 0\\
0 & -\dfrac{2 x}{-9 x-1} & 1\end{bmatrix}\begin{bmatrix}-1 & 0 & 0\\
0 & -9 x-1 & 0\\
0 & 0 & 1-\dfrac{4 {{x}^{2}}}{-9 x-1}\end{bmatrix}
\begin{bmatrix}1 & 0 & 0\\
0 & 1 & 0\\
0 & -\dfrac{2 x}{-9 x-1} & 1\end{bmatrix}^\mathrm{T}
\\
=M_2\begin{bmatrix}-1 & 0 & 0\\
0 & -9 x-1 & 0\\
0 & 0 & \dfrac{-p(x)}{-9 x-1}\end{bmatrix}
M_2^\mathrm{T}\end{multline*}
\end{minipage}}

\noindent and

\noindent\resizebox{0.98\textwidth}{!}{
$\displaystyle P_3=\begin{bmatrix}1 & 0 & 0\\
0 & 1 & 0\\
0 & -\dfrac{2}{9 \left( \dfrac{2 x}{9}+\dfrac{77}{162}\right) } & 1\end{bmatrix}
\begin{bmatrix}-1 & 0 & 0\\
0 & \dfrac{36 x+77}{162} & 0\\
0 & 0 & -\dfrac{162 \left( 4 {{x}^{2}}+9 x+1\right) }{36 x+77}\end{bmatrix}
\begin{bmatrix}1 & 0 & 0\\
0 & 1 & 0\\
0 & -\dfrac{2}{9 \left( \dfrac{2 x}{9}+\dfrac{77}{162}\right) } & 1\end{bmatrix}^\mathrm{T}
$}

There are a few interesting things to note here.  We could not use $P_2$ nor $P_3$ to determine the stability of $p$ and for these note that $D$ has the form
\[\begin{bmatrix}
-1 & 0 & 0\\
0 & q & 0 \\
0 & 0 & -p(x)/q
\end{bmatrix}.\]
However, maybe the linear pencil $P_1$ could determine stability if $D$ had form
\begin{equation}
\begin{bmatrix}
q & 0 & 0 \\
0 & p(x)/q & 0\\
0 & 0 & 1\\
\end{bmatrix}.
\label{DForm}
\end{equation}
Might all such pencils that give the stability of $p$ have a form like this?  A first step towards this line of investigation will be to check if the other pencil \eqref{RelPen} admits this form for $D$.  Before this, another thing to notice is that
\begin{equation}
\begin{bmatrix}
a & -b\\
-b & c
\end{bmatrix}=
\begin{bmatrix}
1 & 0\\
\dfrac{-b}{a} & 1
\end{bmatrix}
\begin{bmatrix}
a & 0\\
0 & c-\dfrac{b^2}{a}
\end{bmatrix}
\begin{bmatrix}
1 & 0\\
\dfrac{-b}{a} & 1
\end{bmatrix}^\mathrm{T},
\label{UsefulTrans}
\end{equation}
and this fact can be exploited to make a pencil for $ca-b^2$.

For example,
\[p(x)=4 {{x}^{2}}+9 x+1=4x^2+4x+1+5x=5x-\frac{(2x+1)^2}{-1}\]
yields a pencil
\begin{equation}\begin{bmatrix}
-1 & 2x+1\\
2x+1 & 5x
\end{bmatrix}
\label{eqn:FirstConsPenEx}
\end{equation}
It may work to try a similar thing for higher degree/number of variables.  
The pencil in \eqref{eqn:FirstConsPenEx} won't give the stability of $p$, since the upper left entry is constant while another entry in the row is nonconstant.  For a second try,
\[p(x)=4x^2+9x+1=(4x+1)(x+2)-(1)^2\]
has pencil
\begin{equation}\begin{bmatrix}
4x+1 & 1\\
1 & x+2
\end{bmatrix}
\label{eqn:SecondConsPenEx}
\end{equation}
which does give that $p$ is stable.

To finish up we present the $LDL^\mathrm{T}$ factorization of 
\[P_4=\begin{bmatrix}5 x+1717/2500 & 11 x+111/100 & 0\\
11 x+111/100 & 25 x+13/4 & 0\\
0 & 0 & 1\end{bmatrix}.\]
It has the form in \eqref{DForm},

\noindent\resizebox{0.98\textwidth}{!}{
$\displaystyle P_4=
\begin{bmatrix}1 & 0 & 0\\
-\dfrac{-11 x-\dfrac{111}{100}}{5 x+\dfrac{1717}{2500}} & 1 & 0\\
0 & 0 & 1\end{bmatrix}
\begin{bmatrix}5 x+\dfrac{1717}{2500} & 0 & 0\\
0 & \dfrac{-\left( 11 x+\dfrac{111}{100}\right)^2 }{5 x+\dfrac{1717}{2500}}+25 x+\dfrac{13}{4} & 0\\
0 & 0 & 1\end{bmatrix}
\begin{bmatrix}1 & 0 & 0\\
-\dfrac{-11 x-\dfrac{111}{100}}{5 x+\dfrac{1717}{2500}} & 1 & 0\\
0 & 0 & 1\end{bmatrix}^\mathrm{T}
$.}

\section{Locating critical points using a Gr{\"o}bner basis}

In this section, we give a few examples of finding the critical points of multivariable polynomials.  The equations for finding the roots of partial derivatives of the polynomial form a system of multivariable polynomial equations.  One method of solving for the roots of univariable polynomials is to compute the eigenvalues of the polynomial's companion matrix.  This idea has an extension in finding solutions to systems of multivariable polynomial equations in higher dimensions---eigenvalues of the multiplication operator matrix are components of the roots for multivariable polynomials.    Throughout this section, we follow the technique outlined in \cite{Cox:1998}.

\noindent The following steps are used to compute the roots of multivariable polynomials:
\begin{enumerate}
\item Divide by leading coefficients to rewrite the system of polynomials as a monic system (leading coef. equals 1).
\item Write a basis for lower order terms.
\item Compute a multiplication operator for one variable, $x_i$ by applying it to each basis element and reducing the result modulo each polynomial in the system.  
\item Based on 2. form a matrix for the full multiplication operator.  
\item Find eigenvalues of the analogous companion matrix.  These are the coordinates of the roots.
\item Substitute each of the eigenvalues into the system to eliminate variable $x_i$.  Apply this algorithm again to each of these reduced systems to eliminate the remaining variables.
\end{enumerate}
Among the ideas explored in the project was to apply Raleigh quotient iteration to each multiplication operator in order to get a lower bound for the minimum distance between critical points and the origin.  

\subsection{Univariable example}

\begin{example}
\label{ex:UniVarSys}
Consider the polynomial 
\[p(x)=(x-2)(x-5)(x+9)=x^3+2x^2-53x+90.\]
Step 1: Already monic.  Step 2: A basis for lower order terms is $\{x^2,x,1\}$.  Step 3: Multiplication operator applied to basis.
\begin{align*}
M_x(1)&=x(1)=x\\
M_x(x)&=x(x)=x^2\\
M_x(x^2)&=x^3\equiv x^3-p(x) \equiv -2x^2+53x-90
\end{align*}
Step 4: Matrix based on multiplication operator.
\[M_x=
\begin{blockarray}{*{4}{c}}
& 1 & x & x^2 \\
\begin{block}{c[*{3}{c}]}
M_x(1) & 0 & 1 & 0\\
M_x(x) & 0 & 0 & 1\\
M_x(x^2) & -90 & 53 & -2\\
\end{block}
\end{blockarray}
\]
This is the companion matrix.  When $\lambda$ is a root of $p(x)$,
\[\left[\begin{array}{*{3}{c}}
0 & 1 & 0\\
0 & 0 & 1\\
-90 & 53 & -2
\end{array} \right]\left[\begin{array}{c}
1\\
\lambda\\
\lambda^2
\end{array}
\right]
=\left[\begin{array}{c}
\lambda\\
\lambda^2\\
-90+53\lambda-2\lambda^2
\end{array}\right]=\left[\begin{array}{c}
\lambda\\
\lambda^2\\
\lambda^3
\end{array}\right]=\lambda \left[\begin{array}{c}
1\\
\lambda\\
\lambda^2
\end{array}
\right]\]

\end{example}

\subsection{Multivariable example}

We give two multivariable examples.  The first uses the standard basis.  The second requires a Gr{\"o}bner basis.

\begin{example}
\label{ex:MultiVarSysNoGrob}

The function
\[f(x,y)=-4xy-\frac{1}{4}x^4-\frac{1}{4}y^4\]
has partial derivatives
\[f_x(x,y)=-4y-x^3\]
and
\[f_y(x,y)=-4x-y^3.\]
A system of equations whose solution gives the set of critical points of $f$ is
\[\left\{
\begin{array}{l}
-4y-x^3=0\\
-4x-y^3=0
\end{array}
\right..
\]
An equivalent system with monic polynomials is
\begin{equation}\left\{
\begin{array}{l}
x^3+4y=0\\
y^3+4x=0
\end{array}
\right..
\label{eqn:MonicMultiEqs}
\end{equation}
The leading monomials of the above are $x^3$ and $y^3$, respectively.

We introduce an ordering
\[\cdots>x^3>x^2y>xy^2>y^3>x^2>xy>y^2>x>y>1\]
Standard monomials not divisible by the leading monomials of \eqref{eqn:MonicMultiEqs} are
\[S=\{1,y,x,y^2,xy,x^2\}\]

Next, we compute the multiplication operator $M_x$ applied to $g$, where $g$ is a standard monomial not divisible by the leading monnomials of the system \eqref{eqn:MonicMultiEqs}:
\begin{align*}
M_x(x^2y^2)&=x^3y^2 \equiv x^3y^2-y^2(x^3+4y)\equiv-4y^3+4(y^3+4x)=16x\\
M_x(x^2y)&=x^3y \equiv x^3y-y(x^3+4y)=-4y^2\\
M_x(xy^2)&=x^2y^2\\
M_x(x^2)&=x^3 \equiv x^3-(x^3+4y)=-4y \numberthis{eqn:MThings} \\
M_x(xy)&=x^2y\\
M_x(y^2)&=xy^2\\
M_x(x)&=x^2\\
M_x(y)&=xy\\
M_x(1)&=x
\end{align*}
The next step is to construct the full multiplication operator matrix
\[M_x=
\begin{blockarray}{*{10}{c}}
& 1 & y & x & y^2 & xy & x^2 & xy^2 & x^2y & x^2y^2\\
\begin{block}{c[*{9}{c}]}
M_x(1) & 0 & 0 & 1 & 0 & 0 & 0 & 0 & 0 & 0\\
M_x(y) & 0 & 0 & 0 & 0 & 1 & 0 & 0 & 0 & 0\\
M_x(x) & 0 & 0 & 0 & 0 & 0 & 1 & 0 & 0 & 0\\
M_x(y^2) & 0 & 0 & 0 & 0 & 0 & 0 & 1 & 0 & 0\\
M_x(xy) & 0 & 0 & 0 & 0 & 0 & 0 & 0 & 1 & 0\\
M_x(x^2) & 0 & -4 & 0 & 0 & 0 & 0 & 0 & 0 & 0\\
M_x(xy^2) & 0 & 0 & 0 & 0 & 0 & 0 & 0 & 0 & 1\\
M_x(x^2y) & 0 & 0 & 0 & -4 & 0 & 0 & 0 & 0 & 0\\
M_x(x^2y^2) & 0 & 0 & 16 & 0 & 0 & 0 & 0 & 0 & 0\\
\end{block}
\end{blockarray}.\]
that captures the relationships \eqref{eqn:MThings}.  We verify the equivalences  \eqref{eqn:MThings} and matrix multiplication show $M_x$ is a mutliplication operator

\noindent\resizebox{0.98\textwidth}{!}{
$\displaystyle \left[\begin{array}{*{9}{c}}
0 & 0 & 1 & 0 & 0 & 0 & 0 & 0 & 0\\
0 & 0 & 0 & 1 & 0 & 0 & 0 & 0 & 0\\
0 & 0 & 0 & 0 & 1 & 0 & 0 & 0 & 0\\
0 & 0 & 0 & 0 & 0 & 1 & 0 & 0 & 0\\
0 & 0 & 0 & 0 & 0 & 0 & 1 & 0 & 0\\
0 & -4 & 0 & 0 & 0 & 0 & 0 & 0 & 0\\
0 & 0 & 0 & 0 & 0 & 0 & 0 & 0 & 1\\
0 & 0 & 0 & -4 & 0 & 0 & 0 & 0 & 0\\
0 & 0 & 16 & 0 & 0 & 0 & 0 & 0 & 0\\
\end{array}\right]
\left[
\begin{array}{c}
1 \\
y \\
x \\
y^2 \\
xy \\
x^2 \\
xy^2 \\
x^2y \\
x^2y^2\\
\end{array}
\right]=\left[ \begin{array}{{c}} x \\ xy \\ x^2 \\ xy^2 \\ x^2y \\ -4y \\ x^2y^2 \\ -4y^2  \\ 16x \end{array} \right]\equiv
\left[ \begin{array}{c} x \\ xy \\ x^2 \\ xy^2 \\ x^2y \\ x^3 \\ x^2y^2 \\ x^3y \\ x^3y^2 \end{array} \right]
=
x \left[ \begin{array}{c} 1 \\ y \\ x \\ y^2 \\ xy \\ x^2 \\ xy^2 \\ x^2y \\ x^2y^2 \end{array} \right].
$}

The eigenvalues of $M_x$ are $\pm 2$, $\pm 2i$, $-\sqrt{2}\pm i \sqrt{2}$, $\sqrt{2} \pm i \sqrt{2}$ and $0$.  These are the $x$-coordinates of the critical points of $f(x,y)$.  These can be substituted into \eqref{eqn:MonicMultiEqs} and the critical points can be found by using the same procedure used in Example \ref{eqn:MonicMultiEqs}.  For example when $x=\pm 2$, the system \eqref{eqn:MonicMultiEqs} reduces to two univariate systems

\[
x=2 \quad \left\{
\begin{array}{l}
8+4y=0\\
y^3+8=0
\end{array}
\right.
\qquad
\text{and}
\qquad x=-2 \quad 
\left\{
\begin{array}{l}
-8+4y=0\\
y^3-8=0
\end{array}
\right..
\]

In the case where there are more than two variables, the critical points can be found by repeating the steps in this example.  After finding $M_{x_1}$ and the $x_1$-coordinates of the critical values, each of these are substituted into the system and $M_{x_2}$ is found yielding $x_2$-coordinates of the critical values corresponding to the substituted $x_1$-coordinate.

This approach for root finding is only valid for special polynomials, but it gives the framework of a more general method.  The polynomials in the system must satisfy the following:
\begin{enumerate}
\item $S$ must be finite.
\item The lower degree terms must be a scalar times an element in $S$.
\end{enumerate}

\end{example}

This method does not work when there are terms in more than one variable with the same degree as the highest term.  However, solving a system of equations of Gr{\"o}bner basis polynomials including the partial derivatives removes the deficiencies of the method in Example \ref{ex:MultiVarSysNoGrob}.

\begin{example} 
\label{ex:MultiVarSysGrob}
In this example, we will find the critical points of the polynomial
\[f(x,y)=\frac{{{x}^{3}}}{3}-5 {{x}^{2}} y+3 x\, {{y}^{2}}+\frac{{{y}^{3}}}{3}-2 {{x}^{2}}+3 x y-5 {{y}^{2}}+7 x-10 y+3.\]
The system we solve is from a Gr{\"o}bner basis including $f_x$ and $f_y$:
\[\left\{\begin{array}{l}
f_1:=f_x=x^2-10xy+3y^2-4x+3y+7\\
f_2:=f_y=-5x^2+6xy+y^2+3x-10y-10\\
f_3=xy-16/44y^2-5/44y+17/44x-25/44\\
f_4=y^3-34068/10736 xy - 47411/10736 y^2 - 12321/10736 x  \\
\hphantom{f_4=}- 12321/10736 y- 8871/10736
\end{array}\right..\]
A basis of standard monomials not divisible by the leading monomials of the system is $S=\{1,y,x,y^2\}$.  Applying multiplication operator $M_x$ to each element of $S$ gives the following:
\begin{align*}
M_x(y^2)&=xy^2-yf_3-16/44f_4+17/44f_3=275/244 y^2+ 553/976 x   \\
&\hphantom{=}+ 2079/976 y+ 79/976\\
M_x(x)&=x^2-f_1-10f_3=7/11y^2+3/22x-41/22y-29/22\\
M_x(y)&=xy-f_3=16/44y^2-17/44x+5/44y+25/44
\end{align*}
Thus the multiplication operator is
\[ M_x=
\begin{blockarray}{*{5}{c}}
 & 1 & y & x & y^2\\
\begin{block}{c[*{4}{c}]}
M_x(1) & 0 &  0 & 1 & 0 \\
M_x(y) & 25/44 &  5/44 & -17/44 & 16/44\\
M_x(x) & -29/22 & -41/22 & 3/22 & 7/11 \\
M_x(y^2) & 79/976 & 2079/976 & 553/976 & 275/244\\
\end{block}
\end{blockarray}\]

The eigenvalues of $M_x$ are $0.45708406078832 \pm 1.07342808539298i$,\newline $1.67284387963102$, and $-1.20996282087980$, which are the $x$-coordinates of the critical points of $f$.

\end{example}

\section{Summary}
\label{sec:Summary}

In this study, we made progress towards solving the main research problem:
\begin{adjustwidth}{0.5in}{0.5in}
  \textit{For a given (EXPO) generating function $g:\mathbb{R}^{2n} \rightarrow \mathbb{R}$, closely approximate $g$ by a polynomial $\hat{g}$ such that the gradient of $\hat{g}$ is nonzero within an elliptical region $E\subset \mathbb{R}^{2n}$, which is as large as possible.  
  }
\end{adjustwidth}
The approach we choose is to construct the polynomial $\hat{g}:E \mapsto \mathbb{R}$ as a composition of mappings from $E$ to $\mathbb{H}^{2n+1}$ and a stable polynomial $\hat{p}:\mathbb{H}^{2n+1}\mapsto \mathbb{R}$.

The mapping from $B$ to $\mathbb{H}^{2n+1}$, given by equations \eqref{eqn:FirstMap1} and \eqref{eqn:FirstMap2} was shown to produce an approximating function with nonzero gradient when the conditions of Theorem \ref{thm:nonzerograd} are satisfied.  An approximating stable polynomial for $g(\mathbf{u})$ may be found following a few steps.  First, find a symmetric determinantal representation for $g(\mathbf{u})$ using Quarrez's construction.  Second, compute the eigendecomposition of each symmetric matrix in the representation and replace negative eigenvalues by zeros.  Third, the determinant of the resulting matrix pencil in the second step is the approximating stable polynomial.  

A drawback to this approximation by stable polynomials is that an inputted stable polynomial will not equal the outputted stable polynomial.  This can be observed in several tests done in Section \ref{sec:StabApproxQuarrez}.  The reason for this drawback is that symmetric matrices with different signed eigenvalues are inherent to Quarrez's construction.  In order to demonstrate stablity, all of these matrices multiplied by a variable must have same-signed eigenvalues.  Thus, we searched for new symmetric determinatal representations that demonstrate stability.

In our search for a new symmetric determinantal representation, we explored reducing the size of the matrix pencil in Quarez's construction.  Based on several examples, a size reduction of $2 {n+\lfloor d/2 \rfloor \choose n+1}$ seems possible.  A systematic description of this size reduction would yield the smallest sized general purpose construction for a symemtric determinantal representation than currently available in the literature \cite{grenet_symmetric_2011, quarez_symmetric_2012}.  

After attempts to reduce the size of the pencils did not yield determinatnal representations that demonstrate stability, we attempted to find a matrix pencil such as \eqref{eqn:StabPen} generated by our original idea for an approximating scheme.  If any matrix pencil from Quarrez's construction such as \eqref{eqn:QuarPen} could be transformed to one with the same form as \eqref{eqn:StabPen}, the resulting approximation scheme would keep some or possibly all input stable polynomials fixed.

Two future improvements that would lead to a more practical solution to the main problem are finding a mapping with less strict conditions than those of Theorem \ref{thm:nonzerograd} and more importantly finding a general purpose symmetric determinantal representation of polynomials that tests for stability.  A place to begin the search for this representation is deriving a transformation as mentioned in the preceding paragraph.  Hopefully, this can be accomplished through a better understanding of matrix pencil size reduction to create an efficient method for enlarging the dynamic aperture of particle accelerators.

\appendix

\section{Large sized determinantal representation reduction examples}
\label{app:LargeRed}

Two larger symmetric matrix pencils from Quarrez's formula are reduced in size to show that a reduction in size by at least ${n+\lfloor d/2 \rfloor \choose n}$ is possible for these cases.

\begin{example}
\label{ex:LargeRedN1D6}
This example shows that a degree six univariable polynomial has a symmetric determinantal representation that has size less than Quarez's construction by 13.  This is one more than the estimate $2 {1+\lfloor 6 /2 \rfloor \choose 1+1}=12$.

The polynomial $7 {{x}^{6}}-3 {{x}^{5}}+4 {{x}^{3}}-2 {{x}^{2}}+7 x-4$ has a size $N=2 {1+1+\lfloor 6/2 \rfloor \choose 1+1}=20$ symmetric determinantal representation below:

\renewcommand{\arraystretch}{2}
\noindent\resizebox{\textwidth}{!}{
$\left[\begin{array}{*{20}{c}}
0.5 & -0.5 & -0.5 x & 0 & 0 & 0 & 0 & 0 & 0 & 0 & 0 & 0 & 0 & 0 & 0 & 0 & 0 & 0.5 x & 0.5 & 0.5\\
-0.5 & 1 & 0 & -0.5 & -0.5 x & 0 & 0 & 0 & 0 & 0 & 0 & 0 & 0 & 0 & 0 & 0.5 x & 0.5 & 0 & 0 & -0.5\\
-0.5 x & 0 & 1 & 0 & 0 & -0.5 x & 0 & 0 & 0 & 0 & 0 & 0 & 0 & 0 & 0.5 x & 0 & 0 & 0 & 0 & -0.5 x\\
0 & -0.5 & 0 & 1 & 0 & 0 & -0.5 & -0.5 x & 0 & 0 & 0 & 0 & 0.5 x & 0.5 & 0 & 0 & 0 & 0 & -0.5 & 0\\
0 & -0.5 x & 0 & 0 & 1 & 0 & 0 & 0 & -0.5 x & 0 & 0 & 0.5 x & 0 & 0 & 0 & 0 & 0 & 0 & -0.5 x & 0\\
0 & 0 & -0.5 x & 0 & 0 & 1 & 0 & 0 & 0 & -0.5 x & 0.5 x & 0 & 0 & 0 & 0 & 0 & 0 & -0.5 x & 0 & 0\\
0 & 0 & 0 & -0.5 & 0 & 0 & 3.5 x-1.5 & -0.5 x & 1 x & 0 & 0 & 1 x & -0.5 x & 3.5 x-2.5 & 0 & 0 & -0.5 & 0 & 0 & 0\\
0 & 0 & 0 & -0.5 x & 0 & 0 & -0.5 x & 1 & 0 & -0.75 x & -0.75 x & 0 & 0 & -0.5 x & 0 & 0 & -0.5 x & 0 & 0 & 0\\
0 & 0 & 0 & 0 & -0.5 x & 0 & 1 x & 0 & 1 & 1.75 x & 1.75 x & 0 & 0 & 1 x & 0 & -0.5 x & 0 & 0 & 0 & 0\\
0 & 0 & 0 & 0 & 0 & -0.5 x & 0 & -0.75 x & 1.75 x & 1 & 0 & 1.75 x & -0.75 x & 0 & -0.5 x & 0 & 0 & 0 & 0 & 0\\
0 & 0 & 0 & 0 & 0 & 0.5 x & 0 & -0.75 x & 1.75 x & 0 & -1 & 1.75 x & -0.75 x & 0 & 0.5 x & 0 & 0 & 0 & 0 & 0\\
0 & 0 & 0 & 0 & 0.5 x & 0 & 1 x & 0 & 0 & 1.75 x & 1.75 x & -1 & 0 & 1 x & 0 & 0.5 x & 0 & 0 & 0 & 0\\
0 & 0 & 0 & 0.5 x & 0 & 0 & -0.5 x & 0 & 0 & -0.75 x & -0.75 x & 0 & -1 & -0.5 x & 0 & 0 & 0.5 x & 0 & 0 & 0\\
0 & 0 & 0 & 0.5 & 0 & 0 & 3.5 x-2.5 & -0.5 x & 1 x & 0 & 0 & 1 x & -0.5 x & 3.5 x-3.5 & 0 & 0 & 0.5 & 0 & 0 & 0\\
0 & 0 & 0.5 x & 0 & 0 & 0 & 0 & 0 & 0 & -0.5 x & 0.5 x & 0 & 0 & 0 & -1 & 0 & 0 & 0.5 x & 0 & 0\\
0 & 0.5 x & 0 & 0 & 0 & 0 & 0 & 0 & -0.5 x & 0 & 0 & 0.5 x & 0 & 0 & 0 & -1 & 0 & 0 & 0.5 x & 0\\
0 & 0.5 & 0 & 0 & 0 & 0 & -0.5 & -0.5 x & 0 & 0 & 0 & 0 & 0.5 x & 0.5 & 0 & 0 & -1 & 0 & 0.5 & 0\\
0.5 x & 0 & 0 & 0 & 0 & -0.5 x & 0 & 0 & 0 & 0 & 0 & 0 & 0 & 0 & 0.5 x & 0 & 0 & -1 & 0 & 0.5 x\\
0.5 & 0 & 0 & -0.5 & -0.5 x & 0 & 0 & 0 & 0 & 0 & 0 & 0 & 0 & 0 & 0 & 0.5 x & 0.5 & 0 & -1 & 0.5\\
0.5 & -0.5 & -0.5 x & 0 & 0 & 0 & 0 & 0 & 0 & 0 & 0 & 0 & 0 & 0 & 0 & 0 & 0 & 0.5 x & 0.5 & -1.5
\end{array}\right]$}

\noindent When multiplied by the matrix below on the left and the transpose of this matrix on the right, the result shows the size of the symmetric linear pencil may be reduced by 12.

\renewcommand{\arraystretch}{1.25}
\noindent\resizebox{\textwidth}{!}{
$\left[\begin{array}{*{20}{c}}
1 & 0 & 0 & 0 & 0 & 0 & 0 & 0 & 0 & 0 & 0 & 0 & 0 & 0 & 0 & 0 & 0 & 0 & 0 & -1\\
1 & 1.25 & 0 & 0 & 0 & 0 & 0 & 0 & 0 & 0 & 0 & 0 & 0 & 0 & 0 & 0 & 0 & 0 & -1.25 & 0\\
0 & 0 & 0 & -0.5 & 0 & 0 & 0 & 0 & 0 & 0 & 0 & 0 & 0 & 0 & 0 & 0 & 1.5 & 0 & 0 & 0\\
1 & 1 & 0 & 0.5 & 0 & 0 & 0 & 0 & 0 & 0 & 0 & 0 & 0 & 0 & 0 & 0 & 0.5 & 0 & 0 & 0\\
2.0 & 2.0 & 0 & 2.0 & 0 & 0 & 1 & 0 & 0 & 0 & 0 & 0 & 0 & -1 & 0 & 0 & 0 & 0 & 0 & 0\\
0 & 0 & -0.25 & 0.5 x & -0.75 & 0 & 0 & 0.5 & 0 & 0 & 0 & 0 & -0.5 & 0 & 0 & 0.25 & -0.5 x & -0.25 & 0 & 0\\
0 & -0.5 {{x}^{2}} & 0.5 x & 0 & -0.5 x & 1 & 0 & 0 & -0.5 & 0 & 0 & 0.5 & 0 & 0 & 0 & 0.5 x & 0 & -0.5 x & 0.5 {{x}^{2}} & 0\\
0 & 0 & 0 & 0 & 0 & 0 & -1 x & 1 & 0 & 0 & 0 & 0 & -1 & 1 x & 0 & 0 & 0 & 0 & 0 & 0\\
0 & 0 & 0 & -1 {{x}^{2}} & 1 x & 0 & 0 & -1 x & 1 & 0 & 0 & -1 & 1 x & 0 & 0 & -1 x & 1 {{x}^{2}} & 0 & 0 & 0\\
0 & -1 {{x}^{3}} & 1 {{x}^{2}} & 0 & -1 {{x}^{2}} & 1 x & 0 & 0 & -1 x & 1 & -1 & 1 x & 0 & 0 & -1 x & 1 {{x}^{2}} & 0 & -1 {{x}^{2}} & 1 {{x}^{3}} & 0\\
0 & 0.5 {{x}^{3}} & -0.5 {{x}^{2}} & 0 & 0.5 {{x}^{2}} & -0.5 x & 0 & 0 & 0.5 x & 0 & 1 & -0.5 x & 0 & 0 & 0.5 x & -0.5 {{x}^{2}} & 0 & 0.5 {{x}^{2}} & -0.5 {{x}^{3}} & 0\\
0 & 0 & 0 & 0 & 0 & 0.5 & 0 & 0 & 0.5 & 0 & 0 & 0.5 & 0 & 0 & 0.5 & 0 & 0 & 0 & 0 & 0\\
0 & 0 & 0.5 & 0 & 0.5 & 0 & 0 & 0.5 & 0 & 0 & 0 & 0 & 0.5 & 0 & 0 & 0.5 & 0 & 0.5 & 0 & 0\\
1 & 1 & 0 & 1 & 0 & 0 & 1 & 0 & 0 & 0 & 0 & 0 & 0 & 0 & 0 & 0 & 0 & 0 & 0 & 0\\
0 & 0.5 {{x}^{2}} & -0.5 x & 0 & 0.5 x & 0 & 0 & 0 & 0.5 & 0 & 0 & -0.5 & 0 & 0 & 1 & -0.5 x & 0 & 0.5 x & -0.5 {{x}^{2}} & 0\\
0 & 0 & 0.5 & 1 x & -0.5 & 0 & 0 & 1 & 0 & 0 & 0 & 0 & -1 & 0 & 0 & 1.5 & -1 x & 0.5 & 0 & 0\\
0 & 0 & 0 & -0.75 & 0 & 0 & 0 & 0 & 0 & 0 & 0 & 0 & 0 & 0 & 0 & 0 & 0.25 & 0 & 0 & 0\\
0 & 0.5 x & 0 & 0 & 0.5 & 0 & 0 & 0 & 0 & 0 & 0 & 0 & 0 & 0 & 0 & -0.5 & 0 & 1 & -0.5 x & 0\\
0 & -0.5 x & 1 & 0 & -0.5 & 0 & 0 & 0 & 0 & 0 & 0 & 0 & 0 & 0 & 0 & 0.5 & 0 & 0 & 0.5 x & 0\\
-0.5 & 0.375 & 0 & 0 & 0 & 0 & 0 & 0 & 0 & 0 & 0 & 0 & 0 & 0 & 0 & 0 & 0 & 0 & -0.375 & 0\end{array}\right]$}

\noindent Indeed, the result of this multiplication is

\noindent\resizebox{\textwidth}{!}{
$\left[\begin{array}{*{20}{c}}-2 & 0 & 0 & 0 & 0 & 0 & 0 & 0 & 0 & 0 & 0 & 0 & 0 & 0 & 0 & 0 & 0 & 0 & 0 & 0\\
0 & -2 & 0 & 0 & 0 & 0 & 0 & 0 & 0 & 0 & 0 & 0 & 0 & 0 & 0 & 0 & 0 & 0 & 0 & 0\\
0 & 0 & -2 & 0 & 0 & 0 & 0 & 0 & 0 & 0 & 0 & 0 & 0 & 0 & 0 & 0 & 0 & 0 & 0 & 0\\
0 & 0 & 0 & 0.5 & 0 & 0 & 0 & 0 & 0 & 0 & 0 & 0 & 0 & 0 & 0 & 0 & 0 & 0 & 0 & 0\\
0 & 0 & 0 & 0 & -2 & 0 & 0 & 0 & 0 & 0 & 0 & 0 & 0 & 0 & 0 & 0 & 0 & 0 & 0 & 0\\
0 & 0 & 0 & 0 & 0 & 0.5 & 0 & 0 & 0 & 0 & 0 & 0 & 0 & 0 & 0 & 0 & 0 & 0 & 0 & 0\\
0 & 0 & 0 & 0 & 0 & 0 & 1 & 0 & 0 & 0 & 0 & 0 & 0 & 0 & -1 & 0 & 0 & 0 & 0 & 0\\
0 & 0 & 0 & 0 & 0 & 0 & 0 & 0 & 0 & 0 & 0 & 0 & 1 & -1 x & 0 & 0 & 0 & 0 & 0 & 0\\
0 & 0 & 0 & 0 & 0 & 0 & 0 & 0 & 0 & 0 & 0 & 1 & -1 x & 0 & 0 & 0 & 0 & 0 & 0 & 0\\
0 & 0 & 0 & 0 & 0 & 0 & 0 & 0 & 0 & 0 & 1 & -1 x & 0 & 0 & 0 & 0 & 0 & 0 & 0 & 0\\
0 & 0 & 0 & 0 & 0 & 0 & 0 & 0 & 0 & 1 & -1 & 2.25 x & -0.75 x & 0 & 0 & 0 & 0 & 0 & 0 & 0\\
0 & 0 & 0 & 0 & 0 & 0 & 0 & 0 & 1 & -1 x & 2.25 x & 0 & 0 & 1 x & 0 & 0 & 0 & 0 & 0 & 0\\
0 & 0 & 0 & 0 & 0 & 0 & 0 & 1 & -1 x & 0 & -0.75 x & 0 & 0 & -0.5 x & 0 & 0 & 0 & 0 & 0 & 0\\
0 & 0 & 0 & 0 & 0 & 0 & 0 & -1 x & 0 & 0 & 0 & 1 x & -0.5 x & 3.5 x-2 & 0 & 0 & 0 & 0 & 0 & 0\\
0 & 0 & 0 & 0 & 0 & 0 & -1 & 0 & 0 & 0 & 0 & 0 & 0 & 0 & 0 & 0 & 0 & 0 & 0 & 0\\
0 & 0 & 0 & 0 & 0 & 0 & 0 & 0 & 0 & 0 & 0 & 0 & 0 & 0 & 0 & -2 & 0 & 0 & 0 & 0\\
0 & 0 & 0 & 0 & 0 & 0 & 0 & 0 & 0 & 0 & 0 & 0 & 0 & 0 & 0 & 0 & 0.5 & 0 & 0 & 0\\
0 & 0 & 0 & 0 & 0 & 0 & 0 & 0 & 0 & 0 & 0 & 0 & 0 & 0 & 0 & 0 & 0 & -1 & 0 & 0\\
0 & 0 & 0 & 0 & 0 & 0 & 0 & 0 & 0 & 0 & 0 & 0 & 0 & 0 & 0 & 0 & 0 & 0 & 1 & 0\\
0 & 0 & 0 & 0 & 0 & 0 & 0 & 0 & 0 & 0 & 0 & 0 & 0 & 0 & 0 & 0 & 0 & 0 & 0 & 0.5\end{array}\right]$,}

\noindent which has eleven diagonal entries in rows whose other entries are zeros.  The determinant of this symmetric linear pencil is $5 {{x}^{2}}\, {{z}^{2}}-8 z-2 y-x-5$.

\end{example}

\begin{example}
\label{ex:LargeRedN3D4}
This example shows that a degree four trivariable polynomial has a symmetric determinantal representation that has size less than Quarez's construction by 11.  This is one more than the estimate $2 {3+\lfloor 4 /2 \rfloor \choose 3+1}=10$.

The polynomial $5 {{x}^{2}}\, {{z}^{2}}-8 z-2 y-x-5$ has a size $N=2 {3+1+\lfloor 4/2 \rfloor \choose 3+1}=30$ symmetric determinantal representation below:

\renewcommand{\arraystretch}{1.5}
\noindent\resizebox{\textwidth}{!}{
$\left[\begin{array}{*{30}{c}}
0.5 & -0.5 & -0.5 z & -0.5 y & -0.5 x & 0 & 0 & 0 & 0 & 0 & 0 & 0 & 0 & 0 & 0 & 0 & 0 & 0 & 0 & 0 & 0 & 0 & 0 & 0 & 0 & 0.5 x & 0.5 y & 0.5 z & 0.5 & 0.5\\
-0.5 & 1 & 0 & 0 & 0 & -0.5 & -0.5 z & 0 & -0.5 y & 0 & 0 & -0.5 x & 0 & 0 & 0 & 0 & 0 & 0 & 0.5 x & 0 & 0 & 0.5 y & 0 & 0.5 z & 0.5 & 0 & 0 & 0 & 0 & -0.5\\
-0.5 z & 0 & 1 & 0 & 0 & 0 & 0 & -0.5 z & 0 & -0.5 y & 0 & 0 & -0.5 x & 0 & 0 & 0 & 0 & 0.5 x & 0 & 0 & 0.5 y & 0 & 0.5 z & 0 & 0 & 0 & 0 & 0 & 0 & -0.5 z\\
-0.5 y & 0 & 0 & 1 & 0 & 0 & 0 & 0 & 0 & 0 & -0.5 y & 0 & 0 & -0.5 x & 0 & 0 & 0.5 x & 0 & 0 & 0.5 y & 0 & 0 & 0 & 0 & 0 & 0 & 0 & 0 & 0 & -0.5 y\\
-0.5 x & 0 & 0 & 0 & 1 & 0 & 0 & 0 & 0 & 0 & 0 & 0 & 0 & 0 & -0.5 x & 0.5 x & 0 & 0 & 0 & 0 & 0 & 0 & 0 & 0 & 0 & 0 & 0 & 0 & 0 & -0.5 x\\
0 & -0.5 & 0 & 0 & 0 & 4 z+y+0.5 x+3 & 0 & 0 & 0 & 0 & 0 & 0 & 0 & 0 & 0 & 0 & 0 & 0 & 0 & 0 & 0 & 0 & 0 & 0 & 4 z+y+0.5 x+2 & 0 & 0 & 0 & -0.5 & 0\\
0 & -0.5 z & 0 & 0 & 0 & 0 & 1 & 0 & 0 & 0 & 0 & 0 & -1.25 x & 0 & 0 & 0 & 0 & -1.25 x & 0 & 0 & 0 & 0 & 0 & 0 & 0 & 0 & 0 & 0 & -0.5 z & 0\\
0 & 0 & -0.5 z & 0 & 0 & 0 & 0 & 1 & 0 & 0 & 0 & 0 & 0 & 0 & 0 & 0 & 0 & 0 & 0 & 0 & 0 & 0 & 0 & 0 & 0 & 0 & 0 & -0.5 z & 0 & 0\\
0 & -0.5 y & 0 & 0 & 0 & 0 & 0 & 0 & 1 & 0 & 0 & 0 & 0 & 0 & 0 & 0 & 0 & 0 & 0 & 0 & 0 & 0 & 0 & 0 & 0 & 0 & 0 & 0 & -0.5 y & 0\\
0 & 0 & -0.5 y & 0 & 0 & 0 & 0 & 0 & 0 & 1 & 0 & 0 & 0 & 0 & 0 & 0 & 0 & 0 & 0 & 0 & 0 & 0 & 0 & 0 & 0 & 0 & 0 & -0.5 y & 0 & 0\\
0 & 0 & 0 & -0.5 y & 0 & 0 & 0 & 0 & 0 & 0 & 1 & 0 & 0 & 0 & 0 & 0 & 0 & 0 & 0 & 0 & 0 & 0 & 0 & 0 & 0 & 0 & -0.5 y & 0 & 0 & 0\\
0 & -0.5 x & 0 & 0 & 0 & 0 & 0 & 0 & 0 & 0 & 0 & 1 & 0 & 0 & 0 & 0 & 0 & 0 & 0 & 0 & 0 & 0 & 0 & 0 & 0 & 0 & 0 & 0 & -0.5 x & 0\\
0 & 0 & -0.5 x & 0 & 0 & 0 & -1.25 x & 0 & 0 & 0 & 0 & 0 & 1 & 0 & 0 & 0 & 0 & 0 & 0 & 0 & 0 & 0 & 0 & -1.25 x & 0 & 0 & 0 & -0.5 x & 0 & 0\\
0 & 0 & 0 & -0.5 x & 0 & 0 & 0 & 0 & 0 & 0 & 0 & 0 & 0 & 1 & 0 & 0 & 0 & 0 & 0 & 0 & 0 & 0 & 0 & 0 & 0 & 0 & -0.5 x & 0 & 0 & 0\\
0 & 0 & 0 & 0 & -0.5 x & 0 & 0 & 0 & 0 & 0 & 0 & 0 & 0 & 0 & 1 & 0 & 0 & 0 & 0 & 0 & 0 & 0 & 0 & 0 & 0 & -0.5 x & 0 & 0 & 0 & 0\\
0 & 0 & 0 & 0 & 0.5 x & 0 & 0 & 0 & 0 & 0 & 0 & 0 & 0 & 0 & 0 & -1 & 0 & 0 & 0 & 0 & 0 & 0 & 0 & 0 & 0 & 0.5 x & 0 & 0 & 0 & 0\\
0 & 0 & 0 & 0.5 x & 0 & 0 & 0 & 0 & 0 & 0 & 0 & 0 & 0 & 0 & 0 & 0 & -1 & 0 & 0 & 0 & 0 & 0 & 0 & 0 & 0 & 0 & 0.5 x & 0 & 0 & 0\\
0 & 0 & 0.5 x & 0 & 0 & 0 & -1.25 x & 0 & 0 & 0 & 0 & 0 & 0 & 0 & 0 & 0 & 0 & -1 & 0 & 0 & 0 & 0 & 0 & -1.25 x & 0 & 0 & 0 & 0.5 x & 0 & 0\\
0 & 0.5 x & 0 & 0 & 0 & 0 & 0 & 0 & 0 & 0 & 0 & 0 & 0 & 0 & 0 & 0 & 0 & 0 & -1 & 0 & 0 & 0 & 0 & 0 & 0 & 0 & 0 & 0 & 0.5 x & 0\\
0 & 0 & 0 & 0.5 y & 0 & 0 & 0 & 0 & 0 & 0 & 0 & 0 & 0 & 0 & 0 & 0 & 0 & 0 & 0 & -1 & 0 & 0 & 0 & 0 & 0 & 0 & 0.5 y & 0 & 0 & 0\\
0 & 0 & 0.5 y & 0 & 0 & 0 & 0 & 0 & 0 & 0 & 0 & 0 & 0 & 0 & 0 & 0 & 0 & 0 & 0 & 0 & -1 & 0 & 0 & 0 & 0 & 0 & 0 & 0.5 y & 0 & 0\\
0 & 0.5 y & 0 & 0 & 0 & 0 & 0 & 0 & 0 & 0 & 0 & 0 & 0 & 0 & 0 & 0 & 0 & 0 & 0 & 0 & 0 & -1 & 0 & 0 & 0 & 0 & 0 & 0 & 0.5 y & 0\\
0 & 0 & 0.5 z & 0 & 0 & 0 & 0 & 0 & 0 & 0 & 0 & 0 & 0 & 0 & 0 & 0 & 0 & 0 & 0 & 0 & 0 & 0 & -1 & 0 & 0 & 0 & 0 & 0.5 z & 0 & 0\\
0 & 0.5 z & 0 & 0 & 0 & 0 & 0 & 0 & 0 & 0 & 0 & 0 & -1.25 x & 0 & 0 & 0 & 0 & -1.25 x & 0 & 0 & 0 & 0 & 0 & -1 & 0 & 0 & 0 & 0 & 0.5 z & 0\\
0 & 0.5 & 0 & 0 & 0 & 4 z+y+0.5 x+2 & 0 & 0 & 0 & 0 & 0 & 0 & 0 & 0 & 0 & 0 & 0 & 0 & 0 & 0 & 0 & 0 & 0 & 0 & 4 z+y+0.5 x+1 & 0 & 0 & 0 & 0.5 & 0\\
0.5 x & 0 & 0 & 0 & 0 & 0 & 0 & 0 & 0 & 0 & 0 & 0 & 0 & 0 & -0.5 x & 0.5 x & 0 & 0 & 0 & 0 & 0 & 0 & 0 & 0 & 0 & -1 & 0 & 0 & 0 & 0.5 x\\
0.5 y & 0 & 0 & 0 & 0 & 0 & 0 & 0 & 0 & 0 & -0.5 y & 0 & 0 & -0.5 x & 0 & 0 & 0.5 x & 0 & 0 & 0.5 y & 0 & 0 & 0 & 0 & 0 & 0 & -1 & 0 & 0 & 0.5 y\\
0.5 z & 0 & 0 & 0 & 0 & 0 & 0 & -0.5 z & 0 & -0.5 y & 0 & 0 & -0.5 x & 0 & 0 & 0 & 0 & 0.5 x & 0 & 0 & 0.5 y & 0 & 0.5 z & 0 & 0 & 0 & 0 & -1 & 0 & 0.5 z\\
0.5 & 0 & 0 & 0 & 0 & -0.5 & -0.5 z & 0 & -0.5 y & 0 & 0 & -0.5 x & 0 & 0 & 0 & 0 & 0 & 0 & 0.5 x & 0 & 0 & 0.5 y & 0 & 0.5 z & 0.5 & 0 & 0 & 0 & -1 & 0.5\\
0.5 & -0.5 & -0.5 z & -0.5 y & -0.5 x & 0 & 0 & 0 & 0 & 0 & 0 & 0 & 0 & 0 & 0 & 0 & 0 & 0 & 0 & 0 & 0 & 0 & 0 & 0 & 0 & 0.5 x & 0.5 y & 0.5 z & 0.5 & -1.5\end{array}\right]$}

\noindent When multiplied by the matrix below on the left and the transpose of this matrix on the right, the result shows the size of the symmetric linear pencil may be reduced by 11.
\begin{center}
\resizebox{0.8 \textwidth}{!}{
$\left[\begin{array}{*{30}{c}}
1 & 0 & 0 & 0 & 0 & 0 & 0 & 0 & 0 & 0 & 0 & 0 & 0 & 0 & 0 & 0 & 0 & 0 & 0 & 0 & 0 & 0 & 0 & 0 & 0 & 0 & 0 & 0 & 0 & -1\\
1 & 0.5 & 0 & 0 & 0.5 x & 0 & 0 & 0 & 0.5 y & 0 & 0 & 0.5 x & 0 & 0 & 0.5 {{x}^{2}} & 0.5 {{x}^{2}} & 0 & 0 & 0.5 x & 0 & 0 & 0.5 y & 0 & 0 & 0 & 0.5 x & 0 & 0 & 0.5 & 0\\
0 & 0 & 1 & 0 & 0 & 0 & 0 & 0 & 0 & 0 & 0 & 0 & 0 & 0 & 0 & 0 & 0 & 0 & 0 & 0 & 0 & 0 & 0 & 0 & 0 & 0 & 0 & 5.0 & 0 & 0\\
0 & 0 & 0 & 1 & 0 & 0 & 0 & 0 & 0 & 0 & 0 & 0 & 0 & 0 & 0 & 0 & 0 & 0 & 0 & 0 & 0 & 0 & 0 & 0 & 0 & 0 & 0 & 0 & 0 & 0\\
0 & -0.5 & 0 & 0 & 0 & 0 & 0 & 0 & 0 & 0 & 0 & 0 & 0 & 0 & 0 & 0 & 0 & 0 & 0 & 0 & 0 & 0 & 0 & 0 & 1 & 0 & 0 & 0 & 0.5 & 0\\
0 & 0.5 & 0 & 0 & 0 & 1 & 0 & 0 & 0 & 0 & 0 & 0 & 0 & 0 & 0 & 0 & 0 & 0 & 0 & 0 & 0 & 0 & 0 & 0 & 0 & 0 & 0 & 0 & -0.5 & 0\\
0 & 0.5 z & 0 & 0 & 0 & 0 & 1 & 0 & 0 & 0 & 0 & 0 & 0 & 0 & 0 & 0 & 0 & 0 & 0 & 0 & 0 & 0 & 0 & 0 & 0 & 0 & 0 & 0 & -0.5 z & 0\\
0 & 0 & 0 & 0 & 0 & 0 & 0 & 1 & 0 & 0 & 0 & 0 & 0 & 0 & 0 & 0 & 0 & 0 & 0 & 0 & 0 & 0 & 0 & 0 & 0 & 0 & 0 & 0 & 0 & 0\\
0 & 0 & 0 & 0 & 0 & 0 & 0 & 0 & 0 & 0 & 0 & 0 & 0 & 0 & 0 & 0 & 0 & 0 & 0 & 0 & 1 & 0 & 0 & 0 & 0 & 0 & 0 & 0 & 0 & 0\\
0 & 0 & 0 & 0 & 0 & 0 & 0 & 0 & 0 & 1 & 0 & 0 & 0 & 0 & 0 & 0 & 0 & 0 & 0 & 0 & 0 & 0 & 0 & 0 & 0 & 0 & 0 & 0 & 0 & 0\\
0 & 0 & 0 & 0 & 0 & 0 & 0 & 0 & 0 & 0 & 1 & 0 & 0 & 0 & 0 & 0 & 0 & 0 & 0 & 0 & 0 & 0 & 0 & 0 & 0 & 0 & 0 & 0 & 0 & 0\\
0 & -0.5 z & 0 & 0 & 0 & 0 & 0 & 0 & 0 & 0 & 0 & 0 & 0 & 0 & 0 & 0 & 0 & 0 & 0 & 0 & 0 & 0 & 0 & 1 & 0 & 0 & 0 & 0 & 0.5 z & 0\\
0 & 0 & 0 & 0 & 0 & 0 & 0 & 0 & 0 & 0 & 0 & 0 & 1 & 0 & 0 & 0 & 0 & 0 & 0 & 0 & 0 & 0 & 0 & 0 & 0 & 0 & 0 & 0 & 0 & 0\\
0 & 0 & 0 & 0 & 0 & 0 & 0 & 0 & 0 & 0 & 0 & 0 & 0 & 1 & 0 & 0 & 0 & 0 & 0 & 0 & 0 & 0 & 0 & 0 & 0 & 0 & 0 & 0 & 0 & 0\\
0 & 0 & 0 & 0 & 0 & 0 & 0 & 0 & 0 & 0 & 0 & 0 & 0 & 0 & 0 & 0 & 0 & 0 & 0 & 0 & 0 & 0 & 0 & 0 & 0 & 0 & 1 & 0 & 0 & 0\\
0 & 0 & 0 & 0 & 0 & 0 & 0 & 0 & 0 & 0 & 0 & 0 & 0 & 0 & 0 & 0 & 0 & 0 & 0 & 0 & 0 & 0 & 0 & 0 & 0 & 0 & 0 & 1 & 0 & 0\\
0 & 0 & 0 & 0 & 0 & 0 & 0 & 0 & 0 & 0 & 0 & 0 & 0 & 0 & 0 & 0 & 1 & 0 & 0 & 0 & 0 & 0 & 0 & 0 & 0 & 0 & 0 & 0 & 0 & 0\\
0 & 0 & 0 & 0 & 0 & 0 & 0 & 0 & 0 & 0 & 0 & 0 & 0 & 0 & 0 & 0 & 0 & 1 & 0 & 0 & 0 & 0 & 0 & 0 & 0 & 0 & 0 & 0 & 0 & 0\\
0 & 0 & 0 & 0 & 0 & 0 & 0 & 0 & 0 & 0 & 0 & 0 & 0 & 0 & 0 & 0 & 0 & 0 & 0 & 0 & 0 & 0 & 1 & 0 & 0 & 0 & 0 & 0 & 0 & 0\\
0 & 0 & 0 & 0 & 0 & 0 & 0 & 0 & 0 & 0 & 0 & 0 & 0 & 0 & 0 & 0 & 0 & 0 & 0 & 1 & 0 & 0 & 0 & 0 & 0 & 0 & 0 & 0 & 0 & 0\\
0 & 0.5 y & 0 & 0 & 0 & 0 & 0 & 0 & 1.5 & 0 & 0 & 0 & 0 & 0 & 0 & 0 & 0 & 0 & 0 & 0 & 0 & 0.5 & 0 & 0 & 0 & 0 & 0 & 0 & -0.5 y & 0\\
0 & -0.25 y & 0 & 0 & 0 & 0 & 0 & 0 & 0.25 & 0 & 0 & 0 & 0 & 0 & 0 & 0 & 0 & 0 & 0 & 0 & 0 & 0.75 & 0 & 0 & 0 & 0 & 0 & 0 & 0.25 y & 0\\
0 & -0.25 x & 0 & 0 & 0 & 0 & 0 & 0 & 0 & 0 & 0 & 0.25 & 0 & 0 & 0 & 0 & 0 & 0 & 0.75 & 0 & 0 & 0 & 0 & 0 & 0 & 0 & 0 & 0 & 0.25 x & 0\\
0 & 0.5 x & 0 & 0 & 0 & 0 & 0 & 0 & 0 & 0 & 0 & 1.5 & 0 & 0 & 0 & 0 & 0 & 0 & 0.5 & 0 & 0 & 0 & 0 & 0 & 0 & 0 & 0 & 0 & -0.5 x & 0\\
0 & 0 & 0 & 0 & -0.25 & 0 & 0 & 0 & 0 & 0 & 0 & 0 & 0 & 0 & -0.5 x & -0.5 x & 0 & 0 & 0 & 0 & 0 & 0 & 0 & 0 & 0 & -0.75 & 0 & 0 & 0 & 0\\
0 & 0 & 0 & 0 & 1.5 & 0 & 0 & 0 & 0 & 0 & 0 & 0 & 0 & 0 & 1 x & 1 x & 0 & 0 & 0 & 0 & 0 & 0 & 0 & 0 & 0 & 0.5 & 0 & 0 & 0 & 0\\
0 & 0 & 0 & 0 & 0 & 0 & 0 & 0 & 0 & 0 & 0 & 0 & 0 & 0 & 1.5 & 0.5 & 0 & 0 & 0 & 0 & 0 & 0 & 0 & 0 & 0 & 0 & 0 & 0 & 0 & 0\\
0 & 0 & 0 & 0 & 0 & 0 & 0 & 0 & 0 & 0 & 0 & 0 & 0 & 0 & 0.25 & 0.75 & 0 & 0 & 0 & 0 & 0 & 0 & 0 & 0 & 0 & 0 & 0 & 0 & 0 & 0\\
0 & -0.25 & 0 & 0 & 0 & 0 & 0 & 0 & 0 & 0 & 0 & 0 & 0 & 0 & 0 & 0 & 0 & 0 & 0 & 0 & 0 & 0 & 0 & 0 & 0 & 0 & 0 & 0 & 0.75 & 0\\
0 & 1.5 & 0 & 0 & 0 & 0 & 0 & 0 & 0 & 0 & 0 & 0 & 0 & 0 & 0 & 0 & 0 & 0 & 0 & 0 & 0 & 0 & 0 & 0 & 0 & 0 & 0 & 0 & -0.5 & 0
\end{array}\right]$}
\end{center}
\noindent Indeed, the result of the multiplication is

\renewcommand{\arraystretch}{2.1}
\noindent\resizebox{0.99 \textwidth}{!}{
$\left[\begin{array}{*{30}{c}}-2 & 0 & 0 & 0 & 0 & 0 & 0 & 0 & 0 & 0 & 0 & 0 & 0 & 0 & 0 & 0 & 0 & 0 & 0 & 0 & 0 & 0 & 0 & 0 & 0 & 0 & 0 & 0 & 0 & 0\\
0 & 0.5 & 2 z & -0.5 y & 0.5 & -0.5 & -0.5 z & 0 & 0 & 0 & 0 & 0.5 z & 0 & 0 & 0.5 y & 0.5 z & 0 & 0 & 0 & 0 & 0 & 0 & 0 & 0 & 0 & 0 & 0 & 0 & 0 & 0\\
0 & 2 z & -24.0 & 0 & 0 & 0 & 0 & -3 z & 3 y & -3 y & 0 & 0 & -3 x & 0 & 0 & -5.0 & 0 & 3 x & 3 z & 0 & 0 & 0 & 0 & 0 & 0 & 0 & 0 & 0 & 0 & 0\\
0 & -0.5 y & 0 & 1 & 0 & 0 & 0 & 0 & 0 & 0 & -0.5 y & 0 & 0 & -0.5 x & 0 & 0 & 0.5 x & 0 & 0 & 0.5 y & 0 & 0 & 0 & 0 & 0 & 0 & 0 & 0 & 0 & 0\\
0 & 0.5 & 0 & 0 & 4.0 z+1 y+0.5 x+1 & 4.0 z+1 y+0.5 x+2 & 0 & 0 & 0 & 0 & 0 & 0 & 0 & 0 & 0 & 0 & 0 & 0 & 0 & 0 & 0 & 0 & 0 & 0 & 0 & 0 & 0 & 0 & 0 & 0\\
0 & -0.5 & 0 & 0 & 4.0 z+1 y+0.5 x+2 & 4.0 z+1 y+0.5 x+3 & 0 & 0 & 0 & 0 & 0 & 0 & 0 & 0 & 0 & 0 & 0 & 0 & 0 & 0 & 0 & 0 & 0 & 0 & 0 & 0 & 0 & 0 & 0 & 0\\
0 & -0.5 z & 0 & 0 & 0 & 0 & 1 & 0 & 0 & 0 & 0 & 0 & -1.25 x & 0 & 0 & 0 & 0 & -1.25 x & 0 & 0 & 0 & 0 & 0 & 0 & 0 & 0 & 0 & 0 & 0 & 0\\
0 & 0 & -3 z & 0 & 0 & 0 & 0 & 1 & 0 & 0 & 0 & 0 & 0 & 0 & 0 & -0.5 z & 0 & 0 & 0 & 0 & 0 & 0 & 0 & 0 & 0 & 0 & 0 & 0 & 0 & 0\\
0 & 0 & 3 y & 0 & 0 & 0 & 0 & 0 & -1 & 0 & 0 & 0 & 0 & 0 & 0 & 0.5 y & 0 & 0 & 0 & 0 & 0 & 0 & 0 & 0 & 0 & 0 & 0 & 0 & 0 & 0\\
0 & 0 & -3 y & 0 & 0 & 0 & 0 & 0 & 0 & 1 & 0 & 0 & 0 & 0 & 0 & -0.5 y & 0 & 0 & 0 & 0 & 0 & 0 & 0 & 0 & 0 & 0 & 0 & 0 & 0 & 0\\
0 & 0 & 0 & -0.5 y & 0 & 0 & 0 & 0 & 0 & 0 & 1 & 0 & 0 & 0 & -0.5 y & 0 & 0 & 0 & 0 & 0 & 0 & 0 & 0 & 0 & 0 & 0 & 0 & 0 & 0 & 0\\
0 & 0.5 z & 0 & 0 & 0 & 0 & 0 & 0 & 0 & 0 & 0 & -1 & -1.25 x & 0 & 0 & 0 & 0 & -1.25 x & 0 & 0 & 0 & 0 & 0 & 0 & 0 & 0 & 0 & 0 & 0 & 0\\
0 & 0 & -3 x & 0 & 0 & 0 & -1.25 x & 0 & 0 & 0 & 0 & -1.25 x & 1 & 0 & 0 & -0.5 x & 0 & 0 & 0 & 0 & 0 & 0 & 0 & 0 & 0 & 0 & 0 & 0 & 0 & 0\\
0 & 0 & 0 & -0.5 x & 0 & 0 & 0 & 0 & 0 & 0 & 0 & 0 & 0 & 1 & -0.5 x & 0 & 0 & 0 & 0 & 0 & 0 & 0 & 0 & 0 & 0 & 0 & 0 & 0 & 0 & 0\\
0 & 0.5 y & 0 & 0 & 0 & 0 & 0 & 0 & 0 & 0 & -0.5 y & 0 & 0 & -0.5 x & -1 & 0 & 0.5 x & 0 & 0 & 0.5 y & 0 & 0 & 0 & 0 & 0 & 0 & 0 & 0 & 0 & 0\\
0 & 0.5 z & -5.0 & 0 & 0 & 0 & 0 & -0.5 z & 0.5 y & -0.5 y & 0 & 0 & -0.5 x & 0 & 0 & -1 & 0 & 0.5 x & 0.5 z & 0 & 0 & 0 & 0 & 0 & 0 & 0 & 0 & 0 & 0 & 0\\
0 & 0 & 0 & 0.5 x & 0 & 0 & 0 & 0 & 0 & 0 & 0 & 0 & 0 & 0 & 0.5 x & 0 & -1 & 0 & 0 & 0 & 0 & 0 & 0 & 0 & 0 & 0 & 0 & 0 & 0 & 0\\
0 & 0 & 3 x & 0 & 0 & 0 & -1.25 x & 0 & 0 & 0 & 0 & -1.25 x & 0 & 0 & 0 & 0.5 x & 0 & -1 & 0 & 0 & 0 & 0 & 0 & 0 & 0 & 0 & 0 & 0 & 0 & 0\\
0 & 0 & 3 z & 0 & 0 & 0 & 0 & 0 & 0 & 0 & 0 & 0 & 0 & 0 & 0 & 0.5 z & 0 & 0 & -1 & 0 & 0 & 0 & 0 & 0 & 0 & 0 & 0 & 0 & 0 & 0\\
0 & 0 & 0 & 0.5 y & 0 & 0 & 0 & 0 & 0 & 0 & 0 & 0 & 0 & 0 & 0.5 y & 0 & 0 & 0 & 0 & -1 & 0 & 0 & 0 & 0 & 0 & 0 & 0 & 0 & 0 & 0\\
0 & 0 & 0 & 0 & 0 & 0 & 0 & 0 & 0 & 0 & 0 & 0 & 0 & 0 & 0 & 0 & 0 & 0 & 0 & 0 & 2 & 0 & 0 & 0 & 0 & 0 & 0 & 0 & 0 & 0\\
0 & 0 & 0 & 0 & 0 & 0 & 0 & 0 & 0 & 0 & 0 & 0 & 0 & 0 & 0 & 0 & 0 & 0 & 0 & 0 & 0 & -0.5 & 0 & 0 & 0 & 0 & 0 & 0 & 0 & 0\\
0 & 0 & 0 & 0 & 0 & 0 & 0 & 0 & 0 & 0 & 0 & 0 & 0 & 0 & 0 & 0 & 0 & 0 & 0 & 0 & 0 & 0 & -0.5 & 0 & 0 & 0 & 0 & 0 & 0 & 0\\
0 & 0 & 0 & 0 & 0 & 0 & 0 & 0 & 0 & 0 & 0 & 0 & 0 & 0 & 0 & 0 & 0 & 0 & 0 & 0 & 0 & 0 & 0 & 2 & 0 & 0 & 0 & 0 & 0 & 0\\
0 & 0 & 0 & 0 & 0 & 0 & 0 & 0 & 0 & 0 & 0 & 0 & 0 & 0 & 0 & 0 & 0 & 0 & 0 & 0 & 0 & 0 & 0 & 0 & -0.5 & 0 & 0 & 0 & 0 & 0\\
0 & 0 & 0 & 0 & 0 & 0 & 0 & 0 & 0 & 0 & 0 & 0 & 0 & 0 & 0 & 0 & 0 & 0 & 0 & 0 & 0 & 0 & 0 & 0 & 0 & 2 & 0 & 0 & 0 & 0\\
0 & 0 & 0 & 0 & 0 & 0 & 0 & 0 & 0 & 0 & 0 & 0 & 0 & 0 & 0 & 0 & 0 & 0 & 0 & 0 & 0 & 0 & 0 & 0 & 0 & 0 & 2 & 0 & 0 & 0\\
0 & 0 & 0 & 0 & 0 & 0 & 0 & 0 & 0 & 0 & 0 & 0 & 0 & 0 & 0 & 0 & 0 & 0 & 0 & 0 & 0 & 0 & 0 & 0 & 0 & 0 & 0 & -0.5 & 0 & 0\\
0 & 0 & 0 & 0 & 0 & 0 & 0 & 0 & 0 & 0 & 0 & 0 & 0 & 0 & 0 & 0 & 0 & 0 & 0 & 0 & 0 & 0 & 0 & 0 & 0 & 0 & 0 & 0 & -0.5 & 0\\
0 & 0 & 0 & 0 & 0 & 0 & 0 & 0 & 0 & 0 & 0 & 0 & 0 & 0 & 0 & 0 & 0 & 0 & 0 & 0 & 0 & 0 & 0 & 0 & 0 & 0 & 0 & 0 & 0 & 2
\end{array}\right],$}
\renewcommand{\arraystretch}{1}

\noindent which has eleven diagonal entries in rows whose other entries are zeros.  The determinant of this symmetric linear pencil is $5 {{x}^{2}}\, {{z}^{2}}-8 z-2 y-x-5$.

\end{example}

\section{Transformation of ellipsoid}
\label{app:Ellipsoid}

\subsection{Ellipsoid transformation overview}

The goal is to transform an elliptical domain to one in $\mathbb{H}$.  This can be achieved by translating the elliptical domain's center to the origin, rotating the elliptical domain, 

\subsection{Rotate Ellipsoid}

The general form of an ellipsoid centered at $(0,0,0)$ is
\[\frac{x^2}{a^2}+\frac{y^2}{b^2}+\frac{z^2}{c^2}=1.\]
An example is $\frac{x^2}{16}+\frac{y^2}{1}+\frac{z^2}{4}=1$.  Its graph is below.

\hspace*{\fill}\includegraphics[scale=0.6]{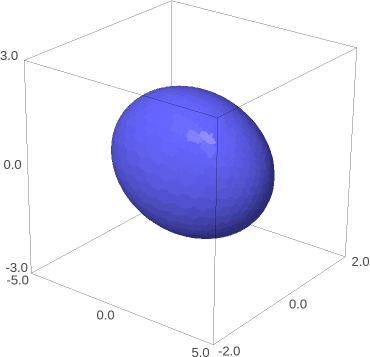}\hspace*{\fill}

We will rotate this using a series of rotation matrices.  $R_x$ is a rotation in the $yz$ plane about the $x$ axis.

\[R_x(\alpha) \mathbf{x}=\left[ \begin{array}{ccc}
1 & 0 & 0\\
0 & \cos(\alpha) & -\sin(\alpha)\\
0 & \sin(\alpha) & \cos(\alpha)\\
\end{array}
\right] \left[ \begin{array}{c}
x\\
y\\
z
\end{array}
\right]=\left[ \begin{array}{c}
x\\
y\cos(\alpha) - z \sin(\alpha)\\
y\sin(\alpha) + z \cos(\alpha)
\end{array}\right]
\]
For $x=0$ and $\alpha=45^\circ$, the cross section of the example ellipsoid is shown below.

\hspace*{\fill} \includegraphics[scale=1]{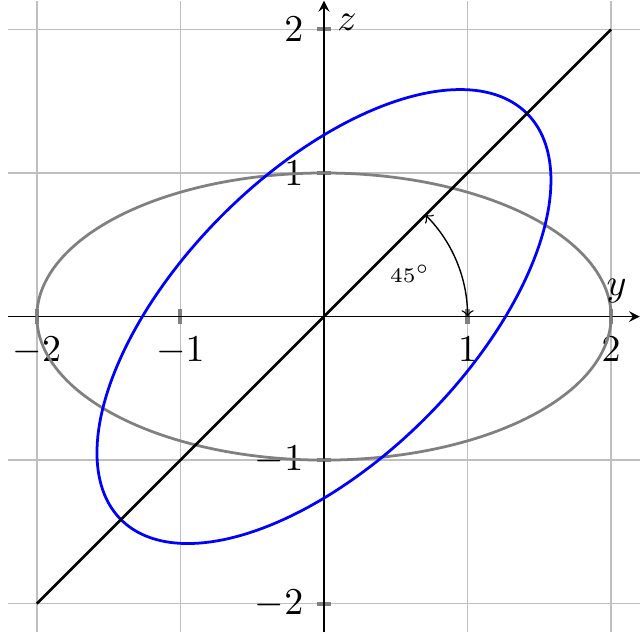} \hspace*{\fill}

Next we apply a rotation in the $xz$ plane about the $y$ axis.
\begin{multline*}R_y(\beta)R_x(\alpha) \mathbf{x}=\left[ \begin{array}{ccc}
\cos(\beta) & 0 & -\sin(\beta)\\
0 & 1 & 0\\
\sin(\beta) & 0 & \cos(\beta)\\
\end{array}
\right]
\left[ \begin{array}{c}
x\\
y\cos(\alpha) - z \sin(\alpha)\\
y\cos(\alpha) + z \sin(\alpha)
\end{array}\right]\\=\left[\begin{array}{c}
x\cos(\beta)-y\cos(\alpha)\sin(\beta)-z\sin(\alpha)\sin(\beta)\\
y\cos(\alpha)-z \sin(\alpha)\\
x\sin(\beta)+y\cos(\alpha)\cos(\beta)+z\sin(\alpha)\cos(\beta)
\end{array}\right]
\end{multline*}
For $y=0$, $\alpha=45^\circ$, and $\beta=120^\circ$, the cross section of the example ellipsoid is shown below.

\hspace*{\fill} \includegraphics[scale=1]{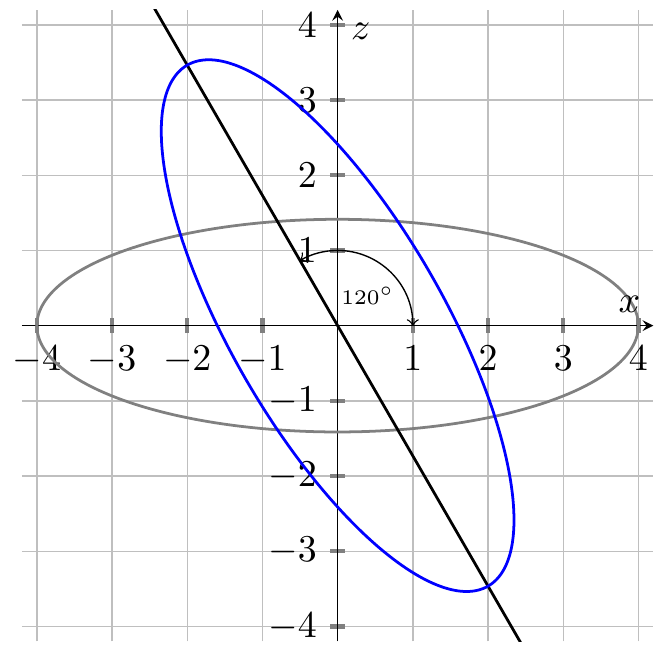} \hspace*{\fill}

Last, we apply a rotation in the $xy$ plane about the $z$ axis.  

\noindent\resizebox{0.98\textwidth}{!}{
$\displaystyle R_z(\gamma)R_y(\beta)R_x(\alpha) \mathbf{x}=\left[ \begin{array}{ccc}
\cos(\gamma) & -\sin(\gamma) & 0\\
\sin(\gamma) & \cos(\gamma) & 0\\
0 & 0 & 1\\
\end{array}
\right]\left[\begin{array}{c}
x\cos(\beta)-y\cos(\alpha)\sin(\beta)-z\sin(\alpha)\sin(\beta)\\
y\cos(\alpha)-z \sin(\alpha)\\
x\sin(\beta)+y\cos(\alpha)\cos(\beta)+z\sin(\alpha)\cos(\beta)
\end{array}\right]$}
\[
=\left[\begin{array}{c}
\begin{minipage}{0.9\textwidth}
\vspace*{-14pt}
\begin{multline*}
x\cos(\beta)\cos(\gamma)-y\cos(\alpha)\sin(\beta)\cos(\gamma)-z\sin(\alpha)\sin(\beta)\cos(\gamma)\\-y\cos(\alpha)\sin(\gamma)+z\sin(\alpha)\sin(\gamma)
\end{multline*}\end{minipage}\\
\\
\begin{minipage}{0.8\textwidth}\vspace*{-14pt}\begin{multline*}x\cos(\beta)\sin(\gamma)-y\cos(\alpha)\sin(\beta)\sin(\gamma)-z\sin(\alpha)\sin(\beta)\sin(\gamma)\\
+y\cos(\alpha)\cos(\gamma)-z\sin(\alpha)\cos(\gamma)\end{multline*}\end{minipage}\\
\\
x\sin(\beta)+y\cos(\alpha)\cos(\beta)+z\sin(\alpha)\cos(\beta)
\end{array}\right]
\]

For the example, take $z=0$ and $\gamma=210$.

\hspace*{\fill} \includegraphics[scale=1]{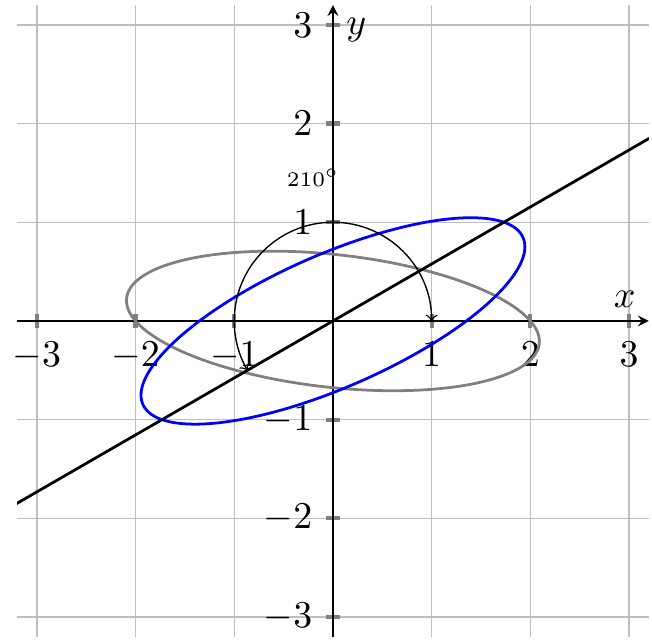} \hspace*{\fill}

Note that the inverse of each rotation is obtained by substituting the oposite angle ($R_x(\alpha)^{-1}=R_x(-\alpha)$).  After rotation, the center of the ellipsoid can be shifted to $(x_0,y_0,z_0)$ in the usual way.

\subsection{Linear algebra interpretation}

Another thing to observe about the rotation matrices is that their transpose is their inverse.  One interpretation for an ellipsoid is linear alebra equation
\[ \left[\begin{array}{ccc} x & y & z \end{array}\right]^\mathrm{T} \left[\begin{array}{ccc}
a^{-2} & 0 & 0\\
0 & b^{-2} & 0\\
0 & 0 & c^{-2}\\
\end{array}\right]
\left[ \begin{array}{c}
x\\
y\\
z
\end{array}\right]=1
\]
The rotated ellipsoid is given by

\noindent\resizebox{0.98\textwidth}{!}{
$\displaystyle \left[\begin{array}{ccc} x & y & z \end{array}\right] R_x(\alpha)^\mathrm{T} R_y(\beta)^\mathrm{T}R_q(\gamma)^\mathrm{T}\left[\begin{array}{ccc}
a^{-2} & 0 & 0\\
0 & b^{-2} & 0\\
0 & 0 & c^{-2}\\
\end{array}\right]R_z(\gamma)R_y(\beta)R_x(\alpha)
\left[ \begin{array}{c}
x\\
y\\
z
\end{array}\right]=1
$}

\noindent If we have
\[\left[\begin{array}{ccc} x & y & z \end{array}\right] A \left[\begin{array}{ccc} x & y & z \end{array}\right]^\mathrm{T}=1\]
We can use the SVD to obtain to obtain $a^{-2}$, $b^{-2}$, and $c^{-2}$ and the product $R_z(\gamma)R_y(\beta)R_x(\alpha)$.  An ellipse with center $\mathbf{v}$ would have equation
\[(\mathbf{x}-\mathbf{v})^\mathrm{T}A(\mathbf{x}-\mathbf{v})=1,\]
where $A$ is positive definite.

\printbibliography


\end{document}